\documentclass[fleqn,usenatbib,aastex,preprintnumbers]{mnras}

\usepackage{newtxtext,newtxmath}
\usepackage[T1]{fontenc}
\usepackage{amsmath}
\usepackage{booktabs}
\usepackage{xcolor}
\usepackage{hyperref}

\DeclareRobustCommand{\VAN}[3]{#2}
\let\VANthebibliography\thebibliography
\def\thebibliography{\DeclareRobustCommand{\VAN}[3]{##3}\VANthebibliography}

 \usepackage{comment}
 \usepackage{titlesec}
 \usepackage{caption}

\titleformat{\paragraph}
{\normalfont\normalsize\bfseries}{\theparagraph}{1em}{}
\titlespacing*{\paragraph}
{0pt}{3.25ex plus 1ex minus .2ex}{1.5ex plus .2ex}

\usepackage{graphicx}	
\usepackage{amsmath}	

\title[Optically Informed Searches of Supernova Neutrinos]{Optically Informed Searches of High-Energy Neutrinos from Interaction-Powered Supernovae}

\author[Pitik et al.]{
	Tetyana Pitik,$^{1,2}$\thanks{E-mail: tetyana.pitik@nbi.ku.dk}
	Irene Tamborra,$^{1}$\thanks{E-mail: tamborra@nbi.ku.dk}
	Massimiliano Lincetto,$^{3}$
	Anna Franckowiak $^{3}$
	\\
	$^{1}$Niels Bohr International Academy and DARK, Niels Bohr Institute, University of Copenhagen, Blegdamsvej 17, 2100 Copenhagen, Denmark\\
    $^{2}$ Dipartimento di Fisica e Geologia, Universit\`{a} di Perugia, I.N.F.N. Sezione di Perugia, Via Pascoli, 06123 Perugia, Italy\\
	$^{3}$Ruhr University Bochum, Faculty of Physics and Astronomy, Astronomical Institute (AIRUB),  Universit\"atsstra{\ss}e 150, 44801 Bochum, Germany}

\pubyear{2023}

\begin{document}
	\label{firstpage}
	\pagerange{\pageref{firstpage}--\pageref{lastpage}}
	\maketitle
	
	\begin{abstract}

The interaction between the ejecta of supernovae (SNe) of Type IIn and a dense circumstellar medium (CSM) can efficiently generate thermal UV/optical radiation and lead to the emission of neutrinos in the $1$-$10^{3}$~TeV range. We investigate the connection between the neutrino signal detectable at the IceCube Neutrino Observatory and the electromagnetic signal observable by optical wide-field, high-cadence surveys to outline the best strategy for upcoming follow-up searches. 
We outline a semi-analytical model that connects the optical lightcurve properties to the SN parameters and find that a large peak luminosity  (${L_{\rm{peak}}\gtrsim 10^{43}-10^{44}\,\mathrm{erg \,s^{-1}}}$) and an average rise time ($t_{\rm{rise}}\gtrsim 10-40$~days) are necessary for copious neutrino emission.
Nevertheless, the most promising $ L_{\rm{peak}}$ and $t_{\rm{rise}}$  can be obtained for SN configurations that are not optimal for neutrino emission. Such ambiguous correspondence between the optical lightcurve properties and the number of IceCube neutrino events implies that relying on optical observations only,  a range of expected neutrino events should be considered (e.g.~the expected number of neutrino events can vary up to two orders of magnitude for some among the brightest SNe IIn observed by the Zwicky Transient Facility up to now, SN 2020usa and SN 2020in). In addition, the peak in the high-energy neutrino curve should be expected a few $t_{\rm{rise}}$ after the peak in the optical lightcurve.
Our findings highlight that it is crucial to infer the SN  properties from multi-wavelength observations rather than focusing on the optical band only to enhance upcoming neutrino searches.
  
\end{abstract}
	

\section{Introduction}
\label{sec:introduction}
Astrophysical neutrinos with TeV--PeV energy  are routinely observed by the IceCube Neutrino Observatory~\citep{Halzen:2022pez,Ahlers:2018fkn,Abbasi:2020jmh}. 
While the sources of the observed neutrino flux are not yet known~\citep{Meszaros:2017fcs,Vitagliano:2019yzm}, a number of follow-up programs aims to link the observed neutrinos to their electromagnetic counterparts. 
In this context, the All-Sky Automated Survey for SuperNovae~\citep[ASAS-SN,][]{ 2014ApJ...788...48S,2017PASP..129j4502K} the Zwicky Transient Facility~\citep[ZTF,][]{2019PASP..131a8002B,2020PASP..132c8001D} and the Panoramic Survey Telescope and Rapid Response System 1~\citep[Pan-STARRS1,][]{2016arXiv161205560C} perform dedicated target-of-opportunity searches for  optical counterparts of neutrino events~\citep{Stein:2022rvc,Pan-STARRS:2019szg,2022MNRAS.516.2455N}, and vice versa the IceCube Neutrino Observatory looks for neutrinos in the direction of the sources discovered by optical surveys~\citep[see e.g.][]{IceCube:2023esf,IceCube:2020mzw}. 
The importance of such multi-messenger searches will be strengthened as large-scale transient facilities come online, such as the Rubin Observatory~\citep{LSST:2022kad}.

The   putative coincidence of the high-energy neutrino event IC200530A with the  candidate superluminous supernova (SLSN) AT2019fdr~\citep{Pitik:2021dyf}~\footnote{Note that the identification of the nature AT2019fdr is still under debate; it has been suggested that its properties might be compatible with the ones of a tidal distruption event~\citep{Reusch:2021ztx}.} makes  searches of high-energy neutrinos from    SNe  timely. 
SLSNe are $\mathcal{O}(10$--$100)$ times brighter than standard core-collapse SNe~\citep{Gal-Yam:2018out,Moriya:2018sig}, with kinetic energy sometimes larger than  $10^{51}$~erg~\citep{Rest:2009wb,Nicholl:2020mkh}. SLSNe are broadly divided into two different spectral types: the ones with hydrogen emission lines (SLSNe II) and those without (SLSNe I), see e.g.~\citep{Gal-Yam:2012ukv}. The majority of SLSNe II  displays strong and narrow hydrogen emission lines similar to those of the less luminous SNe IIn~\citep{Ofek:2006vt,Rest:2009wb,Smith:2008ez} and often dubbed SLSNe IIn. Type IIn SNe  are a sub-class of core-collapse SNe~\citep{Smith:2010vz,Gal-Yam:2006kfs}  characterized by bright and narrow Balmer lines of hydrogen in their spectra which persist for weeks to years after the explosion~\citep{1990MNRAS.244..269S,Filippenko:1997ub,Gal-Yam:2016yms}. 
Type IIn SNe  are expected to have a dense circumstellar material (CSM) surrounding the exploding star. 
The large luminosity of SNe IIn and the evidence of slowly moving material ahead of the ejecta indicate an efficient interaction of the ejecta with the CSM, which has long been considered  a major energy source of the observed optical radiation~\citep{Smith:2016dnb,2017hsn..book..843B}. Given the similarities of the spectral characteristics, SLSNe IIn are deemed  to be extreme cases of SNe IIn, albeit it is unclear  whether SLSNe IIn are just the most luminous  SNe IIn or they represent a separate population.

The collision between the expanding SN ejecta and the dense CSM gives rise to the forward shock, propagating in the dense SN environment, and the reverse shock moving backward in the SN ejecta. The plasma heated by the  forward shock radiates its energy thermally in the UV/X-ray band. Depending on the column density of the CSM, energetic photons can be reprocessed through photoelectric absorption and/or Compton scattering downwards into the visible waveband, producing the  observed optical lightcurve. Alongside the thermal population, a non-thermal distribution of protons and electrons can be created via diffusive shock acceleration.

Once accelerated, the relativistic protons  undergo inelastic hadronic collisions with the non-relativistic protons of the shocked CSM, possibly leading to copious production of high-energy neutrinos and gamma-rays~\citep{Murase:2010cu,Zirakashvili:2015mua,Petropoulou:2017ymv,Sarmah:2022vra}. While gamma-rays are absorbed and reprocessed to a large extent in the dense medium~\citep[see, e.g.,][]{Sarmah:2022vra}, neutrinos  stream freely and reach  Earth without absorption~\citep{Murase:2010cu,Katz:2011zx,Zirakashvili:2015mua,Cardillo:2015zda,Kheirandish:2022eox,Sarmah:2022vra,Sarmah:2023sds,Brose:2022mbx}. If detected, neutrinos with energies $\gtrsim 100$~TeV from an  interacting SN would represent a smoking gun of acceleration of cosmic rays  up to PeV energies~\citep{Bell:2013vxa,Blasi:2013rva,Cristofari:2020mdf,Cristofari:2021jkl}. 

In this paper,  we consider SNe IIn and SLSNe IIn as belonging to the same population, distinguished primarily by the ejecta energetics and CSM density. We investigate  how  neutrino production depends on the characteristic quantities describing  interaction-powered SNe and connect the main features of the optical lightcurve to the observable neutrino signal in order to optimize joint multi-messenger search strategies.

This work is organized as follows. Section~\ref{Sec:Model_description} outlines the SN model. As for the CSM structure, we mostly focus on the scenario involving  SN ejecta propagating in an extended envelope surrounding the progenitor with a wind-like density profile; we then extend our findings to the case involving SN ejecta propagating into a shell of CSM material with uniform density, which might result from a violent eruption shortly before the death of the star. In Sec.~\ref{Sec:Scaling_relations},  we introduce the scaling relations for the SN lightcurve properties.
Section~\ref{Sec: The maximum proton energy } focuses on investigating the dependence of the maximum proton energy on the SN model parameters. In Sec.~\ref{Sec:Production}, after introducing the method adopted to compute the neutrino spectral energy distribution, the dependence of the total energy emitted in neutrinos is investigated as a function of the SN model parameters. 
Section~\ref{sec:detection} outlines the detection prospects of neutrinos by relying on two benchmark  SLSNe IIn observed by ZTF and discusses the most promising strategies to detect neutrinos by relying on optical observations as well multi-messenger follow-up programs. Finally, our findings are summarized in Sec.~\ref{sec:conclusions}.
In addition, the dependence of the SN  lightcurve properties and maximum proton energy on the SN model parameters are discussed in Appendix~\ref{App:parameters} and~\ref{App:maxEp}, respectively.
Moreover, details on the constant density scenario are provided in Appendix~\ref{Appendix: Constant density case}.

\section{Model for interaction-powered supernovae}
\label{Sec:Model_description}

In this section, we present the theoretical framework of our work. First, we describe the CSM configurations. Then, we focus on  the modeling of the interaction between the SN ejecta and the CSM, leading to the observed electromagnetic radiation. We also outline the SN model parameters and the related uncertainty ranges adopted in this work.

\subsection{Modeling of the circumstellar medium}

\begin{figure*}
	\centering
	\includegraphics[width=0.8\textwidth]{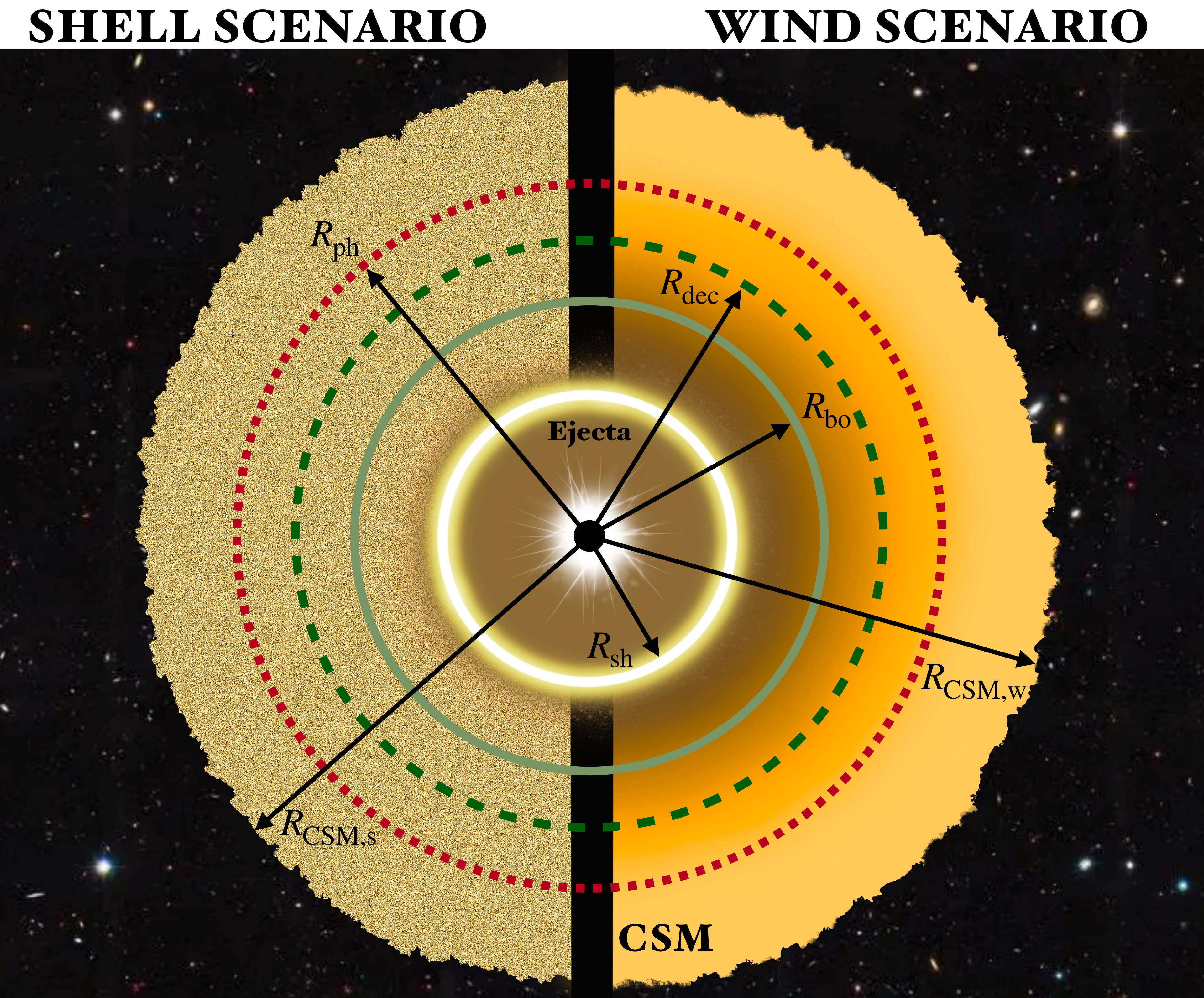}
	\caption{Schematic representation of an interaction-powered SN, under the assumption of spherical symmetry. The central compact object (in black) is surrounded by the SN ejecta (brown) and a compact shell extended up to  $R_{\rm{CSM,s}}$ ($R_{\rm{CSM,w}}$) from the center of explosion for the shell scenario on the left (and the wind scenario, on the right). For the wind density profile,  the color gradient tracks the density gradient (from darker to lighter hues as the density decreases). The region of interaction marked through the yellow-white circle represents the forward shock $R_{\rm{sh}}$ that propagates radially outwards. The solid olive line  marks the position of the breakout radius ($R_{\rm{bo}}$), where the first light leaks out, and the shock becomes collisionless. The dashed dark green line marks the location of the deceleration radius of the ejecta ($R_{\rm{dec}}$). The latter is located at radii smaller  than $R_{\rm{CSM}}$ (as in this sketch) for a relatively large CSM mass compared to the ejecta mass or radii larger than  $R_{\rm{CSM}}$ for  massive ejecta and rarefied CSM; note that  we could have $R_{\rm{dec}}<R_{\rm{bo}}$ for extremely large $M_{\rm{CSM}}/M_{\rm{ej}}$. The  dashed bordeaux line represents the photospheric radius $R_{\rm{ph}}$, where the radiation decouples from the CSM matter and stream  in the outer space freely.
 }
	\label{Fig: SN sketch}
\end{figure*}

Observational data and existing theoretical models indicate that the matter envelope surrounding massive stars could be spherical in shape or exhibit bipolar shells, disks or clumps, with non-trivial density profiles. This is the result of steady or eruptive mass loss episodes, as well as binary interactions of the progenitor prior to the explosion~\citep{Smith:2016dnb}. To this purpose, we consider two  CSM configurations: a uniform shell extended up to a radius $R_{\rm{CSM, s}}$ from the center of the explosion and a spherically symmetric shell with a wind radial profile extending smoothly from the progenitor surface up to an external radius ($R_{\rm{CSM,w}}$), as sketched in Fig.~\ref{Fig: SN sketch}. Henceforth we name the former ``shell scenario'' (s) and the latter ``wind scenario'' (w).

We assume that  the CSM has a mass $M_{\rm{CSM}}$, radial extent $R_{\rm{CSM}}$, and it is spherically distributed around the SN with a density profile described by a power-law function of the radius:
\begin{equation}
\label{Eq:n_CSM}
    n_{\rm{CSM}}(R)=\frac{\rho_{\rm{CSM}}(R)}{m}=\frac{(3-s) M_{\rm{CSM}}}{4\pi R^{3}_{\rm{CSM}}}\bigg(\frac{R}{R_{\rm{CSM}}}\bigg)^{-s}\equiv BR^{-s}\ ,
\end{equation}
where $ m=\mu m_{\rm{H}}, $ with $\mu= 1.3$~\citep{2019arXiv191200844L} being the mean molecular weight for a neutral gas of solar abundance. We neglect the density dependence on the inner radius of the CSM and consider it to  be the same as the progenitor radius $R_{\star}=10^{13}\,\mathrm{cm} \ll R_{\rm{CSM}}$. The case $s=2$ represents the stellar wind scenario, whereas $s=0$ denotes a shell of uniform density. We assume that the density external to the CSM shell ($R>R_{\rm{CSM}}$) is much smaller than the one at $R<R_{\rm{CSM}}$.

\subsection{Shock dynamics}
\label{Sec: Interaction model}

After the SN explodes, and the shock wave  passes through the stellar layers, the ejecta gas evolves to free homologous expansion. Relying  on numerical simulations~\citep[e.g.,][]{Matzner:1998mg}, we assume that during this phase the outer part of the SN ejecta has a power-law density profile~\citep{1994ApJ...420..268C,Moriya:2013hka}:
\begin{equation}
	\rho_{\rm{ej}}(R,t) = g_{n}t^{n-3}R^{-n}\ ,
\end{equation}
with
\begin{equation}
g_{n} = \frac{1}{4\pi (n-\delta)}\frac{[2(5-\delta)(n-5)E_{\rm{k}}]^{(n-3)/2}}{[(3-\delta)(n-3)M_{\rm{ej}}]^{(n-5)/2}}\ ,
\end{equation}
where $E_{\rm{k}}$ is the total SN kinetic energy, $M_{\rm{ej}}$ is the total mass of the SN ejecta, $n$ is the density slope of the outer part of the ejecta, and $\delta$ the slope of the inner one. The parameter $n$ depends on the progenitor properties and the nature  (convective or radiative) of the envelope;  $n\simeq 12$ is typical of red supergiant stars~\citep{Matzner:1998mg}, while lower values are expected for more compact progenitors. In this work, we adopt $n=10$ and $\delta=1$, following~\cite{Suzuki:2020qui}.

The interaction between the SN ejecta and the CSM results in a forward shock moving in the CSM and a reverse shock propagating back in the stellar envelope. For our purposes, only the forward shock is relevant. It is indeed estimated  that the contribution of the reverse shock to the electromagnetic emission, as well as its efficiency in accelerating particles during the timescales of interest, is significantly lower than the one of the forward shock~\citep{2007ApJ...661..879E, Patnaude:2008gq, 2010MNRAS.406.2633S, 2015ApJ...799..238S,2018ApJ...853...46S,Suzuki:2020qui,Zirakashvili:2015mua}. 

Following \cite{1982ApJ...259..302C,Moriya:2013hka}, we assume that the thickness of the shocked region
is much smaller than its radius, $R_{\rm{sh}}$. As long as the mass of the SN ejecta is larger than the swept-up CSM mass, which we define as the ejecta dominated phase (or free expansion phase), the expansion of the forward shock radius is described by~\citep{Moriya:2013hka}: 
\begin{equation}
	\label{Eq: Rsh in free phase}
	R_{\rm{sh}}(t)=R_{\star}+\bigg[\frac{(3-s)(4-s)}{(n-4)(n-3)}\frac{g_{n}}{B}\bigg]^{\frac{1}{n-s}}t^{\frac{n-3}{n-s}}\ ,
\end{equation}
with $B$ defined as in Eq.~\ref{Eq:n_CSM}, and hereafter we assume that the  interaction starts at $t=0$.

When the swept-up CSM mass becomes comparable to the SN ejecta mass, the ejecta start to slow down, entering the CSM dominated phase. This happens at the deceleration radius, defined as the radius $R_{\rm{dec}}$ at which $\int^{R_{\rm{dec}}}_{R_{\star}}4\pi R^{2}\rho_{\rm{CSM}}dR=M_{\rm{ej}}$, namely
\begin{equation}
	\label{eq:Rdec}
	R_{\rm{dec}} = \bigg[\frac{3-s}{4\pi B}M_{\rm{ej}} + R_{\star}^{3-s}\bigg]^{\frac{1}{3-s}}\ .
\end{equation}
According to the relative ratio between $M_{\rm{ej}}$ and $M_{\rm{CSM}}$, the deceleration can occur inside or outside the CSM shell (where a dilute  stellar wind  surrounds the collapsing star). After this transition,  the forward shock evolves as~\citep{Suzuki:2020qui}:
\begin{equation}
	\label{Eq: R_sh in dec phase}
	R_{\rm{sh}}(t) = R_{\rm{dec}} \bigg(\frac{t}{t_{\rm{dec}}}\bigg)^{\frac{2}{5-s}}\ .
\end{equation}
Differentiating Eqs.~\ref{Eq: Rsh in free phase} and \ref{Eq: R_sh in dec phase}, we obtain the forward shock velocity as a function of time:
\begin{equation}
	\label{Eq : v_sh}
	v_{\rm{sh}}(t)=\frac{dR_{\rm{sh}}(t)}{dt}=\begin{cases}
		\frac{n-3}{n-s}\bigg[\frac{(3-s)(4-s)}{(n-4)(n-3)}\frac{g_{n}}{B}\bigg]^{\frac{1}{n-s}}t^{\frac{s-3}{n-s}}\quad R< R_{\rm{dec}}\\
		\frac{2 }{5-s}R_{\rm{dec}}\big(\frac{t}{t_{\rm{dec}}}\big)^{\frac{s-3}{5-s}}\,\,\quad \quad \quad \quad  \quad\ \ R\geq R_{\rm{dec}}\ .
	\end{cases}
\end{equation}
We consider the dynamical evolution under the assumption that the shock is adiabatic for two reasons.  First, we want to compare our results  with the literature on the properties of the SN lightcurves extrapolated by relying on  semi-analytic models for the adiabatic expansion, see e.g.~\cite{Suzuki:2020qui}. 
Second, it has been shown that,  in the radiative regime,  $R_{\rm{sh}}$ has the same temporal dependence as the self-similar solution $\propto t^{{(n-3)}/{(n-s)}}$ in the free expansion phase with radiative losses having a strong impact on the evolution of the shock~\citep{Moriya:2013hka}. 

While the shock propagates in the CSM, the ejecta kinetic energy is dissipated in the interaction and converted into thermal energy. The shock-heated gas behind the forward shock front cools by emitting  thermal energy in the form of free-free radiation (thermal bremsstrahlung). However, if the CSM ahead of the shock is optically thick, such radiation is trapped and remains confined until the  shock breakout, which occurs at the breakout radius ($R_{\rm{bo}}$). The latter is computed  by solving the following equation for the Thomson optical depth (due to photon scattering on electrons)~\footnote{Note that we do not adopt the common approximation $R_{\rm{bo}}\equiv (\kappa_{\rm{es}} K v_{\rm{sh}})/{c}$, valid only when $R_{\rm{bo}}\ll R_{\rm{CSM}}$ and  $v_{\rm{sh}}$ independent on $R$~\citep{Chevalier:2011ha}.}:
\begin{equation}
	\label{Eq: R_BO}
	{\tau_{T}=\int_{R_{\rm{bo}}}^{R_{\rm{CSM}}} \rho_{\rm{CSM}}(R) \kappa_{\rm{es}} dR = \frac{c}{v_{\rm{sh}}}}\ ,
\end{equation}
where  $\kappa_{\rm{es}}$ is the electron scattering opacity, $c$ the speed of light, and $v_{\rm{sh}}$ is defined in Eq.~\ref{Eq : v_sh}. If $R_{\rm{bo}}\leq R_{\star}$,  $R_{\rm{bo}}=R_{\star}$.

We make use of the assumption of constant opacity, valid for electron Compton scattering. The value of $\kappa_{\rm{es}}$, which depends on the composition, typically ranges from $\kappa_{\rm{es}}\sim 0.2\,\mathrm{cm^{2} g^{-1}}$ for hydrogen-free matter to $\kappa_{\rm{es}} \sim 0.4\,\mathrm{cm^{2} g^{-1}}$ for pure hydrogen.
We consider solar composition of the CSM, namely $\kappa_{\rm{es}} = 0.2 (1 + X_{\rm{H}}) \simeq 0.34\, \rm{cm}^{2}~\rm{g}^{-1}$~\citep{1986rpa..book.....R}, where ${X_{\rm{H}}=0.73}$ is the hydrogen mass fraction~\citep{2019arXiv191200844L}.

As long as $\tau_{T}\gg c/v_{\rm{sh}}$, the shock is radiation-mediated (energy density of the radiation is larger than the energy density of the gas) and radiation pressure rather than plasma instabilities mediate the shock. In this regime,  non-thermal particle acceleration is inefficient, since a shock width much larger than the particle gyro-radius hinders standard Fermi acceleration~\citep{1976ApJS...32..233W,Levinson:2007rj,Katz:2011zx,Murase:2010cu}. Furthermore,  diffusion can be neglected. 
When $\tau_{T}< c/v_{\rm{sh}}$, the shock becomes collisionless,  and efficient particle acceleration begins.

\subsection{Interaction-powered supernova emission}
\label{Sec: Interaction-powered SN emission}

When the forward shock propagates in the region with $\tau_{T}<c/v_{\rm{sh}}$, the gas immediately behind the shock is heated to a temperature $T_{\rm{sh}}$. Assuming electron-ion equilibrium, such a temperature can be obtained by the Rankine–Hugoniot conditions:
\begin{equation}
	\label{Eq: T_sh}
	k_{B}T_{\rm{sh}}=2\frac{(\gamma-1)}{(\gamma+1)^{2}}\tilde{\mu} m_{\rm{H}} v^{2}_{\rm{sh}}\approx 118\,{\rm{keV}}\, \bigg(\frac{v_{\rm{sh}}}{10^{9}~ {\rm{cm~s}^{-1}}}\bigg)^{2}\ ,
\end{equation}
where $\gamma=5/3$ is the adiabatic index of the gas. We adopt  mean molecular weight $\tilde{\mu}=0.6$; such a choice is appropriate for  fully ionized CSM with solar composition as it is the case for the  matter right behind the shock (this is different from Eq.~\ref{Eq:n_CSM} where the CSM is assumed to be neutral).
The  thermal emission properties of the  shock-heated material can be fully characterized by the shock velocity $v_{\rm{sh}}$ and the other parameters characterizing the CSM~\citep{Margalit:2021bqe}. 

The observational signatures of the SN lightcurve and spectra depend on the radiative processes, which shape the thermal emission. The main photon production mechanism is free-free emission of the shocked electrons, whose typical timescale is~\citep{2011piim.book.....D}:
\begin{equation}
	\label{Eq: t_cool}
	t_{\rm{cool}}=\frac{3 k_{B} T_{\rm{sh}}}{2n_{\rm{sh}} \Lambda(T)}\ ,
\end{equation}
where $k_{B}$ is the Boltzmann constant, $n_{\rm{sh}}=4 n_{\rm{CSM}}$ is the density of the shocked region. The factor 4 comes from the Rankine–Hugoniot jump conditions across a strong non-relativistic shock. $\Lambda (T)$ is the cooling function (in units of $\rm{erg}\,\rm{cm}^{3}\,\rm{s}^{-1}$) that captures the physics of radiative cooling~\citep{1994ApJ...420..268C}:
\begin{equation}
\label{Eq: Lambda_cool}
    \Lambda(T)=\begin{cases}
    6.2\times 10^{-19} \, T^{-0.6}\quad 10^{5} < T \lesssim T_{\ast}\\
    2.5\times 10^{-27} \,T^{0.5}\,\,\,\,\,\quad\quad\quad T> T_{\ast}\ .
    \end{cases}
\end{equation}
The temperature $T_{\ast} = 4.7\times 10^{7}$~K represents the transition from the regime where free-free emission is dominant ($T\gtrsim T_{\ast}$) to the one where line-emission becomes relevant ($T \lesssim T_{\ast}$).
If the free-free cooling timescale is shorter than the dynamical time, the shock becomes radiative. In this regime, particles behind the shock cool within a layer of width $(t_{\rm{cool}}/t_{\rm{dyn}}) R_{\rm{sh}}$.

Although the radiation created during the interaction could  diffuse from the CSM, the presence of  dense pre-shock material causes the emitted photons to experience multiple scattering episodes before they reach the photosphere (defined as the surface where  $\tau_{T}=1$):
\begin{equation}
	\label{Eq: R_PH}
	R_{\rm{ph}}=\bigg[\frac{s-1}{\kappa_{\rm{es}}B}+R_{\rm{CSM}}^{1-s}\bigg]^{\frac{1}{1-s}}\ .
\end{equation}
The dominant mechanisms responsible for the photon field degradation in the medium are photoelectric absorption and Compton scattering, that generate inelastic energy transfer from photons to electrons during propagation. The result of such energy losses is that the bulk of thermal X-ray photons (see Eq.~\ref{Eq: T_sh}) is absorbed and reprocessed via continuum and line emission in the optical. This  phenomenon is strongly dependent on the CSM mass and extent, as well as on the stage of the shock evolution.

Alongside bremsstrahlung photons, a collisionless shock may produce non-thermal radiation from a relativistic population of electrons accelerated through diffusive shock acceleration. Synchrotron emission of these electrons is mainly expected  in the radio band;  it has been shown that the CSM mass and radius play an important role in defining the radio peak time and luminosity~\citep[see, e.g.,][]{Petropoulou:2016zar}.

\begin{table*}
	\caption{\label{Table: Parameters}Supernova model parameters for the SN wind and shell scenarios. The reference values adopted for our benchmark SN  model are provided together with the uncertainty range for the  most uncertain parameters. }
	\begin{center}
		\hspace{-0.9cm}
		\begin{tabular}[c]{cccc}
			\toprule
			\toprule
			Parameter & Symbol & Benchmark value & Parameter range\\
			\toprule
			Accelerated proton energy fraction & $\varepsilon_{\rm{p}}$ & $0.1$ & $-$\\
			Magnetic energy density fraction & $\varepsilon_{B}$& $3\times 10^{-4}$& $-$\\
			Proton spectral index & $k$ & $2$& $-$\\
			External ejecta density slope & $n$ & $10$ & $-$\\
			Internal ejecta density slope &$\delta$& $1$& $-$\\
			Kinetic energy &  $E_{\rm{k}}$& $ 10^{51}$~erg & $10^{50}$--$10^{53}$~erg  \\
			Ejecta mass &  $ M_{\rm{ej}}$& $ 10\ M_{\odot}$ & $1$--$70\ M_{\odot}$ \\
			CSM mass &$ M_{\rm{CSM}}$& $ 10\ M_{\odot}$ & $1$--$70\ M_{\odot}$\\
			CSM radius & $R_{\rm{CSM}}$ & $10^{16}$~cm& $5\times 10^{15}$--$10^{17}$~cm \\
   \toprule
		\end{tabular}
	\end{center}
	\label{table:SN_param}
\end{table*}
\subsection{Supernova model parameters}

The parameters characterizing SNe/SLSNe of Type IIn carry large uncertainties. For our benchmark SN model, we take into account uncertainty ranges  for the SN energetics, CSM and ejecta masses, as well as the CSM radial extent as summarized in Table~\ref{table:SN_param}. A number of other uncertainties  can significantly impact  the observational features, e.g.~the composition and geometry of the stellar environment or the stellar structure.

The electromagnetic emission of SLSNe IIn  can be explained  invoking a massive CSM shell with enough inertia to decelerate and dissipate most of the kinetic energy of the ejecta: $M_{\rm{CSM}}\gtrsim 40\,M_{\odot}$, $M_{\rm{ej}}\gtrsim 50\,M_{\odot}$, and $E_{\rm{k}}\gtrsim 10^{52}$~erg have been invoked for   SLSNe in the tail of the distribution~\citep[see e.g.][]{Nicholl:2020mkh,Drake:2011kg}, consistent with  pair-instability SN models. On the other hand, SNe IIn may result from the interaction with a less dense surrounding medium, or simply fall in the class of less powerful explosions, with  $M_{\rm{CSM}}\lesssim 5\,M_{\odot}$, $E_{\rm{k}}\sim$~a few $10^{51}$~erg, and $M_{\rm{ej}}\lesssim 50\,M_{\odot}$~\citep[see, e.g.][]{Chatzopoulos:2013vfa}. 

To encompass the wide range of  SN properties and the related uncertainties, we consider the space of parameters summarized in Table~\ref{Table: Parameters}. In the following, we   systematically investigate the dependence  of the lightcurve features, such as the rise time and the peak luminosity on the SN parameters. 
For the sake of completeness, we choose  generous uncertainty ranges, albeit most of the observed SN events do not require kinetic energies larger than $10^{52}$~erg or CSM masses larger than $50\ M_{\odot}$ for example.


\section{Scaling relations for the photometric supernova properties}
\label{Sec:Scaling_relations}
In this section, we introduce the scaling relations for the peak luminosity and the rise time of a SN lightcuve powered by shock interaction. 
Such relations connect these two observable quantities to the SN model parameters.  

\begin{figure}
	\centering
	\includegraphics[width=0.4\textwidth]{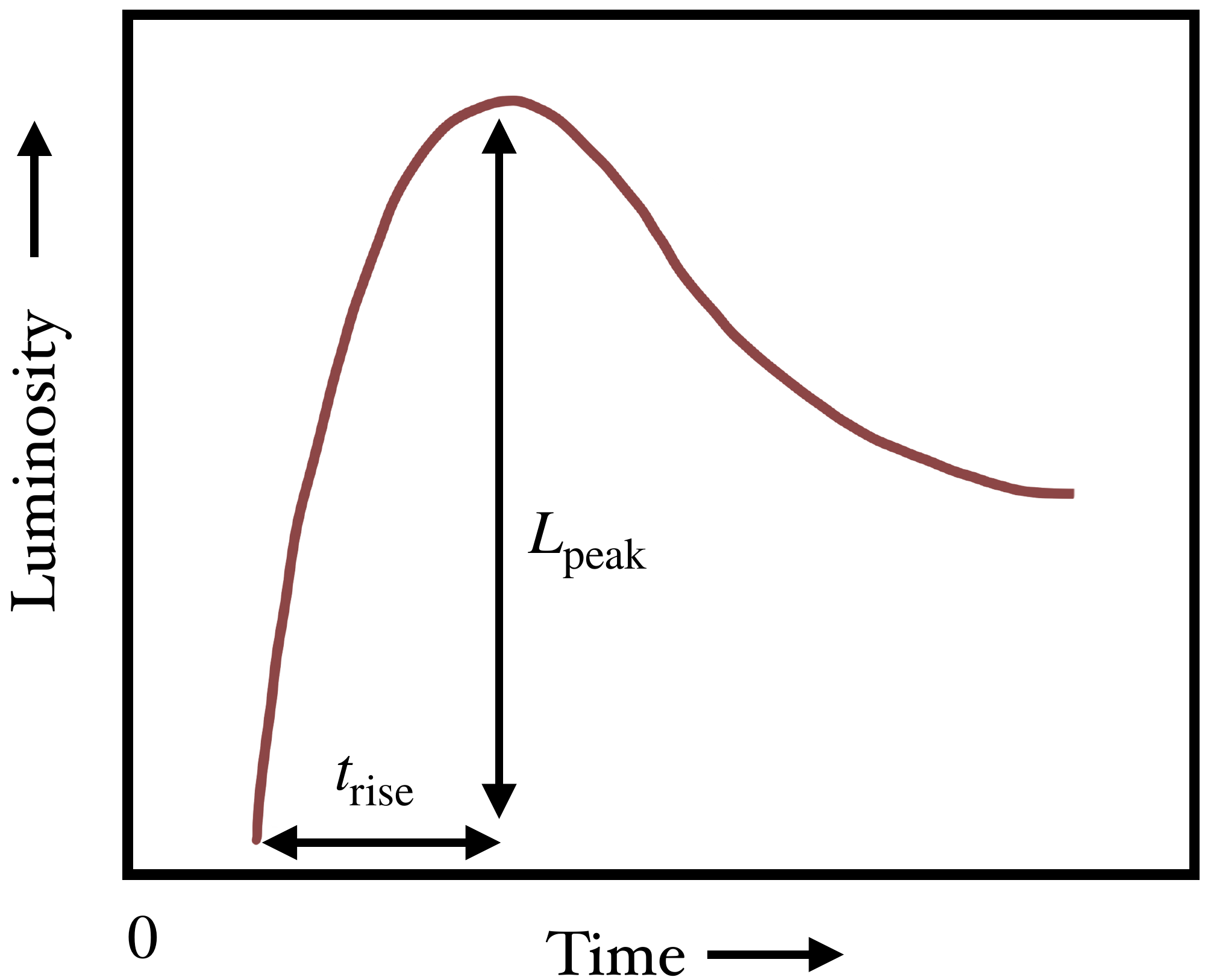}
	\caption{Sketch of the SN  luminosity evolution (in arbitrary units) resulting from  the interaction of the SN  shock with the dense CSM. The origin  ($t=0$) coincides with the SN explosion time. Note that $t_{\rm rise}$ is defined from the time of the shock breakout.
 }
	\label{Fig: LC_sketch}
\end{figure}

\begin{figure*}
	\centering
	\includegraphics[width=1\textwidth]{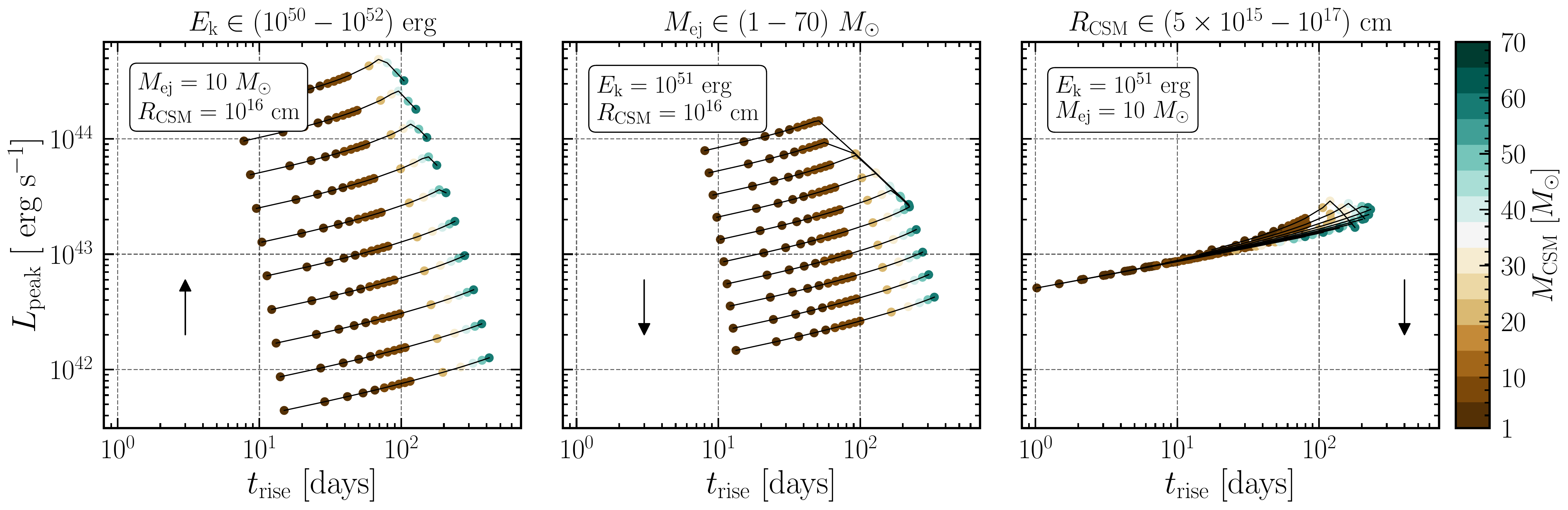}
	\caption{Bolometric peak luminosity as a function of the rise time, for fixed $M_{\rm{ej}}$ and $R_{\rm{CSM}}$ (left panel, and varying $E_{\rm k}$), fixed $E_{\rm k}$ and $R_{\rm{CSM}}$ (middle panel, and varying $M_{\rm{ej}}$), and fixed $E_{\rm k}$ and $M_{\rm{ej}}$ (right panel, and varying $R_{\rm{CSM}}$). In each panel, the arrow points in the direction of increasing values of the parameter indicated on top of the plot (e.g.~in the left panel, the highest curve is obtained with the  the largest kinetic energy, $10^{52}$~erg). For each curve, the color hues mark the variation of  $M_{\rm{CSM}}$. The longest rise times and brightest lightcurves are obtained for  large kinetic energies (left panel), the low ejecta mass (middle panel),  large CSM mass and  small CSM extension (right panel). Models with intermediate $t_{\rm{rise}}$ can reach the largest peak luminosities. The largest dispersion of long-lasting interaction-powered SNe can be achieved by increasing the kinetic energy. By keeping $E_{\rm{k}}$ fixed, an upper limit on $L_{\rm{peak}}$ is expected for different $M_{\rm{ej}}$ and $R_{\rm{CSM}}$.}
	\label{Fig:t_rise-L_peak}
\end{figure*}

We are interested in the shock evolution after shock breakout, when $\tau_{T}< c/v_{\rm{sh}}$. During this regime, the lightcurve is powered by  continuous conversion of the ejecta kinetic energy---see
e.g.~\cite{Chatzopoulos:2011vj,Ginzburg:2012dc,Moriya:2013hka}. Such a phase, however,  reproduces the decreasing-flat part of the SN lightcurve at later times (see Fig.~\ref{Fig: LC_sketch}), while the initial rising part of the optical signal can be explained   considering  photon diffusion in the optically thick region---see e.g.~\cite{Chevalier:2011ha,Chatzopoulos:2011vj}. 

Since we are interested in exploring a broad space of SN model parameters, we  rely on semi-analytical expressions for the characteristic quantities that describe  the optical lightcurve, namely the bolometric luminosity peak $L_{\rm{peak}}$ and the rise time to the peak $t_{\rm{rise}}$ (see Fig.~\ref{Fig: LC_sketch}). 
By performing 1D radiation-hydrodynamic simulations for a large region of the space of parameters, \cite{Suzuki:2020qui} fitted the output of their numerical simulations with  semi-analytical scaling relations, investigating the relation between $L_{\rm{peak}}$ and $t_{\rm{rise}}$. In this way, it is possible to analyze the dependence of the lightcurve properties on the parameters characterizing the SN interaction, i.e.~the kinetic energy  of the ejecta ($E_{\rm{k}}$),  the mass  of the ejecta ($M_{\rm{ej}}$), the mass  of the CSM ($M_{\rm{CSM}}$), and the extent  of the CSM ($R_{\rm{CSM}}$). \cite{Suzuki:2020qui}   found that the semi-analytical scaling relations describe relatively well the numerical results, once accounting for some calibration factors.
In this section, we review the scaling relations we adopt throughout our work.

As the forward shock propagates in the CSM, the post-shock thermal energy per unit radius coming from the dissipation of the kinetic energy is given by
\begin{equation}
\label{Eq: kinetic_E_per_radius}
    \mathcal{E}_{\rm{k}}(R)=\frac{dE_{\rm{k}}}{dR}=\frac{9}{8}\pi R^{2}v_{\rm{sh}}^{2}(R)\rho_{\rm{CSM}}(R)\ .
\end{equation}
where we have used the Rankine–Hugoniot jump conditions across a strong non-relativistic shock that provide a compression ratio $\simeq 4$. 

We define the bolometric peak luminosity as the kinetic power of the shock at  breakout:
\begin{equation}
\label{Eq: L_peak}
    L_{\rm{peak}}=\frac{dE_{\rm{k}}}{dt}\bigg|_{R_{\rm{bo}}}=\frac{9}{8}\pi R_{\rm{bo}}^{2}v_{\rm{sh}}^{3}(R_{\rm{bo}})\rho_{\rm{CSM}}(R_{\rm{bo}})\ .
\end{equation}
When the shock is still crossing the CSM envelope, the radiated photons undergo multiple scatterings before reaching the photosphere. The diffusion coefficient is $D(R)\sim c\lambda(R)$, with $\lambda(R)=1/\kappa_{\rm{es}}\rho_{\rm{CSM}}(R)$ being the photon mean free path. The time required to diffuse  from $R_{\rm{bo}}$ to the photosphere $R_{\rm{ph}}$ represents the rise time of the bolometric lightcurve~\citep{Ginzburg:2012dc}~\footnote{This definition of the rise time is valid as long as the CSM is dense enough to cause  shock breakout in the CSM wind. If this is not the case, the breakout occurs at the surface of the collapsing star; the  CSM masses responsible for this scenario are  not considered in our investigation.}:
\begin{equation}
	\label{Eq: t_rise}
	t_{\rm{rise}}\approx \int_{R_{\rm{\rm{bo}}}}^{R_{\rm{ph}}}\frac{d(R-R_{\rm{bo}})^{2}} {D(R)}\sim \int_{R_{\rm{\rm{bo}}}}^{R_{\rm{ph}}} \frac{2(R-R_{\rm{bo}})\kappa_{\rm{es}}\rho_{\rm{CSM}}(R)dR}{c}\ .
\end{equation}

Furthermore, 
after the forward shock breaks out from the optically thick part of the CSM at $R_{\rm{ph}}$, its luminosity is expected to be primarily emitted in the UV/X-ray region of the spectrum, and not in the optical~\citep{Ginzburg:2012dc}. Hence,  we consider the photospheric radius as the  radius beyond which the optical emission is negligible. 
Distinguishing the free-expansion regime (FE, $M_{\rm{ej}}\gg M_{\rm{CSM}}$) and the blast-wave regime (BW, $M_{\rm{ej}}\ll M_{\rm{CSM}}$)~\citep{Suzuki:2020qui}~\footnote{Note that this distinction should not be confused with the ejecta/CSM-dominated phases introduced in Sec.~\ref{Sec: Interaction model}.}, the kinetic energy dissipated during the shock evolution in the optically thick region is:
\begin{equation}
    E_{\rm{diss, thick}}=\begin{cases}
        \int_{R_{\star}}^{R_{\rm{ph}}} \mathcal{E}_{\rm{k}}(R) dR \quad \quad\quad \mathrm{for\,\, FE}\\
        \\
        \frac{(3-s)(\gamma+1)}{3+9\gamma-2s-2\gamma s}E_{\rm{k}} \,\, \quad\quad \  \mathrm{for\,\, BW}\ .
    \end{cases}
\end{equation}
Part of this energy is converted into thermal energy and radiated. The fraction radiated in the band of interest depends on multiple factors, including the cooling regime of the shock during the evolution, as well as the ionization state and CSM properties. We parametrize these unknowns by introducing the fraction $\varepsilon_{\rm{rad}}$ of the total dissipated energy $E_{\rm{diss, thick}}$ that is emitted in the optical band.
We note that we adopt a definition of the rise time which differs from the Arnett's rule employed in~\cite{Suzuki:2020qui}, leading to comparable results, except for extremely low values of $R_{\rm{CSM}}$ ($\sim 10^{15}$~cm), which we do not consider in this work. In Appendix~\ref{App:parameters} we provide illustrative examples of the dependence of $L_{\rm{peak}}$, $t_{\rm{rise}}$, and $E_{\rm{diss,thick}}$ on the  parameters characterizing the SN lightcurve for the wind CSM configuration ($s=2$).

Figure~\ref{Fig:t_rise-L_peak} shows $L_{\rm{peak}}$ as a function of $t_{\rm{rise}}$, obtained  by adopting the semi-analytic modeling in the FE and BW regimes.
We note that the largest dispersion in the peak luminosity for long-lasting SNe/SLSNe IIn is obtained by varying  the ejecta kinetic energy (left panel).
For fixed kinetic energy,  we see that the SN models corresponding to different ejecta mass (middle panel) all converge to approximately similar peak luminosity for longer $t_{\rm{rise}}$, which corresponds to  the region where the shock evolution is in the BW regime. This means that there is an upper limit on $L_{\rm{peak}}$ for a certain $t_{\rm{rise}}$, and the only way to overcome this limit is by increasing the ejecta energy. 
Changes in $R_{\rm{CSM}}$ (right panel) lead to  the smallest dispersion in $L_{\rm{peak}}$ among all the considered parameters. It is the variation of the kinetic energy that causes the largest spread in $L_{\rm{peak}}$.  Our findings are in agreement with the ones of~\cite{Suzuki:2020qui}.

\section{Maximum proton energy}
\label{Sec: The maximum proton energy }
In order to estimate the number of neutrinos  and their typical energy during the  shock evolution in the CSM, we first need to examine the energy gain and loss mechanisms that determine the maximum energy up to which protons can be accelerated.  We assume first-order Fermi acceleration, which takes place at the shock front with the accelerating particles gaining energy as they cross  the shock front back and forth. 
\begin{figure*}
	\centering
	\includegraphics[width=0.75\textwidth]{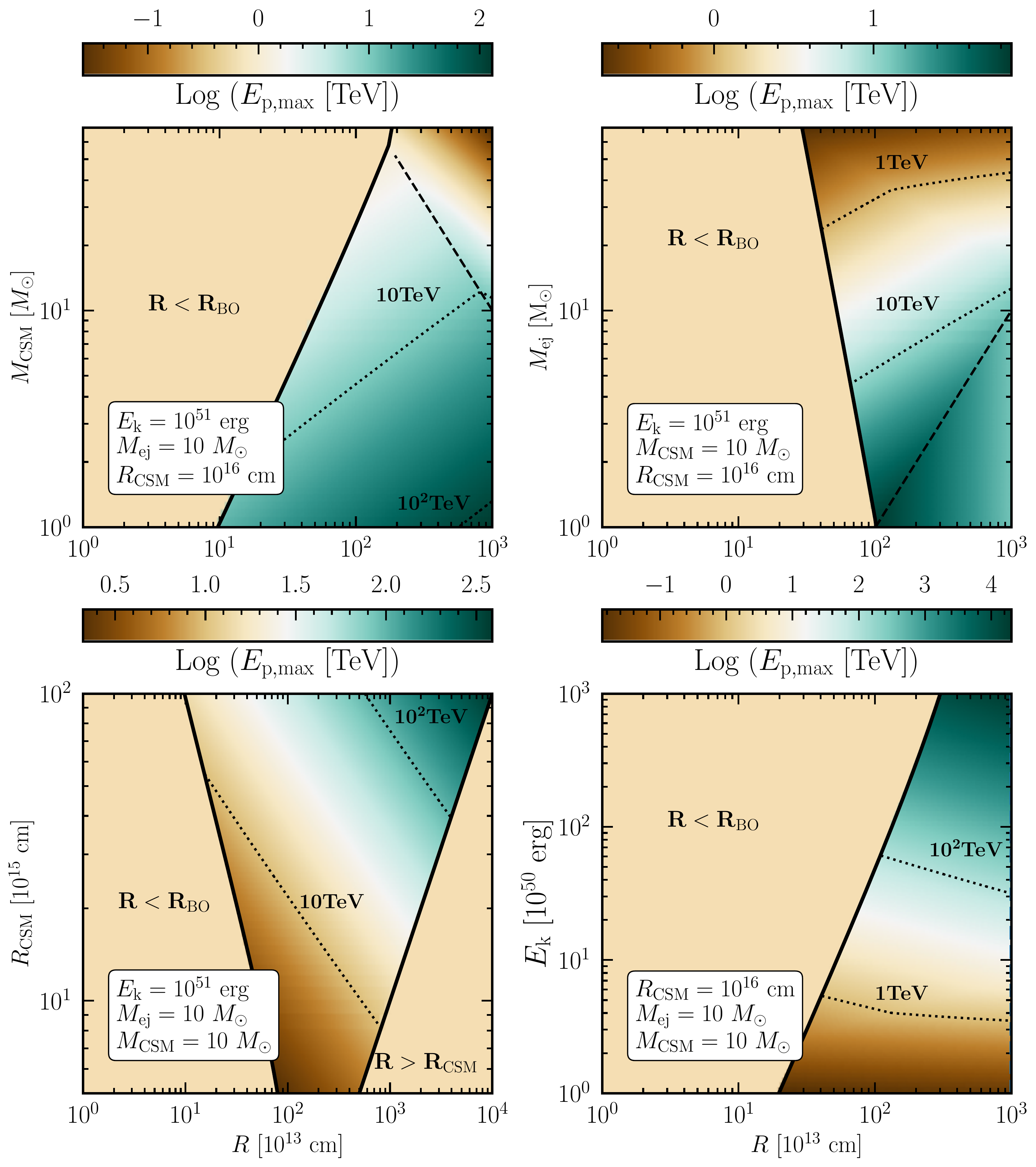}
	\caption{Contour plots of the maximum proton energy for the wind scenario in the plane spanned by the distance from the central engine and $M_{\rm{CSM}}$ (top left panel),  $M_{\rm{ej}}$ (top right panel), $R_{\rm{CSM}}$ (bottom left panel), and $E_{\rm{k}}$ (bottom right panel), while  the  remaining three SN model parameters are  fixed to their benchmark values. In each panel, the dashed line marks the deceleration radius, after which $E_{\rm{p,max}}$  decreases. The maximum proton energy increases with the radius (and therefore with time). Indeed, the largest $E_{\rm{p,max}}$  is obtained in the late stages of the shock evolution. Large $R_{\rm{CSM}}$ and $M_{\rm{ej}}$, and small $M_{\rm{CSM}}$ and $E_{\rm{k}}$ lead to the longest interaction times.  This statement is not true when $M_{\rm{ej}}\ll M_{\rm{CSM}}$ (see left upper and bottom righ panels). The black solid lines define the edges of the interaction region, $R_{\rm{bo}}\leq R\leq R_{\rm{CSM}}$.}
	\label{Fig: Ep_max_R_wind}
\end{figure*}

In the Bohm limit, where the proton mean free path is equal to its gyroradius ${r_{g}=\gamma_p m_{p} c^{2}/eB}$, the proton acceleration timescale is ${t_{\rm{acc}}\sim 6\gamma_{\rm{p}} m_{\rm{p}} c^{3}/eB v^{2}_{\rm{sh}}}$ \citep[see, e.g,]{Protheroe:2003vc,Tammi:2008vg,Caprioli:2013dca}, where $B=\sqrt{9\pi \varepsilon_{B} v_{\rm{sh}}^{2}\rho_{\rm{CSM}}}$ is the turbulent magnetic field in the post-shock region, whose energy density is assumed to be a fraction $\varepsilon_{B}$ of the post-shock thermal energy $ {U_{\rm{th}} = (9/8) v^{2}_{\rm{sh}} \rho_{\rm{CSM}} }$.  

The maximum energy up to which protons can be accelerated is determined by the competition between particle acceleration  and   energy loss mechanisms, such that  $ t_{\rm{acc}} \leq t_{\rm{p,cool}}$, with $t_{\rm{p,cool}}$ being the total proton cooling time. The relevant cooling times  are  the advection time ($t_{\rm{adv}}\sim \Delta R_{\rm{acc}}/v_{\rm{sh}}$, with $\Delta R_{\rm{acc}}$ being the width of the acceleration region) and the proton-proton interaction time (${ t_{\rm{pp}} = (4 k_{\rm{pp}} \sigma_{\rm{pp}} n_{\rm{CSM}} c) ^{-1}}$, where we assume  constant inelasticity $ k_{\rm{pp}} = 0.5 $ and  energy-dependent cross-section $ \sigma_{\rm{pp}} (E_{\rm{p}})$~\citep{Zyla:2020zbs}). 

As pointed out in~\cite{Fang:2020bkm}, taking $\Delta R_{\rm{acc}}\sim R_{\rm{sh}}$ may be appropriate  for adiabatic shocks only. If the shock is radiative, particles in the post-shock region cool via free-free emission within a layer of width $\sim (t_{\rm{cool}}/t_{\rm{dyn}}) R_{\rm{sh}}$ (see Sec.~\ref{Sec: Interaction-powered SN emission}), making the gas far from the shock quasi-neutral, and thus hindering the magnetic field amplification crucial in the acceleration mechanism~\citep{2004MNRAS.353..550B}. Hence, we adopt $\Delta R_{\rm{acc}}=(t_{\rm{cool}}/t_{\rm{dyn}}) R_{\rm{sh}}$ for $t_{\rm{\rm{cool}}}<t_{\rm{dyn}}$, and $\Delta R_{\rm{acc}}= R_{\rm{sh}}$ otherwise. 

The total proton cooling time can thus be written as ${ t^{-1}_{\rm{p,cool}} = t^{-1}_{\rm{pp}} + \textrm{max}[t^{-1}_{\rm{dyn}},t^{-1}_{\rm{cool}}] }$. It is important to note that relativistic protons in the shocked region may also interact with the ambient photons via $p\gamma$ interactions. However, we ignore such an energy loss channel, by relying on the findings of~\cite{Murase:2010cu,Fang:2020bkm} that showed that $p\gamma$ interactions can be  neglected for a wide range of SN parameters.

Figure~\ref{Fig: Ep_max_R_wind} shows contours of $E_{\rm{p,max}}$  for the wind scenario. 
 The black solid lines mark  the edges of the interaction region, hence Fig.~\ref{Fig: Ep_max_R_wind} also provides an idea of the  the typical interaction duration. 
Fixing  three of the SN model parameters to their benchmark values (see Table~\ref{Table: Parameters}), the shortest period of interaction is obtained for small $R_{\rm{CSM}}$ and large $M_{\rm{CSM}}$, or small $M_{\rm{ej}}$ and large $E_{\rm{k}}$. In fact in both cases the shock breakout is delayed.   The maximum proton energy increases with the radius, and  the largest $E_{\rm{p,max}}$  can be obtained in the late stages of the shock evolution, hinting that high-energy neutrino production should be favored at later times after the bolometric peak. 

The breaks observed in the contour lines in the upper and lower right panels of Fig.~\ref{Fig: Ep_max_R_wind} represent the transition between the regimes where free-free and line-emission dominate. From the two upper panels, we see that $E_{\rm{p,max}}$ reaches its maximum value at $R_{\rm{dec}}$, and declines later. But this is not always the case;  as shown in Appendix~\ref{App:maxEp},  when the proton energy loss times are longer than the dynamical time, the maximum proton energy  decreases throughout the evolution.

\section{Expected neutrino emission from interaction-powered supernovae}
\label{Sec:Production}
In this section, the spectral energy distribution of  neutrinos is introduced.  We then present our findings on the dependence of the expected number of neutrinos  on the SN model parameters and link  the   neutrino signal  to the properties of the SN lightcurves.

\subsection{Spectral energy distribution of  neutrinos}
\label{Sec: Spectral energy distributions of protons and neutrinos }

A fraction $\varepsilon_{\rm{p}}$ of the dissipated kinetic energy of the shock (Eq.~\ref{Eq: kinetic_E_per_radius}) is used to accelerate protons swept-up from the CSM; we adopt  $\varepsilon_{\rm{p}}=0.1$, assuming that  the shocks accelerating protons are parallel or quasi-parallel and therefore efficient diffusive shock acceleration occurs~\citep{Caprioli:2013dca}. However, lower values of $\varepsilon_{\rm{p}}$ would be possible for oblique shocks, with poorer particle acceleration efficiency. Given the linear dependence of proton and neutrino spectra on this parameter, it is straightforward to rescale our results. 

Assuming a power-law energy distribution with spectral index $k=2$, the number of protons injected per unit radius and unit Lorenz factor is
\begin{equation}
    Q_{\rm{p}}(\gamma_{\rm{p}},R)=A(R)\gamma_{\rm{p}}^{-2}\log^{-1}\bigg(\frac{\gamma_{\rm{p, max}}}{\gamma_{\rm{p, min}}}\bigg)\ ,
\end{equation}
for $\gamma_{\rm{p,min}}<\gamma<\gamma_{\rm{p,max}}$, and zero otherwise.
We set the minimum Lorentz factor of  accelerated protons  ${\gamma_{\rm{p, min}} = 1}$, while $ \gamma_{\rm{p, max}} $ is obtained by comparing the acceleration and the energy-loss time scales at each radius during the shock evolution, as discussed in Sec.~\ref{Sec: The maximum proton energy }. The normalization factor $A(R)$ is
\begin{equation}
    A(R)=\frac{9}{8}\pi \varepsilon_{\rm{p}} R^{2}v_{\rm{sh}}^{2}(R)\rho_{\rm{CSM}}(R)\propto \begin{cases}
    R^{\frac{2n-sn+5s-12}{n-3}}\quad\textrm{for}\,R\leq R_{\rm{dec}}\\
    R^{-1}\quad\quad\quad \quad \,\, \textrm{for}\,R>R_{\rm{dec}}\ .
    \end{cases}
\end{equation}
The injection rate of protons in the deceleration phase does not depend on the SN density structure nor the CSM density profile.
Since we aim to compute the neutrino emission, we track the temporal evolution of the proton distribution in the shocked region between the shock breakout radius $ R_{\rm{bo}} $ and the outer radius $R_{\rm{CSM}}$.

The evolution of the proton distribution is given by~\citep{1997ApJ...490..619S,2012ApJ...751...65F,Petropoulou:2016zar}:
\begin{equation}
	\frac{\partial N_{\rm{p}}(\gamma_{\rm{p}},R)}{\partial R} - \dfrac{\partial }{\partial \gamma_{\rm{p}}}\bigg[\frac{\gamma_{\rm{p}}}{R} N_{\rm{p}}(\gamma_{\rm{p}}, R)\bigg] + \frac{N_{\rm{p}}(\gamma_{\rm{p}}, R)}{v_{\rm{sh}}(R) t_{\rm{pp}}(R)} = Q_{\rm{p}}(\gamma_{\rm{p}},R)\ ,
\label{Eq:Np}	
\end{equation}
where $ N_{\rm{p}} (\gamma_{\rm{p}}, R)$ represents the total number of protons in the shell at a given radius $ R $ with Lorentz factor between $ \gamma_{\rm{p}} $ and $ \gamma_{\rm{p}} + \rm{d}\gamma_{\rm{p}} $. The second term on the left hand side of Eq.~\ref{Eq:Np} takes into account energy losses due to the adiabatic expansion of the SN shell, while  $ pp $ collisions are  treated as an escape term~\citep{1997ApJ...490..619S}. Other energy loss channels for protons are  negligible~\citep{Murase:2010cu}. Furthermore, in Eq.~\ref{Eq:Np}, the diffusion term has been neglected since the shell is assumed to be homogeneous. 

The neutrino production rates, $ Q_{\nu_{i}+\bar{\nu}_{i}} \, [\rm{GeV}^{-1}\rm{s}^{-1}] $, for muon and electron flavor (anti)neutrinos are given by~\citep{Kelner:2006tc}:
\begin{eqnarray}
\label{Eq:Q_nu_mu}
	Q_{\nu_{\mu} + \bar{\nu}_{\mu}}(E_{\nu}, R) &=& 4 n_{\rm{CSM}}(R) m_{\rm{p}} c^{3} \int_{0}^{1} dx \frac{\sigma_{\rm{pp}}(E_{\nu}/x)}{x}\times \\ \nonumber  & & N_{\rm{p}} \bigg(\frac{E_{\nu}}{x m_{\rm{p}} c^{2}}, R\bigg)\bigg(F^{(1)}_{\nu_{\mu}}(E_{\nu}, x) + F^{(2)}_{\nu_{\mu}}(E_{\nu}, x)\bigg)\ , \\
	\label{Eq:Q_nu_e}
		Q_{\nu_{e} + \bar{\nu}_{e}}(E_{\nu}, R) &=& 4 n_{\rm{CSM}}(R) m_{\rm{p}} c^{3} \int_{0}^{1} dx \frac{\sigma_{\rm{pp}}(E_{\nu}/x)}{x}\times \\ \nonumber  & & N_{\rm{p}} \bigg(\frac{E_{\nu}}{x m_{\rm{p}} c^{2}}, R\bigg) F_{\nu_{e}}(E_{\nu}, x)\ ,
\end{eqnarray}
where $ x = E_{\nu}/E_{\rm{p}} $. The functions $ F^{(1)}_{\nu_{\mu}} $, $ F^{(2)}_{\nu_{\mu}} $ and $ F_{\nu_{e}} $ follow the definitions in~\cite{Kelner:2006tc}. Equations~\ref{Eq:Q_nu_mu} and \ref{Eq:Q_nu_e} are valid for $ E_{\rm{p}} > 0.1$~TeV,  corresponding to  the energy range under investigation. Note that, for the parameters we use in this work, the synchrotron cooling of charged pions and muons produced via $pp$ interactions is negligible. Therefore, the neutrino spectra are not affected by the cooling of mesons.

\subsection{Energy emitted in neutrinos}
\label{Sec. results_neutrinos_wind}

The total energy that goes in neutrinos in the energy range $[E_{\nu,1},E_{\nu,2}]$ during the entire interaction period is given by 
\begin{eqnarray}
	\label{Eq: E_nu_tot}
    \mathcal{E}_{\nu+\bar{\nu}}= \int_{t_{\rm{BO}}}^{t_{\rm{CSM}}}dt\int^{E_{\nu,2}}_{E_{\nu,1}}dE_{\nu} E_{\nu}  [Q_{\nu_{\mu} + \bar{\nu}_{\mu}}(E_{\nu}, R)+Q_{\nu_{e} + \bar{\nu}_{e}}(E_{\nu}, R)]\ ,
\end{eqnarray}
where $t_{\rm{BO}}$ and $t_{\rm{CSM}}$ are expressed  in the progenitor reference frame.

\begin{figure*}
	\centering
 	\includegraphics[width=0.9\textwidth]{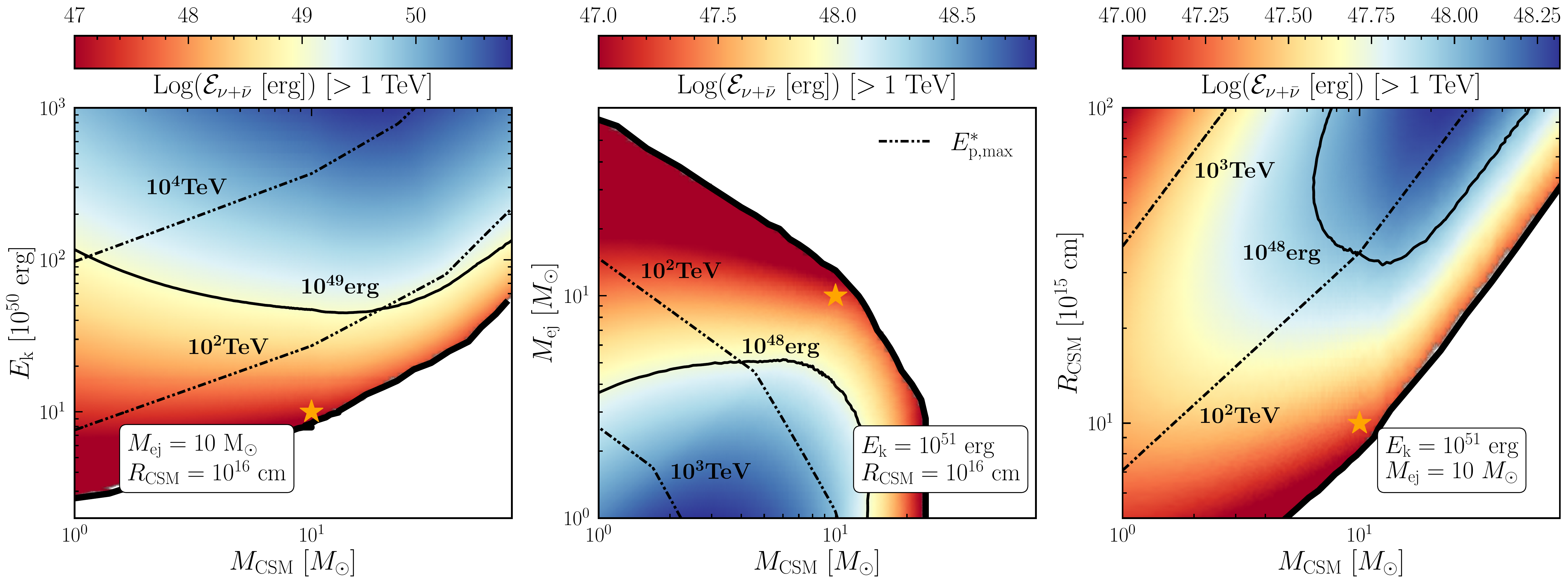}\\
  	\includegraphics[width=0.9\textwidth]{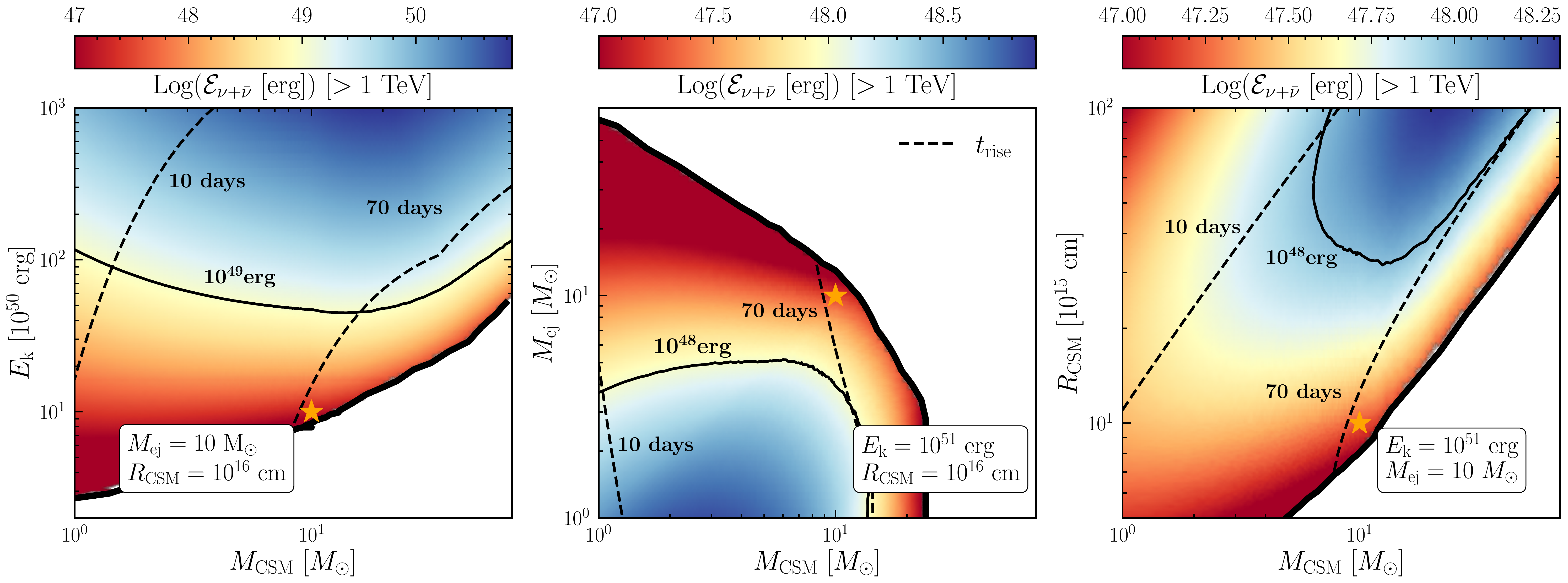}\\
  	\includegraphics[width=0.9\textwidth]{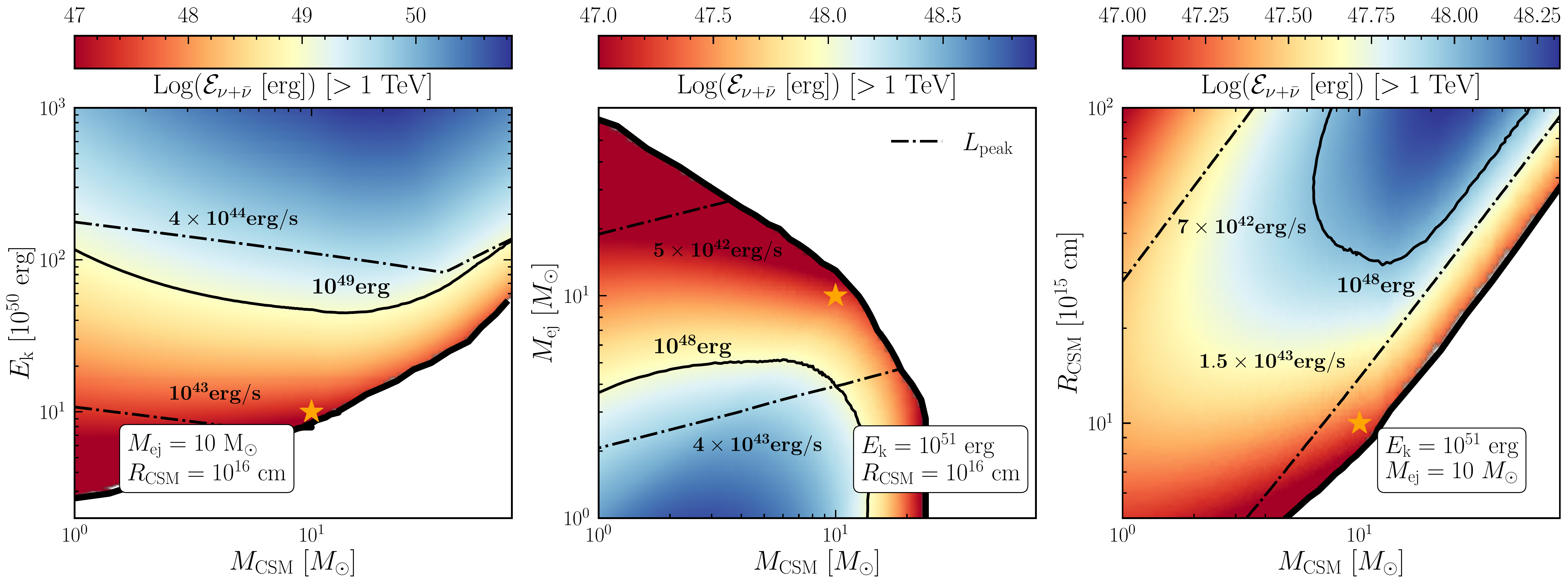}\\
\caption{Contour plots of the total neutrino energy $\mathcal{E}_{\nu+\bar\nu}$ integrated for $E_{\nu}\geq 1\,\mathrm{TeV}$ through the evolution of the shock in the CSM, as a function of $M_{\rm{CSM}}$ and $E_{\rm{k}}$ (left panels), $M_{\rm{ej}}$ (middle panels), and $R_{\rm{CSM}}$ (right panels) for the wind scenario.  In order to highlight the dependence on the SN properties, isocontours of the  maximum proton energy $E^{\ast}_{\rm{p,max}}$ (double-dot dashed contours, top row), $t_{\rm{rise}}$ (dashed contours, middle row), and $L_{\rm{peak}}$ (dot dashed contours, bottom row) are  displayed.    All  quantities are expressed in the SN reference frame. The white regions represent parts of the parameter space with $E_{\rm{p,max}}^{\ast}\lesssim 10$~TeV excluded from our investigation. Our benchmark SN model is marked with an orange star. 
	The SN configurations leading to the largest $\mathcal{E}_{\nu+\bar\nu}$ are given by large SN kinetic energies ($E_{\rm{k}}\gtrsim 10^{51}$~erg), small ejecta masses ($M_{\rm{ej}}\lesssim 10 \, M_{\odot}$), intermediate CSM masses with respect to $M_{\rm{ej}}$ (i.e., $1 M_{\odot}\lesssim M_{\rm{CSM}}\lesssim 30 M_{\odot}$),  and relatively large CSM extent ($R_{\rm{CSM}}\gtrsim 10^{16}$~cm).
	} 
	\label{Fig: Etot_wind}
\end{figure*}

In order to connect the observed properties of the SN lightcurve to the neutrino ones (e.g.,~the total energy that goes in neutrinos or their typical spectral energy), for each configuration of SN model parameters we integrate the neutrino production rate between $t_{\rm{BO}}$ and $t_{\rm{CSM}}$, for  $ E_{\nu}\geq 1\, \mathrm{TeV}$, as in Eq.~\ref{Eq: E_nu_tot}. The results are shown in Fig.~\ref{Fig: Etot_wind}, where  we fix two of the SN parameters at their benchmark values (see Table~\ref{table:SN_param}) and investigate $\mathcal{E}_{\nu+\bar\nu}$ in the plane spanned by the remaining two. Note that we do not consider the regions of the SN parameter space with the maximum achievable proton energy ($E^\ast_{\rm{p,max}}$, see Appendix~\ref{App:maxEp} for more details) smaller than $10$~TeV since they would lead to neutrinos in the energy range dominated by atmospheric events in IceCube (see Sec.~\ref{sec:detection}). If we were to integrate the neutrino rate for $E_{\nu,1}> 1\,\mathrm{TeV}$ (Eq.~\ref{Eq: E_nu_tot}), the contour lines for $\mathcal{E}_{\nu+\bar\nu}$ would be  shifted to the left.  Isocontours of the  maximum achievable proton energy $E^{\ast}_{\rm{p,max}}$ (first row), the rise time $t_{\rm{rise}}$ (second row), and the bolometric peak $L_{\rm{peak}}$ (third row)
are also displayed on top of the $\mathcal{E}_{\nu+\bar\nu}$ colormap in Fig.~\ref{Fig: Etot_wind}.

In all panels of Fig.~\ref{Fig: Etot_wind}, $\mathcal{E}_{\nu+\bar\nu}$ increases with $M_{\rm{CSM}}$, due to the larger  target proton number. Nevertheless, such a trend saturates once the critical $n_{\rm{CSM}}$ (corresponding to a critical $M_{\rm{CSM}}$) is reached, where either $pp$ interactions or the cooling of thermal plasma  significantly limit the maximum proton energy,  thus decreasing the number of neutrinos produced with high energy. For masses larger than the critical CSM mass,  neutrinos could be abundantly produced  either  appreciably increasing the kinetic energy (left panel), or  decreasing the ejecta mass (middle panel), or  increasing the CSM radius (right panel). 
From the contour lines in each panel, we  see that the optimal configuration for what concerns neutrino production  results from  large $E_{\rm{k}}$ and small $M_{\rm{ej}}$, which lead to large shock velocities $v_{\rm{sh}}$,  large $R_{\rm{CSM}}$, and not extremely large $M_{\rm{CSM}}$, compared to a fixed $M_{\rm{ej}}$.
Nevertheless, the panels in the upper row of Fig.~\ref{Fig: Etot_wind} indicate that the configurations with the largest proton energies (and thus spectral neutrino energies) always prefer a balance between $E_{\rm{k}}$, $M_{\rm{ej}}$, and $R_{\rm{CSM}}$ with $M_{\rm{CSM}}$.

It is important to observe the peculiar behavior resulting from  the variation of $R_{\rm{CSM}}$ (right panels of Fig.~\ref{Fig: Etot_wind}). For  fixed $M_{\rm{CSM}}$,  $\mathcal{E}_{\nu+\bar\nu}$ increases, then saturates at a certain $R_{\rm{CSM}}$, and decreases thereafter.  For very small $R_{\rm{CSM}}$, the CSM density is relatively large, and the shock becomes collisionless  close to $R_{\rm{CSM}}$, probing a  low fraction of the total CSM mass and thus producing a small number of neutrinos. This suppression is alleviated by increasing $R_{\rm{CSM}}$. Nevertheless, a  large $R_{\rm{CSM}}$ for fixed $M_{\rm{CSM}}$  leads to a low CSM density, and thus the total neutrino energy drops. For increasing $M_{\rm{CSM}}$, such inversion in $\mathcal{E}_{\nu+\bar\nu}$ happens  at larger $R_{\rm{CSM}}$. We also see that the largest $\mathcal{E}_{\nu+\bar\nu}$ is obtained in the right upper corner of the right panels. This is mainly related to the duration of the shock interaction. The longer the interaction time, the larger the CSM mass swept-up by the collisionless shock.

The panels in the middle row of Fig.~\ref{Fig: Etot_wind} show how the neutrino energy varies as a function of  $t_{\rm{rise}}$. Large neutrino energy is obtained for slow rising lightcurves. In particular, given our choice of the parameters for these contour plots, the most optimistic scenarios for neutrinos lie in the region with  $10\,\mathrm{days}\lesssim t_{\rm{rise}}\lesssim 50\,\mathrm{days}$. Such findings hold  for a wide range of  parameters for interacting SNe. Extremely large $t_{\rm{rise}}$, on the other hand, are expected to be determined by very large $M_{\rm{CSM}}$, which can substantially limit  the production of particles in the high energy regime.

The bottom panels of Fig.~\ref{Fig: Etot_wind} illustrate how  $\mathcal{E}_{\nu+\bar\nu}$ is linked to $L_{\rm{peak}}$. 
In particular,  $L_{\rm{peak}}$ closely tracks $\mathcal{E}_{\nu+\bar\nu}$. However, $L_{\rm{peak}}$ can increase with $M_{\rm{CSM}}$ to larger values than what neutrinos do, given its linear dependence on the CSM density (see Eq.~\ref{Eq: L_peak}). Overall, the regions where the largest $\mathcal{E}_{\nu+\bar\nu}$ (and hence number of neutrinos) is obtained are also the regions where $L_{\rm{peak}}$ is the largest. It is not always true the opposite. Hence,  large $L_{\rm{peak}}$ is a necessary, but not sufficient condition to have large $\mathcal{E}_{\nu+\bar\nu}$.

To summarize, a large  $\mathcal{E}_{\nu+\bar\nu}$ is expected for large SN kinetic energy ($E_{\rm{k}}\gtrsim 10^{51}$~erg), small ejecta mass ($M_{\rm{ej}}\lesssim 10 \, M_{\odot}$), intermediate CSM mass with respect to $M_{\rm{ej}}$ ($1\, M_{\odot}\lesssim M_{\rm{CSM}}\lesssim 30\, M_{\odot}$),  and relatively extended CSM  ($R_{\rm{CSM}}\gtrsim 10^{16}$~cm). These features  imply  large bolometric luminosity peak  ($ L_{\rm{peak}}\gtrsim 10^{43}$--$10^{44}$~erg) and average rise time ($t_{\rm{rise}}\gtrsim 10$--$20$~days). On the other hand, it is important to note that degeneracies are present in the SN parameter space~\citep[see also][]{Pitik:2021dyf} and comparable $ L_{\rm{peak}}$ and $t_{\rm{rise}}$  can be obtained for SN model parameters ($E_{\rm{k}}$, $M_{\rm{ej}}$, $R_{\rm{CSM}}$, and $M_{\rm{ej}}$) that are not optimal for   neutrino emission.

It is important to stress that in this section we have considered $\mathcal{E}_{\nu+\bar\nu}$ as a proxy of the expected number of neutrino events that is investigated in Sec.~\ref{sec:detection}. Moreover, we have compared $\mathcal{E}_{\nu+\bar\nu}$ to the bolometric luminosity expected at the peak and not to the luminosity effectively radiated, $L_{\rm{peak,obs}}$.


\section{Neutrino detection prospects}
\label{sec:detection}
In this section, we investigate the neutrino detection prospects. In order to do so, we select two  especially bright  SNe observed by ZTF, SN 2020usa and SN 2020in. On the basis of our findings, we also discuss  the most promising strategies for neutrino searches and multi-messenger follow-up programs.

\subsection{Expected number of neutrino events at Earth}
\label{sec:nurateflux}

The neutrino and antineutrino flux ($F_{\nu_{\alpha} +\bar{\nu}_{\alpha}}$ with $\alpha=e, \mu, \tau$)  at Earth from a SN at redshift $z$ and as a function of time in the observer frame is [$\mathrm{GeV}^{-1} \mathrm{s}^{-1}\mathrm{cm}^{-2}$]:
\begin{equation}
\label{eq:F}
	F_{\nu_{\alpha}+\bar{\nu}_{\alpha}}(E_{\nu}, t) = \frac{(1 + z)^2}{4 \pi d^{2}_{L}(z)} v_{\rm{sh}}\underset{\beta}{\sum} P_{\nu_{\beta} \rightarrow \nu_{\alpha}}Q_{\nu_{\beta} +\bar{\nu}_{\beta}}\bigg(E_{\nu_{\alpha}}(1 + z),\frac{v_{\rm{sh}} t}{1 + z}\bigg)\ ,
\end{equation}
where $Q_{\nu_{\beta} +\bar{\nu}_{\beta}}$ is defined as in Eqs.~\ref{Eq:Q_nu_mu} and \ref{Eq:Q_nu_e}. Neutrinos change their flavor while propagating, hence the flavor transition probabilities are given by~\citep{Anchordoqui:2013dnh}:
\begin{eqnarray}
	P_{\nu_{e}\rightarrow\nu_{\mu}} &=& P_{\nu_{\mu}\rightarrow\nu_{e}} = P_{\nu_{e}\rightarrow\nu_{\tau}} = \frac{1}{4}\sin^{2}2\theta_{12}\ ,\\
	P_{\nu_{\mu}\rightarrow\nu_{\mu}}&=& P_{\nu_{\mu}\rightarrow\nu_{\tau}} = \frac{1}{8}(4 - \sin^{2}2\theta_{12})\ ,\\
	P_{\nu_{e}\rightarrow\nu_{e}}&=& 1-\frac{1}{2}\sin^{2}2\theta_{12}\ ,
\end{eqnarray}
with $\theta_{12} \simeq 33.5$ deg~\citep{Esteban:2020cvm}, and  $P_{\nu_{\beta}\rightarrow \nu_{\alpha}} = P_{\bar{\nu}_{\beta}\rightarrow \bar{\nu}_{\alpha}}$. The  luminosity distance   $ d_{L}(z) $ is  defined in a flat $\Lambda$CDM  cosmology:
\begin{equation}
	\label{luminosity_distance}
	d_{L}(z) = (1+z) \frac{c}{H_{0}} \int_{0}^{z} \frac{dz^\prime}{\sqrt{\Omega_{\Lambda}+\Omega_{M}(1+z^\prime)^3}}\ ,
\end{equation}
where $\Omega_{M} = 0.315$, $\Omega_{\Lambda} = 0.685$ and the Hubble constant is ${H_{0} = 67.4}$~km s$^{-1}$ Mpc$^{-1}$~\citep{Aghanim:2018eyx}.

Due to the better angular resolution of muon-induced track events compared to cascades, we focus on muon neutrinos and antineutrinos. Therefore, the event rate expected at the IceCube Neutrino Observatory, after taking into account neutrino flavor conversion, is 
\begin{equation}
	\label{eq: neutrino event number}
	\dot{N}_{\nu_{\mu}+\bar{\nu}_{\mu}}= \int_{E_{\nu,1}}^{E_{\nu,2}} dE_{\nu} A_{\rm{eff}}(E_{\nu}, \alpha)  F_{\nu_{\mu}+\bar{\nu}_{\mu}}(E_{\nu}, t)\ ,
\end{equation}
where $ A_{\rm{eff}}(E_{\nu},\alpha)$ is the detector effective area~\citep{IceCube:2021xar} for a SN at declination $\alpha$.

\subsection{Expected number of neutrino events for SN 2020usa and SN 2020in}
\label{sec:SN2020usa-SN 2020in}

\begin{table*}
	\caption{\label{Table: SN2020usa and SN2020in} Characteristic properties of  our representative SLSNe, SN 2020usa and SN 2020in.
 }
	\begin{center}
		\hspace{-0.9cm}
		\begin{tabular}[c]{c|cccccc}
			\toprule
			 & Redshift & $t_{\rm{rise, obs}}$ [days] & $L_{\rm{peak,obs}}$ [$\mathrm{erg\, s^{-1}}$] & $E_{\rm{rad,obs}}$ [erg] & $t_{\rm{dur,obs}}$ [days] & Declination [deg]\\
            \toprule
            SN 2020usa & $0.26$ & $65$ & $8\times 10^{43}$ & $1.3\times 10^{51}$ & $350$&$-2.3$\\
            SN 2020in & $0.11$ & $42$ & $3\times 10^{43}$ & $3.3\times 10^{50}$ & $413$ & $20.2$\\

			\toprule
		\end{tabular}
	\end{center}
\end{table*}
To investigate the  expected number of neutrino events, we select two among the brightest sources observed by ZTF
whose observable properties  are summarized in Table~\ref{Table: SN2020usa and SN2020in}: SN2020usa\footnote{\url{https://lasair-ztf.lsst.ac.uk/objects/ZTF20acbcfaa/}} and SN2020in~\footnote{\url{https://lasair-ztf.lsst.ac.uk/objects/ZTF20aaaweke/}}. We retrieve the  photometry data of the sources in the ZTF-g and ZTF-r bands, and
correct the measured fluxes for Galactic extinction~\citep{SFDmap}. Using linear interpolation of the individual ZTF-r and ZTF-g light curves, we perform a trapezoid integration between the respective center wavelengths to estimate the radiated energy at each time of measurement. The resulting lightcurve is interpolated with Gaussian process regression~\citep{george} and taken as a lower limit to the bolometric SN emission.
From such pseudo-bolometric lightcurve, the rise time and peak luminosity are determined. The rise time is  defined as the difference between the peak time and the estimated SN breakout time. The latter  is determined by taking the average between the time of the first detection in ZTF-g or ZTF-r bands  and the last non-detection in either band.
In what follows, we consider the radiative efficiency ${\varepsilon_{\rm{rad}}=L_{\rm{peak,obs}}/L_{\rm{peak}}}$ as a free parameter 
in the  range  $\varepsilon_{\rm{rad}}\in [0.2, 0.7]$~\citep[see, e.g,][]{Villar:2017oya}. 
Furthermore, we  assume that $\varepsilon_{\rm{rad}}=E_{\rm{rad,obs}}/E_{\rm{diss, thick}}$ also holds. 

For both SNe, we perform a scan over the SN model parameters ($E_{\rm{k}}$, $M_{\rm{ej}}$, $M_{\rm{CSM}}$, and $R_{\rm{CSM}}$) which fulfill the following conditions: 
\begin{itemize}
\item[-] $t_{\rm{rise}}\in [1, 1.5] \times t_{\rm{rise,obs}}$, namely we allow an error  up to $50\%$ on the estimation of $t_{\rm{rise}}$;
\item[-] $L_{\rm{peak}}\geq L_{\rm{peak,obs}}$;
\item[-] $E_{\rm{k}}>E_{\rm{rad, obs}}$. We narrow the investigation range to ${E_{\rm{k}}\in [10^{51}, 2\times 10^{52}]}$~erg for SN2020usa and $E_{\rm{k}}\in [4\times 10^{50}, 5\times 10^{51}]$~erg for SN2020in, assuming that at least  $\sim 10\%$ and at most $80\%$ of the total energy $E_{\rm{k}}$ is radiated. 
\item[-]  $t_{\rm{dur,th}}\geq t_{\rm{dur,obs}}$. Here $t_{\rm{dur,obs}}$ is the observational temporal window available for each SN event, while $t_{\rm{dur,th}}=t(R_{\rm{ph}})-t(R_{\rm{bo}})$ is the time that   the shock takes to cross the optically thick part of the CSM envelope after breakout (as mentioned in Sec.~\ref{Sec: Interaction-powered SN emission}, the shock is expected to peak in the X-ray band once out of the optically thick part). 
\end{itemize}

\begin{figure*}
	\centering
	\includegraphics[width=0.8\textwidth]{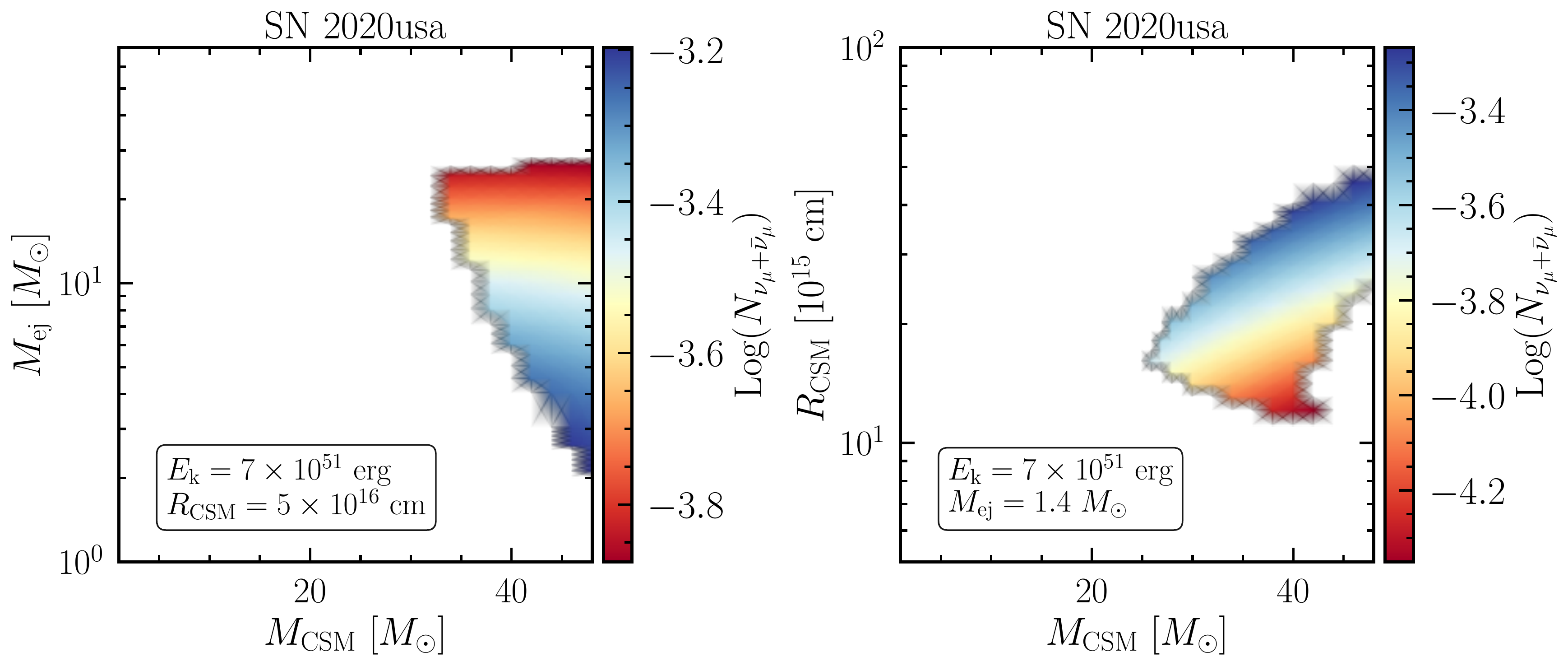}\\
 \includegraphics[width=0.8\textwidth]{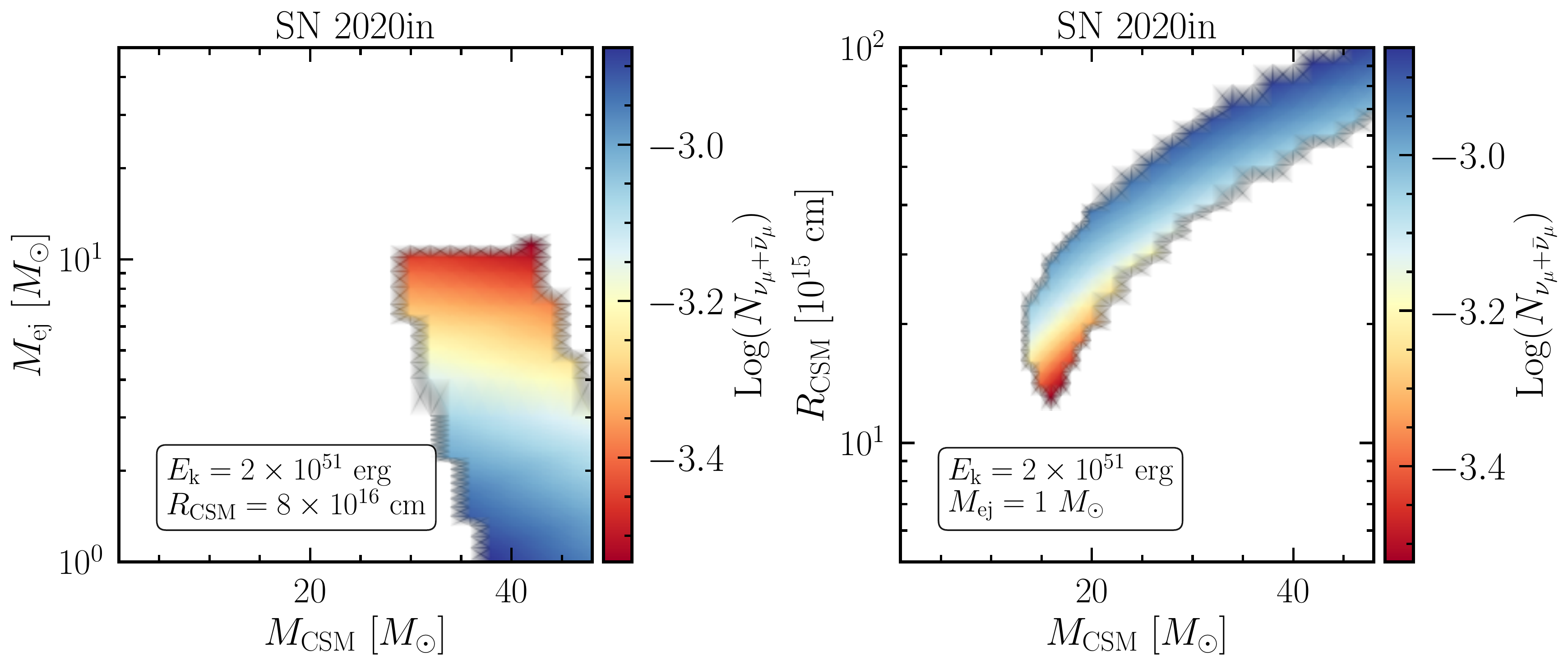}
	\caption{Contour plot of the number of muon neutrino and antineutrino events expected at the IceCube Neutrino Observatory (for the wind scenario and integrated over the  duration of  CSM interaction) in the SN model parameter space compatible with the observation of  SN2020usa (top panels) and SN2020in (bottom panels). Only a fraction of the SN parameter space is compatible with the optical data. Importantly, for fixed $L_{\rm{peak,obs}}$ and $t_{\rm{rise, obs}}$, a different number of neutrino events could be obtained according to the specific combination of $M_{\rm ej}$, $M_{\rm CSM}$, $R_{\rm CSM}$, $E_{\rm k}$ compatible with the observed optical properties. }
	\label{Fig: SN2020 contour}
\end{figure*}
Figure~\ref{Fig: SN2020 contour} shows the  total number of muon neutrino and antineutrino events, integrated over the  duration of the interaction in the CSM for $E_{\nu}\geq 1$~TeV, expected at IceCube in the the wind scenario, for  $E_{\rm{k}}$ selected to  maximize the space of parameters compatible with the conditions mentioned above. Similarly to Fig.~\ref{Fig: Etot_wind},  the regions with the largest number of neutrino events  are those with lower $M_{\rm{ej}}$ and larger $R_{\rm{CSM}}$, for  fixed $M_{\rm{CSM}}$. It is important to note that,  for given observed SN  properties ($L_{\rm{peak,obs}}$ and $t_{\rm{rise, obs}}$), the expected number of neutrino events is not unique; in fact, as shown in Sec.~\ref{Sec:Production}, there is  degeneracy in the SN model parameters that leads to the same $L_{\rm{peak,obs}}$ and $t_{\rm{rise, obs}}$.

\begin{figure}
	\centering
	\includegraphics[width=0.4\textwidth]{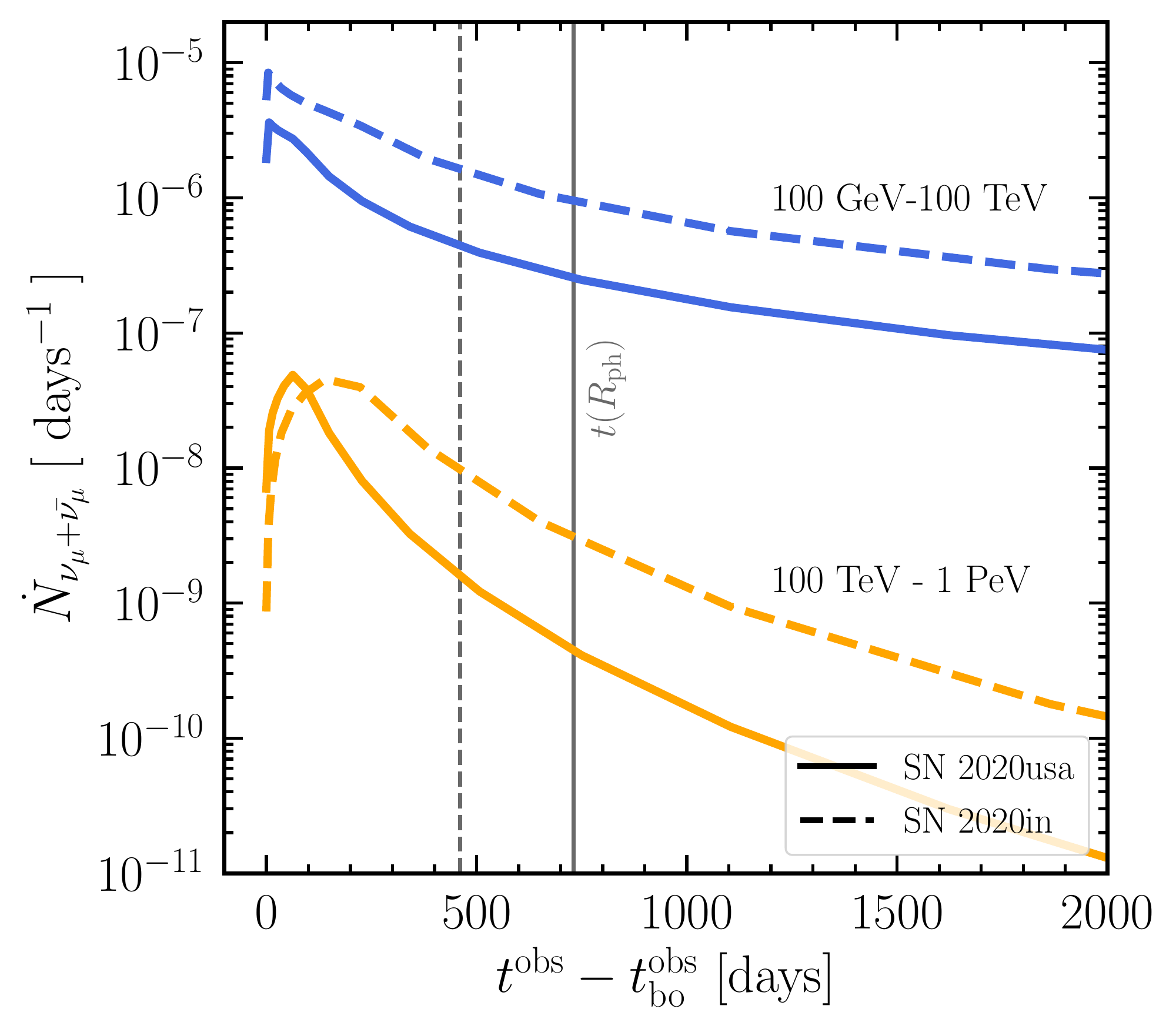}
	\caption{Muon neutrino and antineutrino event rates predicted for SN 2020usa and SN 2020in  at the IceCube Neutrino Observatory as  functions of  time in the observer frame, after the shock breakout, assuming $\varepsilon_{\rm{rad}}=0.2$. The SN model parameters have been chosen to optimize neutrino production [$M_{\rm{ej}}=5.5\,M_{\odot}$, $M_{\rm{CSM}}=48\,M_{\odot}$, $R_{\rm{CSM}}=5.5\times 10^{16}\,\mathrm{cm}$, $E_{\rm{k}}=10^{52}\,\mathrm{erg}$ for SN2020usa;  $M_{\rm{ej}}=5\,M_{\odot}$, $M_{\rm{CSM}}=46 \,M_{\odot}$, $R_{\rm{CSM}}= 10^{17}\,\mathrm{cm}$, $E_{\rm{k}}=5\times 10^{51}\,\mathrm{erg}$ for SN2020in]. The event rate increases slightly more slowly in the high energy band ($100~\mathrm{TeV}$--$1~\mathrm{PeV}$) with respect to the low energy one at early times, and  it declines after peak because of the decreasing trend of $v_{\rm{sh}}$ as a function of time. The gray vertical lines indicate the time at which the shock  reaches the photospheric radius $R_{\rm{ph}}$ (solid and dashed for SN 2020 usa and SN 2020in, respectively). 
 }
	\label{fig: SN2020 rate}
\end{figure}
\begin{figure*}
	\centering
	\includegraphics[width=0.4\textwidth]{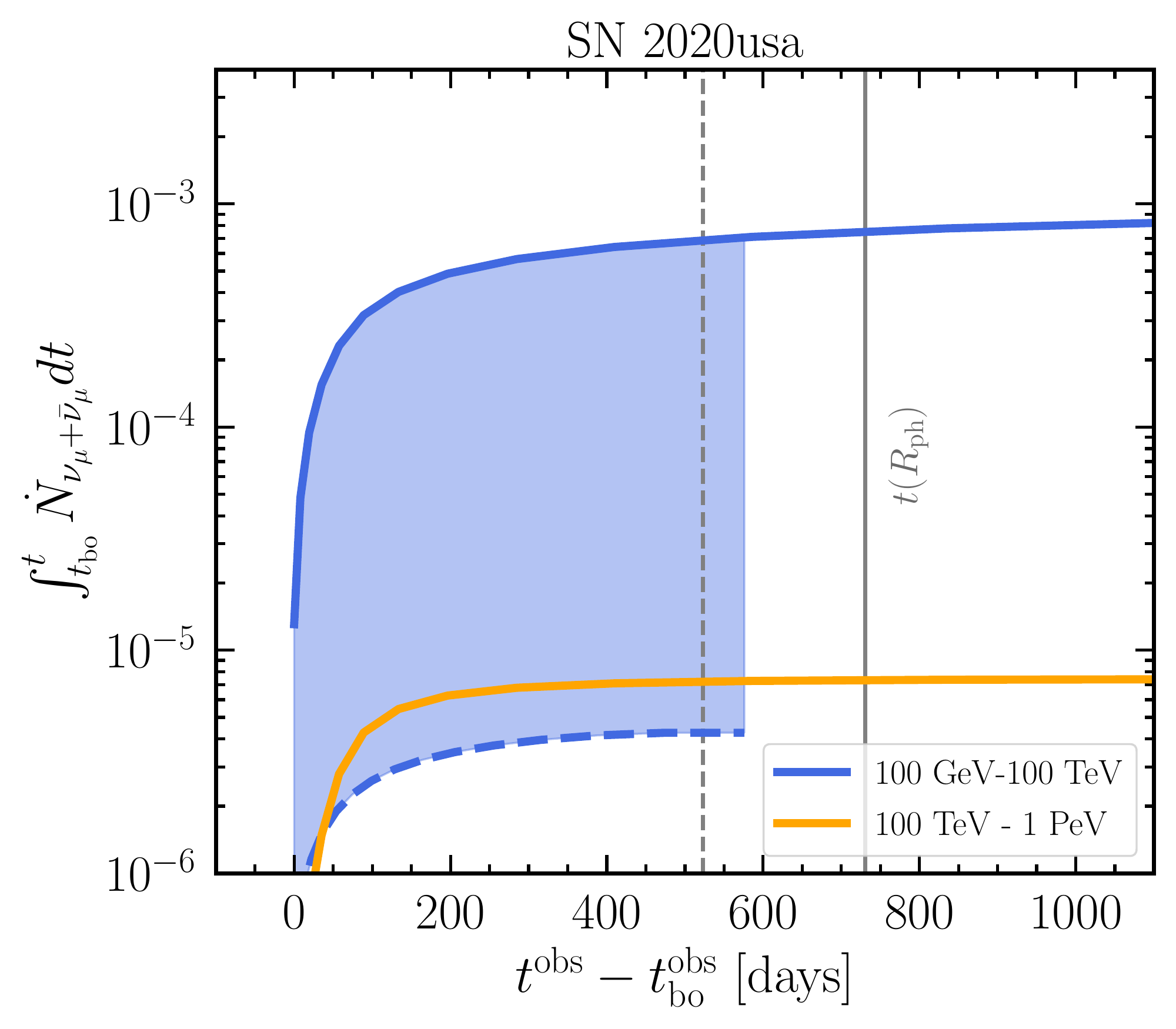}
 	\includegraphics[width=0.4\textwidth]{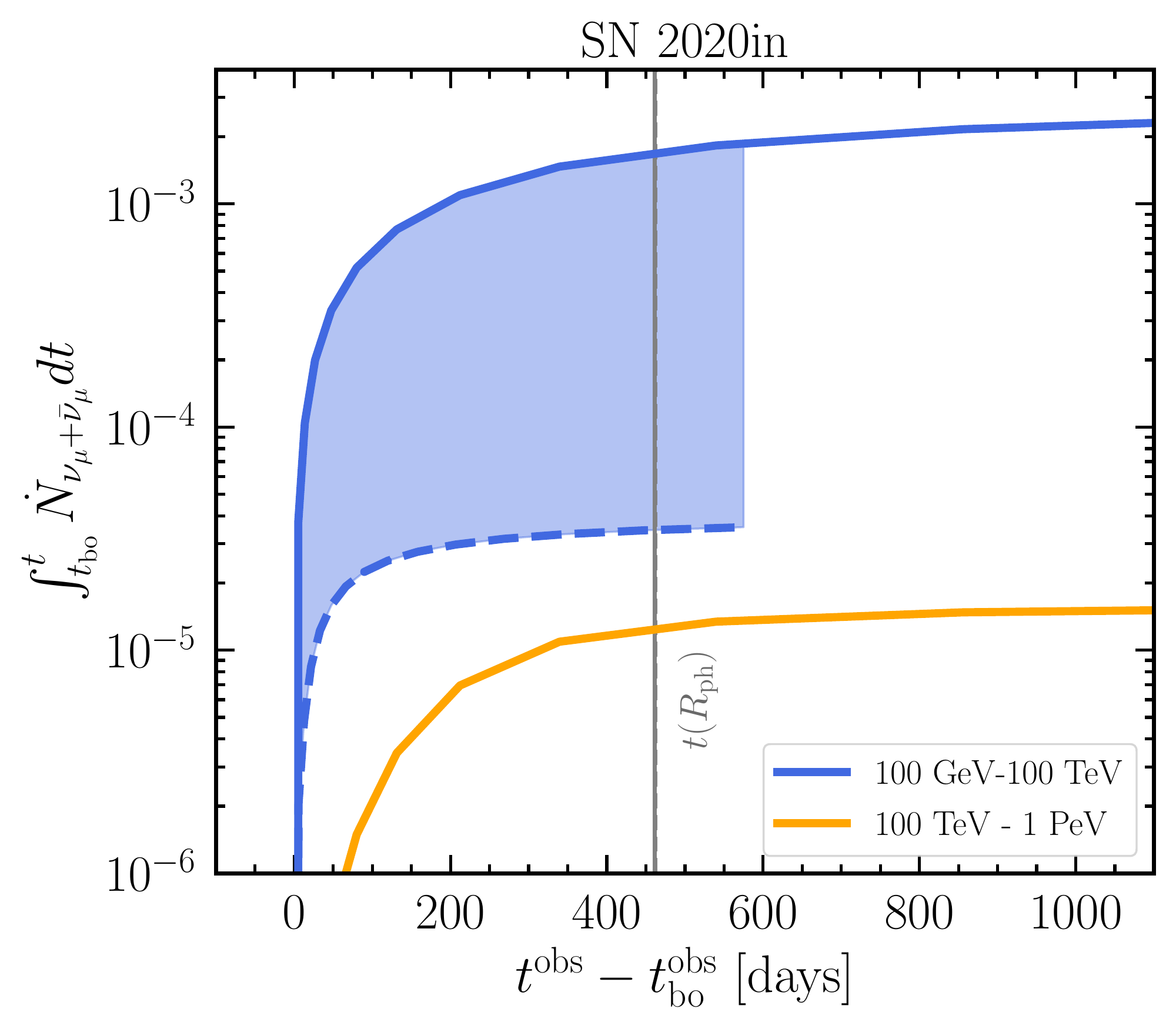}
	\caption{Cumulative number of muon neutrino and antineutrino events for SN 2020usa and SN2020in, as  functions of  time in the observer frame. The solid and dashed lines correspond to the the most optimistic and pessimistic cumulative number of events in the indicated energy range, respectively. The SN model parameters for the most optimistic scenario are the same as the ones in Fig.~\ref{fig: SN2020 rate}, while the parameters leading to the most pessimistic conditions for neutrino production are $M_{\rm{ej}}=1\,M_{\odot}$, $M_{\rm{CSM}}=25\,M_{\odot}$, $R_{\rm{CSM}}=9\times 10^{15}\,\mathrm{cm}$, $E_{\rm{k}}=2\times 10^{51}\,\mathrm{erg}$ for SN 2020usa, and $M_{\rm{ej}}=1.6\,M_{\odot}$, $M_{\rm{CSM}}=10\,M_{\odot}$, $R_{\rm{CSM}}=9\times 10^{15}\,\mathrm{cm}$, $E_{\rm{k}}=7\times 10^{50}\,\mathrm{erg}$, for SN2020in. In both cases $\varepsilon_{\rm{rad}}=0.7$. Neutrinos in the the energy range $[100~\mathrm{TeV}$, $1~\mathrm{PeV}]$ are not produced in the pessimistic scenarios. The gray vertical lines indicate the time at which the shock  reaches the photospheric radius $R_{\rm{ph}}$. 
 }
	\label{fig: SN2020 cumulative}
\end{figure*}
Figure~\ref{fig: SN2020 rate} represents  the  muon neutrino and antineutrino event rate expected at IceCube for SN2020usa and SN2020in, each  as a function of time for two energy ranges, and for the most optimistic scenario. Figure~\ref{fig: SN2020 cumulative} displays the corresponding cumulative neutrino number of events for both most optimistic and pessimistic scenarios. The two cases are selected after scanning over  $\varepsilon_{\rm{rad}}$. The smaller is $\varepsilon_{\rm{rad}}$, the higher $E_{\rm{k}}$ is needed to explain the observations, and since  we adopt a fixed fraction of the shock energy $\varepsilon_{\rm{p}}$ that goes into acceleration of relativistic protons,  the best case for neutrino production is the one with the lowest $\varepsilon_{\rm{rad}}$. 
 Choosing $\varepsilon_{\rm{rad, min}}=0.2$, we only select the SN parameters that satisfy the following conditions:  $t_{\rm{rise}}\in [1, 1.5]\times t_{\rm{rise,obs}}$, $L_{\rm{peak}}\in [1, 1.5] \times L_{\rm{peak,obs}}/\varepsilon_{\rm{rad,min}}$, and $E_{\rm{rad}}\in [1, 1.5] \times E_{\rm{rad,obs}}/\varepsilon_{\rm{rad,min}}$, hence considering an error on $L_{\rm{peak,obs}}$ and $E_{\rm{rad,obs}}$ of at most $50\%$. 
After an initial rise, the neutrino event rate for both considered energy ranges ($100~\mathrm{GeV}$--$100~\mathrm{TeV}$ and $100~\mathrm{~TeV}$--$1~\mathrm{PeV}$) decreases with time, with a steeper rate for the high-energy range, where the slow increase of $E_{\rm{p,max}}$ does not  compensate the drop in the CSM density. 
Note that the cumulative number of neutrino events is relatively small because, although the SN 2020usa and SN 2020in have large $L_{\rm{peak,obs}}$, they occurred at relatively large distance from Earth ($\sim$ Gpc), as evident from Table~\ref{Table: SN2020usa and SN2020in}. If other SNe exhibiting similar photometric properties should be observed at smaller $z$, then the expected neutrino flux should be rescaled with respect to the results shown here by the SN distance squared (see Sec.~\ref{sec:follow-up_strategy} and Fig.~\ref{Fig: SN2020_Nutot_z}).

\begin{figure}
	\centering
 \includegraphics[width=0.45\textwidth]{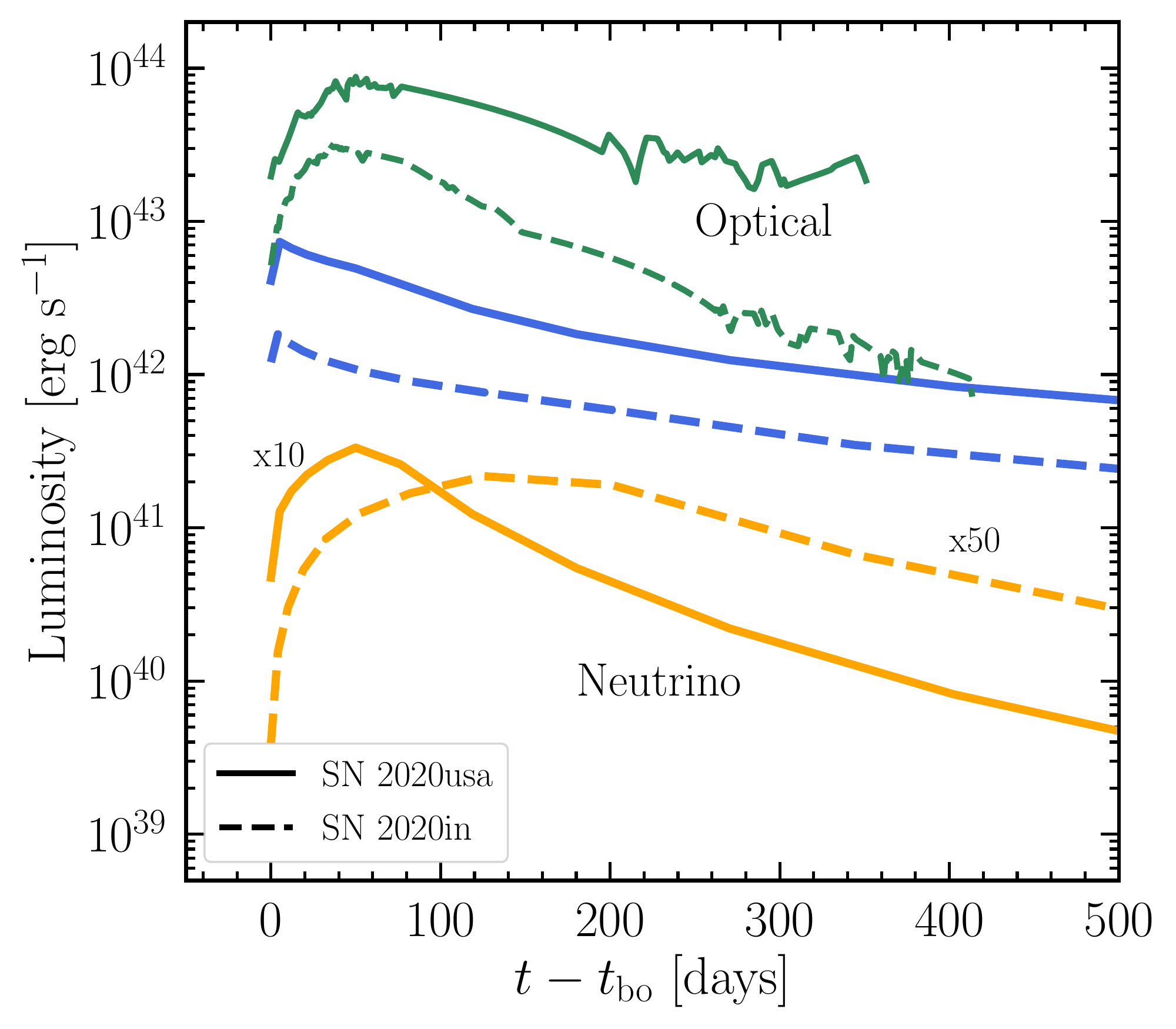}
	\caption{Muon neutrino and antineutrino  (taking into account  flavor oscillation, in blue and orange) and optical luminosities (after interpolation, in green) for SN 2020usa (solid lines) and SN 2020in (dashed lines) as  functions of time in the source frame. The two selected SNe exhibit a comparable evolution of the total neutrino luminosity (blue lines) because  $t_{\rm{rise}}$ and $L_{\rm{peak}}$ for both SNe are such to lead to similar parameters for what concerns the most optimistic prospects for neutrino emission. The blue curves have been obtained by considering the $100~\mathrm{GeV}$--$1~\mathrm{PeV}$ energy range. The orange lines represent the neutrino luminosity in the high energy range $100~\mathrm{TeV}$--$1~\mathrm{PeV}$ and  show how the peak of the high energy neutrinos is shifted [up to $\mathcal{O}(\rm{100\,days})$] with respect to the optical peak. 
 }
	\label{Fig:luminosities}
\end{figure}

Figure~\ref{Fig:luminosities} shows, for the most optimistic SN model parameter configuration, a comparison between $L_{\nu_{\mu}+\bar{\nu}_{\mu}}$ (obtained taking into account  flavor oscillation) and  the optical luminosity for SN 2020usa and SN 2020in. Besides the difference in the intrinsic optical brightness, the two SNe display  comparable evolution in  the neutrino luminosity, with the neutrino luminosity peak being  $\sim 3$ times brighter for  SN 2020usa than SN 2020in. This is  due to the fact that $t_{\rm{rise}}$ and $L_{\rm{peak}}$ for both SNe are such to lead to similar SN model parameters for what concerns the most optimistic prospects for neutrino emission. Note that an investigation that also takes into account  the late evolution of the optical lightcurve might have an impact on this result, but it out of the scope of this work.

\subsection{Characteristics of the detectable neutrino signal}
\label{sec:nudetection}

The  neutrino luminosity curve does not peak at the same time as the optical lightcurve, as visible from Fig.~\ref{Fig:luminosities}.
In fact the position of the optical peak is intrinsically related to propagation effects of photons in the CSM, and thus to the CSM properties, as discussed in Sec.~\ref{Sec:Scaling_relations} and Appendix~\ref{App:parameters}. 
The peak in the neutrino curve, instead, solely depends  on the CSM radial density distribution and the evolution of the maximum spectral energy. Because of this, the neutrino event rate as well as the neutrino luminosity in the high-energy range ($100~\mathrm{TeV}$--$1~\mathrm{PeV}$) peak at $t|_{E^{\ast}_{\rm{p,max}}}$, namely the time at which the maximum proton energy is reached (see Appendix~\ref{App:maxEp} for $E_{\rm{p,max}}$ and Fig.~\ref{Fig: SN2020_Flux_t_Enu} for the trend of the neutrino flux at Earth).

The most favorable time window for detecting energetic neutrinos ($\gtrsim 100$~TeV) would be a few times $t_{\rm rise}$ around the electromagnetic bolometric peak, which corresponds to $\mathcal{O}(100\,\rm{days})$ days for $E_{\rm{k}}\lesssim 10^{52}$~erg, $M_{\rm{ej}}\lesssim 10 M_{\odot}$, $M_{\rm{CSM}}\lesssim 20 M_{\odot}$, and $R_{\rm{CSM}}\lesssim {\rm{few}} \times 10^{16}$~cm (see Fig.~\ref{Fig:EP_MAX_cool_wind}).
Interestingly, the IceCube neutrino event IC200530A associated with the candidate SLSN event AT2019fdr was detected about $300$~days after the optical peak~\citep{Pitik:2021dyf}, in agreement  with our findings~\footnote{AT2019fdr occurred at $z \simeq 0.27$, and the optical lightcurve displayed  $t_{\rm{rise}} = 98$~days and $L_{\rm{peak}} = 2.1 \times 10^{44}$~erg$/$s, considering a radiative efficiency of $18$--$35\%$, \cite{Pitik:2021dyf} estimated that about $4.6 \times 10^{-2}$ muon neutrino and antineutrino events were  expected assuming excellent discrimination of the atmospheric background.}.

\begin{figure}
	\centering
	\includegraphics[width=0.5\textwidth]{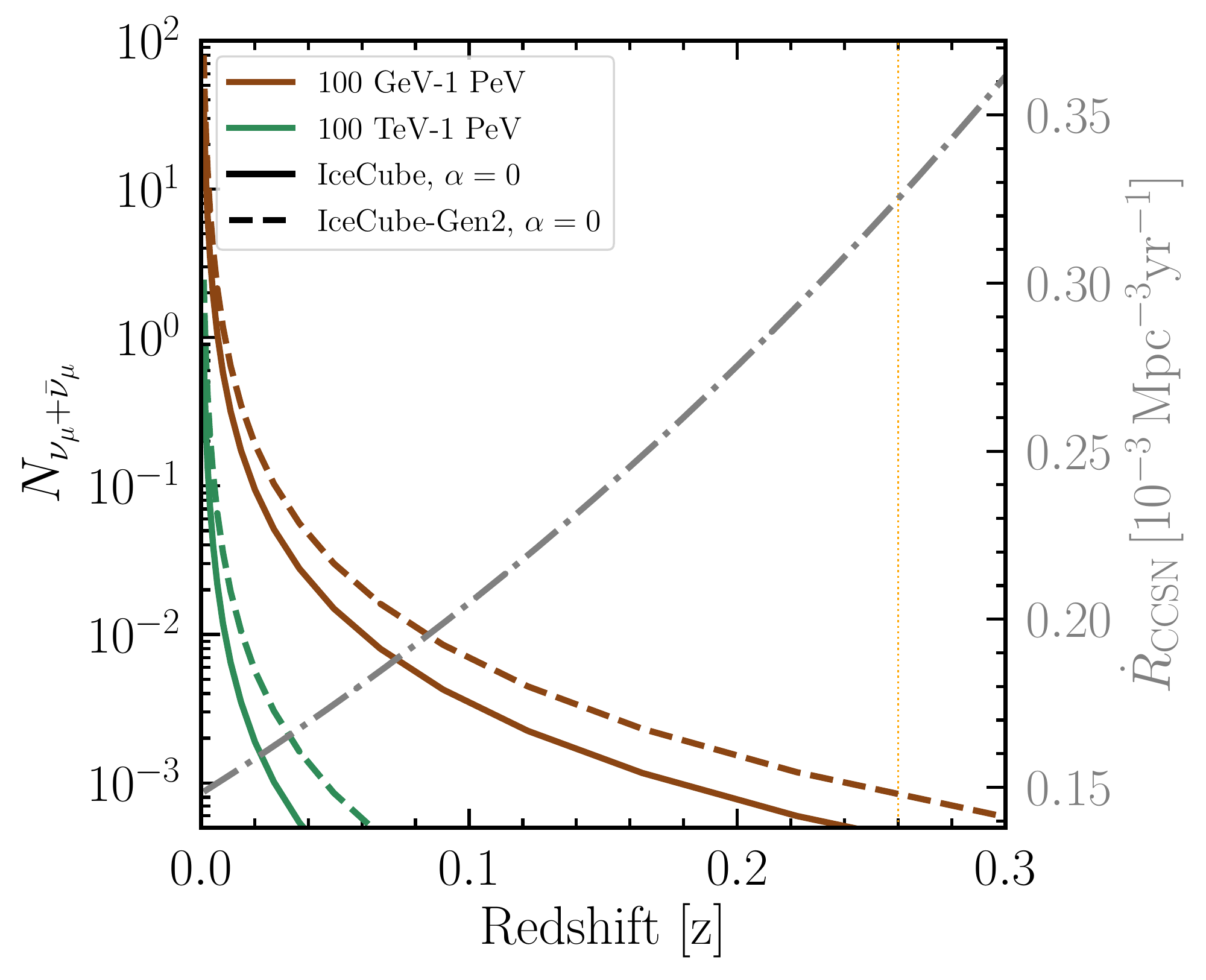}
	\caption{Number of muon neutrino and antineutrino events expected at the IceCube Neutrino Observatory (solid lines) and IceCube Gen2 (dashed lines) as  functions of the redshfit for a benchmark SN with the same properties of SN 2020usa but located at declination $\alpha=0$ and variable $z$. The number of neutrino events is obtained integrating up to $200$~days to optimize the signal discrimination with respect to the background. The redshift of SN 2020usa is marked with a dashed orange line to guide the eye. The core-collapse SN rate is plotted as a dot-dashed line (see y-axis scale on the right), in order to compare the expected number of neutrino events with the probability of  finding SNe at a given $z$; the core-collapse SNe rate should be considered as an upper limit of the rate of interaction-powered SNe and SLSNe (see main text for details).  We expect $N_{\nu_\mu+\bar\nu_\mu} =10$ at $z \simeq 0.002$ ($d_{\rm{L}}\geq9$~Mpc) for IceCube and $z \simeq 0.003$ ($d_{\rm{L}}\geq13$~Mpc) for IceCube-Gen2. }
	\label{Fig: SN2020_Nutot_z}
\end{figure}

Figure~\ref{Fig: SN2020_Nutot_z} shows the dependence of the  number of neutrino events expected in IceCube and IceCube-Gen2 in a temporal window of 200 days and as functions  of the redshift for a  benchmark SN with the same properties of SN 2020usa, but placed at declination $\alpha = 0$~deg and  redshift $z$. We consider the  number of neutrino events expected in a time window of $200$~days in order to optimize the signal over background classification (see Sec.~\ref{sec:MM}). One can see that IceCube expects to detect $N_{\nu_\mu+\bar\nu_\mu} \gtrsim 10$ for SNe at distance $\lesssim 9$~Mpc ($z\lesssim 0.002$); while $N_{\nu_\mu+\bar\nu_\mu} \gtrsim 10$ should be detected for SNe at a distance $ \lesssim 13$~Mpc ($z\lesssim 0.003$) for IceCube-Gen2~\citep{IceCube-Gen2:2020qha}. 

In order to compare the expected number of neutrino events with the likelihood of finding SNe at redshift $z$, Fig.~\ref{Fig: SN2020_Nutot_z} also shows the core-collapse SN rate~\citep{Yuksel:2008cu,Vitagliano:2019yzm} for reference. Note that the rate of interaction-powered SNe is very uncertain and it is not clear whether their redshift evolution follows the star-formation rate~\citep{Smith:2010vz}; hence the core-collapse SN rate should be considered as an upper limit of the rate of interaction-powered SNe and SLSNe, under the assumption that the latter follow the same redshift evolution.

The evolution of the energy flux of neutrinos is displayed in Fig.~\ref{Fig: SN2020_Flux_t_Enu}. One can see that for $E_{\nu}\gtrsim 100$~TeV, the energy flux increases up to around $100$~days, and then decreases. This trend can be explained considering the evolution of $E_{\rm{p,max}}$ (see also Fig.~\ref{Fig:luminosities}). 
\begin{figure}
	\centering
	\includegraphics[width=0.5\textwidth]{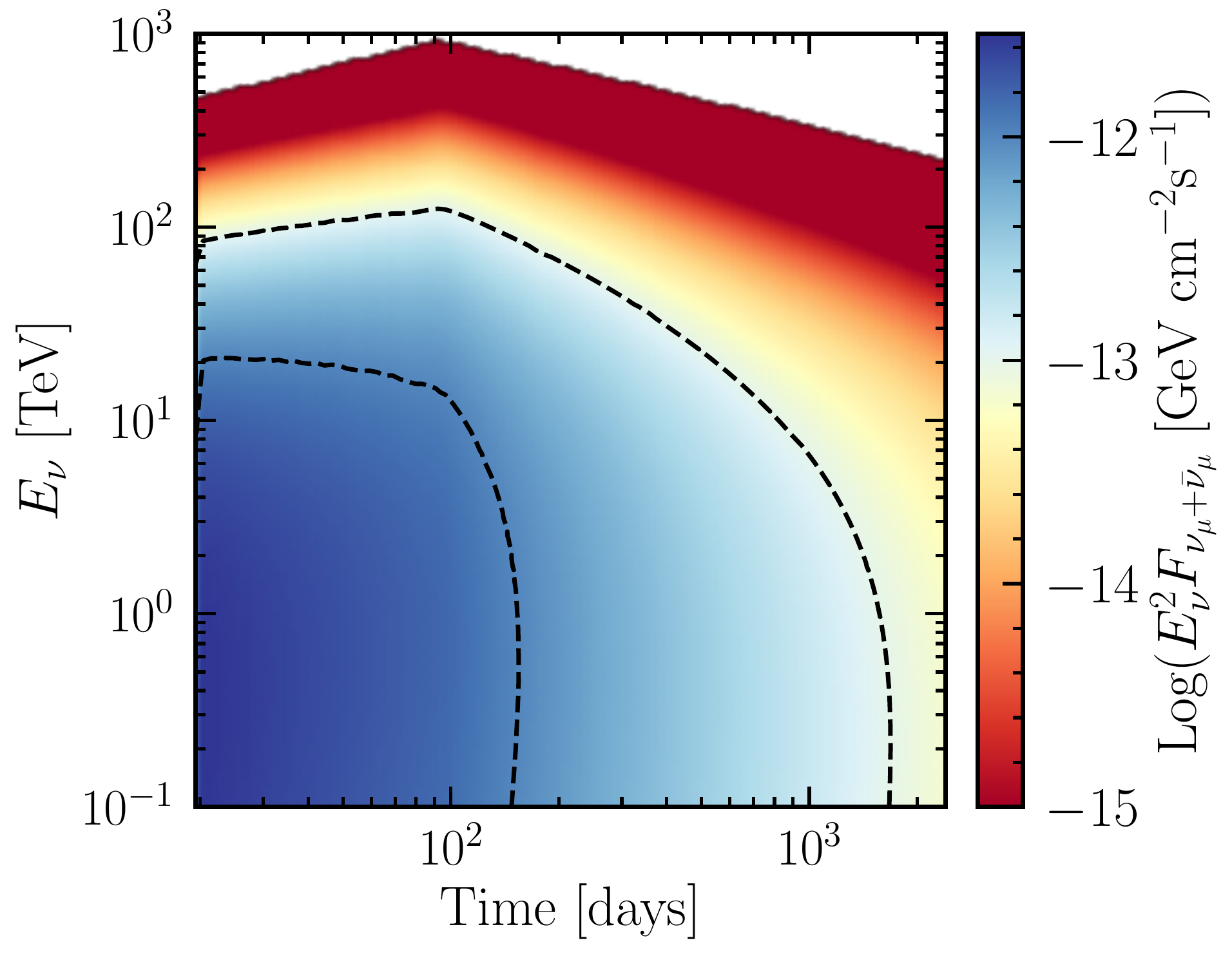}
	\caption{Contour plot of the muon neutrino and antineutrino energy flux expected at Earth for SN2020usa in the best case scenario and in the plane spanned by the arrival time of neutrinos  and the neutrino energy. At low energies the neutrino flux decreases with time after the breakout. At high energies ($E_{\nu}\gtrsim 100$~TeV), instead, it  increases with time, peaks at around $100$~days, and then decreases. This is related to the time of maximum $E_{\rm{p,max}}$ (see also Fig.~\ref{Fig:luminosities}). The white color marks the regions where the flux is zero.}
	\label{Fig: SN2020_Flux_t_Enu}
\end{figure}

\subsection{Follow-up strategy for neutrino searches}
\label{sec:follow-up_strategy}

Our findings in Sec.~\ref{Sec: Spectral energy distributions of protons and neutrinos } suggest that a large $L_{\rm{peak}}$ and average $t_{\rm{rise}}$ are necessary, but not sufficient, to guarantee large neutrino emission. This is due to the large degeneracy existing in the SN model parameter space that could lead to SN lightcurves with comparable properties in the optical, but largely different neutrino emission.

Despite the degeneracy in the SN properties leading to comparable optical signals, the semi-analytical procedure outlined in this work allows to restrict the range of $E_{\rm{k}}$, $M_{\rm{ej}}$, $M_{\rm{CSM}}$, and  $R_{\rm{CSM}}$ that matches the measured  $t_{\rm{rise}}$ and $L_{\rm{peak}}$. This procedure then  forecasts an expectation range for the number of neutrino events  detectable by IceCube to guide upcoming follow-up searches  (see Sec.~\ref{sec:SN2020usa-SN 2020in} for an application to two SNe detected by ZTF), also taking into account the unknown radiative efficiency $\varepsilon_{\rm{rad}}$.  

For measured  $t_{\rm{rise}}$ and $L_{\rm{peak}}$, through the method outlined in this paper, it is possible to predict the largest expected number of neutrino events. On the other hand, if an interaction-powered SN should be detected in the optical, and   no neutrino should be observed, this would imply that the SN model parameters compatible with the measured  $t_{\rm{rise}}$ and $L_{\rm{peak}}$ are not optimal for neutrino production.

Our findings highlight the need to carry out multi-wavelength SN observations to better infer the SN properties and then optimize neutrino searches through the procedure presented in this work. In fact,  relying on radio and X-ray all-sky surveys, one could narrow down the values of  $M_{\rm{ej}}$, $M_{\rm{CSM}}$, and  $R_{\rm{CSM}}$ compatible with the data~\citep{Margalit:2021bqe,Chevalier:2016hzo}. Because CSM interaction signatures appear clearly in the UV, early SN observations by future ultraviolet satellites, such as ULTRASAT~\citep{Shvartzvald:2023ofi}, will be critical to provide insights into the CSM properties. Further information on the CSM can be obtained in the X-ray regime~\citep{Margalit:2021bqe}, e.g.~through surveys such as the extended ROentgen Survey with an Imaging Telescope Array~\citep[eROSITA;][]{10.1117/12.856577}. In addition, the Vera Rubin Observatory~\citep{2009arXiv0912.0201L} will detect numerous  SNe providing a large sample for a neutrino stacking analysis.

\subsection{Multi-messenger follow-up programs}
\label{sec:MM}
There are two ways to search for neutrinos from SNe. 
\begin{itemize}
\item[-] One can compile a catalog of SNe detected by electromagnetic surveys and use archival all-sky neutrino data to search for an excess of neutrinos from the catalogued sources. Such a search is most sensitive when a stacking of all sources is applied~\citep[see e.g.][]{IceCube:2023esf}. The stacking requires a weighting of the sources relative to each other. Previous searches assumed that all sources are neutrino standard candles, i.e.~the neutrino flux at Earth would scale with the inverse of the square of the luminosity distance, or used the optical peak flux as a weight. This work indicates that neither of those assumptions is justified. 
Modeling of the multi-wavelength emission can yield a source-by-source prediction of the neutrino emission, which can be used as a weight. 

Another important analysis choice is the time window to consider for the neutrino search. A too long time window increases the background of atmospheric neutrinos, while a too short time window cuts parts of the signal. The prediction of the temporal evolution of the neutrino signal  by our modeling allows to optimize the neutrino search time window. Finally, also the spectral energy distribution of neutrinos from SNe can be used to optimize the analysis in terms of background rejection.

\item[-] One can utilize electromagnetic follow-up observations of neutrino alerts released by neutrino telescopes~\citep[see e.g.,][]{IceCube:2016cqr}. Also here, defining a time window in order to assess the coincidence between an electromagnetic counterpart and the neutrino alert will be crucial. Once a potential counterpart is identified further follow-up observations (e.g.~spectroscopy and multiple wavelength) can be scheduled to ensure classification of the source as SN and allow for a characterization of the CSM properties.

In order to forecast the expected neutrino signal reliably and better guide neutrino searches, in addition to optical data, input from  X-ray and radio surveys would allow to  characterize the CSM properties (see Sec.~\ref{sec:follow-up_strategy}). In addition, it would be helpful to guide neutrino searches  relying on the optical spectra at different times to characterize the duration of the interaction.
\end{itemize}
In summary, the modeling of particle emission from SNe presented in this paper will be crucial to guide targeted multi-messenger searches.

\section{Conclusions}
\label{sec:conclusions}
Supernovae and SLSNe of Type IIn show in their spectra strong signs of circumstellar interaction with a hydrogen-rich medium.
The interaction between the SN ejecta and the CSM  powers  thermal UV/optical emission as well as high-energy neutrino emission. This work aims to explore the connection between the  energy emitted in neutrinos detectable at the IceCube Neutrino Observatory (and its successor IceCube-Gen2) and the photometric properties of the optical signals observable by wide-field, high-cadence surveys. Our  main goal is to outline the best follow-up strategy for upcoming multi-messenger searches.

We rely on a semi-analytical model that connects the optical lightcurve observables to the SN properties and the correspondent neutrino emission, we find that the largest  energy emitted in neutrinos and antineutrino is expected for large SN kinetic energy ($E_{\rm{k}}\gtrsim 10^{51}$~erg), small ejecta mass ($M_{\rm{ej}}\lesssim 10 \, M_{\odot}$), intermediate CSM mass ($1\, M_{\odot}\lesssim M_{\rm{CSM}}\lesssim 30\, M_{\odot}$),  and  extended CSM  ($R_{\rm{CSM}}\gtrsim 10^{16}$~cm). Such parameters lead to large bolometric peak luminosity ($ L_{\rm{peak}}\gtrsim 10^{43}$--$10^{44}$~erg) and  average rise time ($t_{\rm{rise}}\gtrsim 10$--$40$~days). However, these lightcurve features are necessary but not sufficient to guarantee ideal conditions for neutrino detection.
In fact, different  configurations of the SN model parameters could lead to comparable optical lightcurves, but vastly different neutrino emission. 
 
The degeneracy between the optical lightcurve properties and the SN model parameters challenges the possibility of outlining a simple procedure to determine the expected  number of IceCube neutrino events by solely relying on SN observations in the optical. While our method allows to foresee the largest possible number of neutrino events for given $ L_{\rm{peak}}$ and $t_{\rm{rise}}$, the eventual lack of neutrino detection for upcoming nearby SNe  could hint towards SN properties that are different with respect to the ones maximizing the neutrino signal, therefore  constraining the SN model parameter space compatible with neutrino and optical observations.

We also find that the peak of the neutrino curve does not coincide with the one of the optical lightcurve. Hence, one should consider a time window of a few $t_{\rm rise}$ around $L_{\rm peak}$ when looking for neutrinos. The time window should indeed be optimized to guarantee a fair signal discrimination with respect to the background. 

Our findings suggest that previous neutrino stacking searches that assumed all SNe as neutrino
standard candles, or used weights based on optical peak flux,  might have not been optimal, as they do not take into account the diversity in the SN properties leading to a large variation in the  number of neutrinos expected at Earth.  Importantly, multi-wavelength observations are  necessary to break the  degeneracy between the optical lightcurve and the SN properties and will be essential to forecast the expected neutrino signal and optimize multi-messenger searches.

\section*{Acknowledgments}
We are grateful to Takashi Moriya for insightful discussions, Steve Schulze for discussions on the ZTF lightcurve data, and Jakob van Santen for exchanges concerning the IceCube-Gen2 effective areas. We acknowledge support from the Villum Foundation (Project No.~13164), the Carlsberg Foundation (CF18-0183), as well as the Deutsche Forschungsgemeinschaft through Sonderforschungbereich
SFB~1258 ``Neutrinos and Dark Matter in Astro- and
Particle Physics'' (NDM) and
the Collaborative Research Center SFB~1491 ``Cosmic Interacting Matters - from Source to Signal.'' T.P.~also acknowledges support from Fondo Ricerca di Base 2020 (MOSAICO) of the University of Perugia.

\section*{Data Availability}
Data can be shared upon reasonable request to the authors.

\bibliographystyle{mnras}
\bibliography{myreferences_mnras}

\begin{thebibliography}{}
\makeatletter
\relax
\def\mn@urlcharsother{\let\do\@makeother \do\$\do\&\do\#\do\^\do\_\do\%\do\~}
\def\mn@doi{\begingroup\mn@urlcharsother \@ifnextchar [ {\mn@doi@}
  {\mn@doi@[]}}
\def\mn@doi@[#1]#2{\def\@tempa{#1}\ifx\@tempa\@empty \href
  {http://dx.doi.org/#2} {doi:#2}\else \href {http://dx.doi.org/#2} {#1}\fi
  \endgroup}
\def\mn@eprint#1#2{\mn@eprint@#1:#2::\@nil}
\def\mn@eprint@arXiv#1{\href {http://arxiv.org/abs/#1} {{\tt arXiv:#1}}}
\def\mn@eprint@dblp#1{\href {http://dblp.uni-trier.de/rec/bibtex/#1.xml}
  {dblp:#1}}
\def\mn@eprint@#1:#2:#3:#4\@nil{\def\@tempa {#1}\def\@tempb {#2}\def\@tempc
  {#3}\ifx \@tempc \@empty \let \@tempc \@tempb \let \@tempb \@tempa \fi \ifx
  \@tempb \@empty \def\@tempb {arXiv}\fi \@ifundefined
  {mn@eprint@\@tempb}{\@tempb:\@tempc}{\expandafter \expandafter \csname
  mn@eprint@\@tempb\endcsname \expandafter{\@tempc}}}

\bibitem[\protect\citeauthoryear{Aartsen et~al.}{Aartsen
  et~al.}{2017}]{IceCube:2016cqr}
Aartsen M.~G.,  et~al., 2017, \mn@doi [Astropart. Phys.]
  {10.1016/j.astropartphys.2017.05.002}, 92, 30

\bibitem[\protect\citeauthoryear{Aartsen et~al.}{Aartsen
  et~al.}{2021}]{IceCube-Gen2:2020qha}
Aartsen M.~G.,  et~al., 2021, \mn@doi [J. Phys. G] {10.1088/1361-6471/abbd48},
  48, 060501

\bibitem[\protect\citeauthoryear{Abbasi et~al.}{Abbasi
  et~al.}{2021a}]{Abbasi:2020jmh}
Abbasi R.,  et~al., 2021a, \mn@doi [Phys. Rev. D]
  {10.1103/PhysRevD.104.022002}, 104, 022002

\bibitem[\protect\citeauthoryear{Abbasi et~al.}{Abbasi
  et~al.}{2021b}]{IceCube:2020mzw}
Abbasi R.,  et~al., 2021b, \mn@doi [Astrophys. J.] {10.3847/1538-4357/abe123},
  910, 4

\bibitem[\protect\citeauthoryear{{Abbasi} et~al.,}{{Abbasi}
  et~al.}{2023}]{IceCube:2023esf}
{Abbasi} R.,  et~al., 2023, \mn@doi [Astrophys. J. Lett.]
  {10.3847/2041-8213/acd2c9}, \href
  {https://ui.adsabs.harvard.edu/abs/2023ApJ...949L..12A} {949, L12}

\bibitem[\protect\citeauthoryear{Aghanim et~al.}{Aghanim
  et~al.}{2020}]{Aghanim:2018eyx}
Aghanim N.,  et~al., 2020, \mn@doi [Astron. Astrophys.]
  {10.1051/0004-6361/201833910}, 641, A6

\bibitem[\protect\citeauthoryear{Ahlers \& Halzen}{Ahlers \&
  Halzen}{2018}]{Ahlers:2018fkn}
Ahlers M.,  Halzen F.,  2018, \mn@doi [Prog. Part. Nucl. Phys.]
  {10.1016/j.ppnp.2018.05.001}, 102, 73

\bibitem[\protect\citeauthoryear{{Ambikasaran}, {Foreman-Mackey}, {Greengard},
  {Hogg}  \& {O'Neil}}{{Ambikasaran} et~al.}{2015}]{george}
{Ambikasaran} S.,  {Foreman-Mackey} D.,  {Greengard} L.,  {Hogg} D.~W.,
  {O'Neil} M.,  2015, \mn@doi [IEEE Transactions on Pattern Analysis and
  Machine Intelligence] {10.1109/TPAMI.2015.2448083}, \href
  {https://ui.adsabs.harvard.edu/abs/2015ITPAM..38..252A} {38, 252}

\bibitem[\protect\citeauthoryear{Anchordoqui et~al.}{Anchordoqui
  et~al.}{2014}]{Anchordoqui:2013dnh}
Anchordoqui L.~A.,  et~al., 2014, \mn@doi [JHEAp]
  {10.1016/j.jheap.2014.01.001}, 1-2, 1

\bibitem[\protect\citeauthoryear{{Bell}}{{Bell}}{2004}]{2004MNRAS.353..550B}
{Bell} A.~R.,  2004, \mn@doi [Mon. Not. Roy. Astron. Soc.]
  {10.1111/j.1365-2966.2004.08097.x}, \href
  {https://ui.adsabs.harvard.edu/abs/2004MNRAS.353..550B} {353, 550}

\bibitem[\protect\citeauthoryear{Bell}{Bell}{2013}]{Bell:2013vxa}
Bell A.~R.,  2013, \mn@doi [Astropart. Phys.]
  {10.1016/j.astropartphys.2012.05.022}, 43, 56

\bibitem[\protect\citeauthoryear{{Bellm} et~al.,}{{Bellm}
  et~al.}{2019}]{2019PASP..131a8002B}
{Bellm} E.~C.,  et~al., 2019, \mn@doi [\pasp] {10.1088/1538-3873/aaecbe}, \href
  {https://ui.adsabs.harvard.edu/abs/2019PASP..131a8002B} {131, 018002}

\bibitem[\protect\citeauthoryear{Blasi}{Blasi}{2013}]{Blasi:2013rva}
Blasi P.,  2013, \mn@doi [Astron. Astrophys. Rev.] {10.1007/s00159-013-0070-7},
  21, 70

\bibitem[\protect\citeauthoryear{{Blinnikov}}{{Blinnikov}}{2017}]{2017hsn..book..843B}
{Blinnikov} S.,  2017, in {Alsabti} A.~W.,  {Murdin} P.,  eds, , Handbook of
  Supernovae.
p.~843, \mn@doi{10.1007/978-3-319-21846-5\_31}

\bibitem[\protect\citeauthoryear{{Brose}, {Sushch}  \& {Mackey}}{{Brose}
  et~al.}{2022}]{Brose:2022mbx}
{Brose} R.,  {Sushch} I.,   {Mackey} J.,  2022, \mn@doi [Mon. Not. Roy. Astron.
  Soc.] {10.1093/mnras/stac2234}, \href
  {https://ui.adsabs.harvard.edu/abs/2022MNRAS.516..492B} {516, 492}

\bibitem[\protect\citeauthoryear{Caprioli \& Spitkovsky}{Caprioli \&
  Spitkovsky}{2014}]{Caprioli:2013dca}
Caprioli D.,  Spitkovsky A.,  2014, \mn@doi [Astrophys. J.]
  {10.1088/0004-637X/783/2/91}, 783, 91

\bibitem[\protect\citeauthoryear{Cardillo, Amato  \& Blasi}{Cardillo
  et~al.}{2015}]{Cardillo:2015zda}
Cardillo M.,  Amato E.,   Blasi P.,  2015, \mn@doi [Astropart. Phys.]
  {10.1016/j.astropartphys.2015.03.002}, 69, 1

\bibitem[\protect\citeauthoryear{{Chambers} et~al.,}{{Chambers}
  et~al.}{2016}]{2016arXiv161205560C}
{Chambers} K.~C.,  et~al., 2016, \mn@doi [arXiv e-prints]
  {10.48550/arXiv.1612.05560}, \href
  {https://ui.adsabs.harvard.edu/abs/2016arXiv161205560C} {p. arXiv:1612.05560}

\bibitem[\protect\citeauthoryear{Chatzopoulos, Wheeler  \& Vinko}{Chatzopoulos
  et~al.}{2012}]{Chatzopoulos:2011vj}
Chatzopoulos E.,  Wheeler J.~C.,   Vinko J.,  2012, \mn@doi [Astrophys. J.]
  {10.1088/0004-637X/746/2/121}, 746, 121

\bibitem[\protect\citeauthoryear{Chatzopoulos, Wheeler, Vinko, Horvath  \&
  Nagy}{Chatzopoulos et~al.}{2013}]{Chatzopoulos:2013vfa}
Chatzopoulos E.,  Wheeler J.~C.,  Vinko J.,  Horvath Z.~L.,   Nagy A.,  2013,
  \mn@doi [Astrophys. J.] {10.1088/0004-637X/773/1/76}, 773, 76

\bibitem[\protect\citeauthoryear{{Chevalier}}{{Chevalier}}{1982}]{1982ApJ...259..302C}
{Chevalier} R.~A.,  1982, \mn@doi [\apj] {10.1086/160167}, \href
  {https://ui.adsabs.harvard.edu/abs/1982ApJ...259..302C} {259, 302}

\bibitem[\protect\citeauthoryear{{Chevalier} \& {Fransson}}{{Chevalier} \&
  {Fransson}}{1994}]{1994ApJ...420..268C}
{Chevalier} R.~A.,  {Fransson} C.,  1994, \mn@doi [Astrophys. J.]
  {10.1086/173557}, \href
  {https://ui.adsabs.harvard.edu/abs/1994ApJ...420..268C} {420, 268}

\bibitem[\protect\citeauthoryear{{Chevalier} \& {Fransson}}{{Chevalier} \&
  {Fransson}}{2017}]{Chevalier:2016hzo}
{Chevalier} R.~A.,  {Fransson} C.,  2017, in {Alsabti} A.~W.,  {Murdin} P.,
  eds, , Handbook of Supernovae.
p.~875, \mn@doi{10.1007/978-3-319-21846-5_34}

\bibitem[\protect\citeauthoryear{Chevalier \& Irwin}{Chevalier \&
  Irwin}{2011}]{Chevalier:2011ha}
Chevalier R.~A.,  Irwin C.~M.,  2011, \mn@doi [Astrophys. J. Lett.]
  {10.1088/2041-8205/729/1/L6}, 729, L6

\bibitem[\protect\citeauthoryear{Cristofari}{Cristofari}{2021}]{Cristofari:2021jkl}
Cristofari P.,  2021, \mn@doi [Universe] {10.3390/universe7090324}, 7, 324

\bibitem[\protect\citeauthoryear{Cristofari, Blasi  \& Amato}{Cristofari
  et~al.}{2020}]{Cristofari:2020mdf}
Cristofari P.,  Blasi P.,   Amato E.,  2020, \mn@doi [Astropart. Phys.]
  {10.1016/j.astropartphys.2020.102492}, 123, 102492

\bibitem[\protect\citeauthoryear{{Dekany} et~al.,}{{Dekany}
  et~al.}{2020}]{2020PASP..132c8001D}
{Dekany} R.,  et~al., 2020, \mn@doi [\pasp] {10.1088/1538-3873/ab4ca2}, \href
  {https://ui.adsabs.harvard.edu/abs/2020PASP..132c8001D} {132, 038001}

\bibitem[\protect\citeauthoryear{{Draine}}{{Draine}}{2011}]{2011piim.book.....D}
{Draine} B.~T.,  2011, {Physics of the Interstellar and Intergalactic Medium}

\bibitem[\protect\citeauthoryear{Drake et~al.}{Drake
  et~al.}{2011}]{Drake:2011kg}
Drake A.~J.,  et~al., 2011, \mn@doi [Astrophys. J.]
  {10.1088/0004-637X/735/2/106}, 735, 106

\bibitem[\protect\citeauthoryear{{Ellison}, {Patnaude}, {Slane}, {Blasi}  \&
  {Gabici}}{{Ellison} et~al.}{2007}]{2007ApJ...661..879E}
{Ellison} D.~C.,  {Patnaude} D.~J.,  {Slane} P.,  {Blasi} P.,   {Gabici} S.,
  2007, \mn@doi [Astrophys. J.] {10.1086/517518}, \href
  {https://ui.adsabs.harvard.edu/abs/2007ApJ...661..879E} {661, 879}

\bibitem[\protect\citeauthoryear{Esteban, Gonzalez-Garcia, Maltoni, Schwetz  \&
  Zhou}{Esteban et~al.}{2020}]{Esteban:2020cvm}
Esteban I.,  Gonzalez-Garcia M.,  Maltoni M.,  Schwetz T.,   Zhou A.,  2020,
  \mn@doi [JHEP] {10.1007/JHEP09(2020)178}, 09, 178

\bibitem[\protect\citeauthoryear{Fang, Metzger, Vurm, Aydi  \& Chomiuk}{Fang
  et~al.}{2020}]{Fang:2020bkm}
Fang K.,  Metzger B.~D.,  Vurm I.,  Aydi E.,   Chomiuk L.,  2020, \mn@doi
  [Astrophys. J.] {10.3847/1538-4357/abbc6e}, 904, 4

\bibitem[\protect\citeauthoryear{Filippenko}{Filippenko}{1997}]{Filippenko:1997ub}
Filippenko A.~V.,  1997, \mn@doi [Ann. Rev. Astron. Astrophys.]
  {10.1146/annurev.astro.35.1.309}, 35, 309

\bibitem[\protect\citeauthoryear{{Finke} \& {Dermer}}{{Finke} \&
  {Dermer}}{2012}]{2012ApJ...751...65F}
{Finke} J.~D.,  {Dermer} C.~D.,  2012, \mn@doi [Astrophys. J.]
  {10.1088/0004-637X/751/1/65}, \href
  {https://ui.adsabs.harvard.edu/abs/2012ApJ...751...65F} {751, 65}

\bibitem[\protect\citeauthoryear{Gal-Yam}{Gal-Yam}{2012}]{Gal-Yam:2012ukv}
Gal-Yam A.,  2012, \mn@doi [Science] {10.1126/science.1203601}, 337, 927

\bibitem[\protect\citeauthoryear{{Gal-Yam}}{{Gal-Yam}}{2017}]{Gal-Yam:2016yms}
{Gal-Yam} A.,  2017, in {Alsabti} A.~W.,  {Murdin} P.,  eds, , Handbook of
  Supernovae.
p.~195, \mn@doi{10.1007/978-3-319-21846-5_35}

\bibitem[\protect\citeauthoryear{Gal-Yam}{Gal-Yam}{2019}]{Gal-Yam:2018out}
Gal-Yam A.,  2019, \mn@doi [Ann. Rev. Astron. Astrophys.]
  {10.1146/annurev-astro-081817-051819}, 57, 305

\bibitem[\protect\citeauthoryear{Gal-Yam et~al.}{Gal-Yam
  et~al.}{2007}]{Gal-Yam:2006kfs}
Gal-Yam A.,  et~al., 2007, \mn@doi [Astrophys. J.] {10.1086/510523}, 656, 372

\bibitem[\protect\citeauthoryear{Ginzburg \& Balberg}{Ginzburg \&
  Balberg}{2012}]{Ginzburg:2012dc}
Ginzburg S.,  Balberg S.,  2012, \mn@doi [Astrophys. J.]
  {10.1088/0004-637X/757/2/178}, 757, 178

\bibitem[\protect\citeauthoryear{{Halzen} \& {Kheirandish}}{{Halzen} \&
  {Kheirandish}}{2022}]{Halzen:2022pez}
{Halzen} F.,  {Kheirandish} A.,  2022, arXiv e-prints, \href
  {https://ui.adsabs.harvard.edu/abs/2022arXiv220200694H} {p. arXiv:2202.00694}

\bibitem[\protect\citeauthoryear{{Hambleton} et~al.,}{{Hambleton}
  et~al.}{2022}]{LSST:2022kad}
{Hambleton} K.~M.,  et~al., 2022, arXiv e-prints, \href
  {https://ui.adsabs.harvard.edu/abs/2022arXiv220804499H} {p. arXiv:2208.04499}

\bibitem[\protect\citeauthoryear{{IceCube Collaboration} et~al.,}{{IceCube
  Collaboration} et~al.}{2021}]{IceCube:2021xar}
{IceCube Collaboration} et~al., 2021, \mn@doi [arXiv e-prints]
  {10.48550/arXiv.2101.09836}, \href
  {https://ui.adsabs.harvard.edu/abs/2021arXiv210109836I} {p. arXiv:2101.09836}

\bibitem[\protect\citeauthoryear{Kankare et~al.}{Kankare
  et~al.}{2019}]{Pan-STARRS:2019szg}
Kankare E.,  et~al., 2019, \mn@doi [Astron. Astrophys.]
  {10.1051/0004-6361/201935171}, 626, A117

\bibitem[\protect\citeauthoryear{{Katz}, {Sapir}  \& {Waxman}}{{Katz}
  et~al.}{2011}]{Katz:2011zx}
{Katz} B.,  {Sapir} N.,   {Waxman} E.,  2011, arXiv e-prints, \href
  {https://ui.adsabs.harvard.edu/abs/2011arXiv1106.1898K} {p. arXiv:1106.1898}

\bibitem[\protect\citeauthoryear{Kelner, Aharonian  \& Bugayov}{Kelner
  et~al.}{2006}]{Kelner:2006tc}
Kelner S.~R.,  Aharonian F.~A.,   Bugayov V.~V.,  2006, \mn@doi [Phys. Rev. D]
  {10.1103/PhysRevD.74.034018}, 74, 034018

\bibitem[\protect\citeauthoryear{{Kheirandish} \& {Murase}}{{Kheirandish} \&
  {Murase}}{2022}]{Kheirandish:2022eox}
{Kheirandish} A.,  {Murase} K.,  2022, arXiv e-prints, \href
  {https://ui.adsabs.harvard.edu/abs/2022arXiv220408518K} {p. arXiv:2204.08518}

\bibitem[\protect\citeauthoryear{{Kochanek} et~al.,}{{Kochanek}
  et~al.}{2017}]{2017PASP..129j4502K}
{Kochanek} C.~S.,  et~al., 2017, \mn@doi [\pasp] {10.1088/1538-3873/aa80d9},
  \href {https://ui.adsabs.harvard.edu/abs/2017PASP..129j4502K} {129, 104502}

\bibitem[\protect\citeauthoryear{{LSST Science Collaboration} et~al.,}{{LSST
  Science Collaboration} et~al.}{2009}]{2009arXiv0912.0201L}
{LSST Science Collaboration} et~al., 2009, \mn@doi [arXiv e-prints]
  {10.48550/arXiv.0912.0201}, \href
  {https://ui.adsabs.harvard.edu/abs/2009arXiv0912.0201L} {p. arXiv:0912.0201}

\bibitem[\protect\citeauthoryear{Levinson \& Bromberg}{Levinson \&
  Bromberg}{2008}]{Levinson:2007rj}
Levinson A.,  Bromberg O.,  2008, \mn@doi [Phys. Rev. Lett.]
  {10.1103/PhysRevLett.100.131101}, 100, 131101

\bibitem[\protect\citeauthoryear{{Lodders}}{{Lodders}}{2019}]{2019arXiv191200844L}
{Lodders} K.,  2019, arXiv e-prints, \href
  {https://ui.adsabs.harvard.edu/abs/2019arXiv191200844L} {p. arXiv:1912.00844}

\bibitem[\protect\citeauthoryear{Margalit, Quataert  \& Ho}{Margalit
  et~al.}{2022}]{Margalit:2021bqe}
Margalit B.,  Quataert E.,   Ho A. Y.~Q.,  2022, \mn@doi [Astrophys. J.]
  {10.3847/1538-4357/ac53b0}, 928, 122

\bibitem[\protect\citeauthoryear{Matzner \& McKee}{Matzner \&
  McKee}{1999}]{Matzner:1998mg}
Matzner C.~D.,  McKee C.~F.,  1999, \mn@doi [Astrophys. J.] {10.1086/306571},
  510, 379

\bibitem[\protect\citeauthoryear{M\'esz\'aros}{M\'esz\'aros}{2017}]{Meszaros:2017fcs}
M\'esz\'aros P.,  2017, \mn@doi [Ann. Rev. Nucl. Part. Sci.]
  {10.1146/annurev-nucl-101916-123304}, 67, 45

\bibitem[\protect\citeauthoryear{Moriya, Maeda, Taddia, Sollerman, Blinnikov
  \& Sorokina}{Moriya et~al.}{2013}]{Moriya:2013hka}
Moriya T.~J.,  Maeda K.,  Taddia F.,  Sollerman J.,  Blinnikov S.~I.,
  Sorokina E.~I.,  2013, \mn@doi [Mon. Not. Roy. Astron. Soc.]
  {10.1093/mnras/stt1392}, 435, 1520

\bibitem[\protect\citeauthoryear{Moriya, Sorokina  \& Chevalier}{Moriya
  et~al.}{2018}]{Moriya:2018sig}
Moriya T.~J.,  Sorokina E.~I.,   Chevalier R.~A.,  2018, \mn@doi [Space Sci.
  Rev.] {10.1007/s11214-018-0493-6}, 214, 59

\bibitem[\protect\citeauthoryear{Murase, Thompson, Lacki  \& Beacom}{Murase
  et~al.}{2011}]{Murase:2010cu}
Murase K.,  Thompson T.~A.,  Lacki B.~C.,   Beacom J.~F.,  2011, \mn@doi [Phys.
  Rev. D] {10.1103/PhysRevD.84.043003}, 84, 043003

\bibitem[\protect\citeauthoryear{{Necker} et~al.,}{{Necker}
  et~al.}{2022}]{2022MNRAS.516.2455N}
{Necker} J.,  et~al., 2022, \mn@doi [Mon. Not. Roy. Astron.]
  {10.1093/mnras/stac2261}, \href
  {https://ui.adsabs.harvard.edu/abs/2022MNRAS.516.2455N} {516, 2455}

\bibitem[\protect\citeauthoryear{{Nicholl} et~al.,}{{Nicholl}
  et~al.}{2020}]{Nicholl:2020mkh}
{Nicholl} M.,  et~al., 2020, \mn@doi [Nature Astronomy]
  {10.1038/s41550-020-1066-7}, \href
  {https://ui.adsabs.harvard.edu/abs/2020NatAs...4..893N} {4, 893}

\bibitem[\protect\citeauthoryear{Ofek et~al.}{Ofek et~al.}{2007}]{Ofek:2006vt}
Ofek E.~O.,  et~al., 2007, \mn@doi [Astrophys. J. Lett.] {10.1086/516749}, 659,
  L13

\bibitem[\protect\citeauthoryear{Patnaude \& Fesen}{Patnaude \&
  Fesen}{2009}]{Patnaude:2008gq}
Patnaude D.~J.,  Fesen R.~A.,  2009, \mn@doi [Astrophys. J.]
  {10.1088/0004-637X/697/1/535}, 697, 535

\bibitem[\protect\citeauthoryear{Petropoulou, Kamble  \& Sironi}{Petropoulou
  et~al.}{2016}]{Petropoulou:2016zar}
Petropoulou M.,  Kamble A.,   Sironi L.,  2016, \mn@doi [Mon. Not. Roy. Astron.
  Soc.] {10.1093/mnras/stw920}, 460, 44

\bibitem[\protect\citeauthoryear{Petropoulou, Coenders, Vasilopoulos, Kamble
  \& Sironi}{Petropoulou et~al.}{2017}]{Petropoulou:2017ymv}
Petropoulou M.,  Coenders S.,  Vasilopoulos G.,  Kamble A.,   Sironi L.,  2017,
  \mn@doi [Mon. Not. Roy. Astron. Soc.] {10.1093/mnras/stx1251}, 470, 1881

\bibitem[\protect\citeauthoryear{Pitik, Tamborra, Angus  \& Auchettl}{Pitik
  et~al.}{2022}]{Pitik:2021dyf}
Pitik T.,  Tamborra I.,  Angus C.~R.,   Auchettl K.,  2022, \mn@doi [Astrophys.
  J.] {10.3847/1538-4357/ac5ab1}, 929, 163

\bibitem[\protect\citeauthoryear{Predehl et~al.,}{Predehl
  et~al.}{2010}]{10.1117/12.856577}
Predehl P.,  et~al., 2010, in Arnaud M.,  Murray S.~S.,   Takahashi T.,  eds,
  Vol. 7732, Space Telescopes and Instrumentation 2010: Ultraviolet to Gamma
  Ray. SPIE, p. 77320U, \mn@doi{10.1117/12.856577}, \url
  {https://doi.org/10.1117/12.856577}

\bibitem[\protect\citeauthoryear{Protheroe \& Clay}{Protheroe \&
  Clay}{2004}]{Protheroe:2003vc}
Protheroe R.~J.,  Clay R.~W.,  2004, \mn@doi [Publ. Astron. Soc. Pac.]
  {10.1071/AS03047}, 21, 1

\bibitem[\protect\citeauthoryear{Rest et~al.}{Rest et~al.}{2011}]{Rest:2009wb}
Rest A.,  et~al., 2011, \mn@doi [Astrophys. J.] {10.1088/0004-637X/729/2/88},
  729, 88

\bibitem[\protect\citeauthoryear{Reusch et~al.}{Reusch
  et~al.}{2022}]{Reusch:2021ztx}
Reusch S.,  et~al., 2022, \mn@doi [Phys. Rev. Lett.]
  {10.1103/PhysRevLett.128.221101}, 128, 221101

\bibitem[\protect\citeauthoryear{{Rybicki} \& {Lightman}}{{Rybicki} \&
  {Lightman}}{1986}]{1986rpa..book.....R}
{Rybicki} G.~B.,  {Lightman} A.~P.,  1986, {Radiative Processes in
  Astrophysics}

\bibitem[\protect\citeauthoryear{{Sarmah}, {Chakraborty}, {Tamborra}  \&
  {Auchettl}}{{Sarmah} et~al.}{2022}]{Sarmah:2022vra}
{Sarmah} P.,  {Chakraborty} S.,  {Tamborra} I.,   {Auchettl} K.,  2022, \mn@doi
  [\jcap] {10.1088/1475-7516/2022/08/011}, \href
  {https://ui.adsabs.harvard.edu/abs/2022JCAP...08..011S} {2022, 011}

\bibitem[\protect\citeauthoryear{{Sarmah}, {Chakraborty}, {Tamborra}  \&
  {Auchettl}}{{Sarmah} et~al.}{2023}]{Sarmah:2023sds}
{Sarmah} P.,  {Chakraborty} S.,  {Tamborra} I.,   {Auchettl} K.,  2023, \mn@doi
  [arXiv e-prints] {10.48550/arXiv.2303.13576}, \href
  {https://ui.adsabs.harvard.edu/abs/2023arXiv230313576S} {p. arXiv:2303.13576}

\bibitem[\protect\citeauthoryear{{Sato}, {Katsuda}, {Morii}, {Bamba}, {Hughes},
  {Maeda}, {Ishida}  \& {Fraschetti}}{{Sato}
  et~al.}{2018}]{2018ApJ...853...46S}
{Sato} T.,  {Katsuda} S.,  {Morii} M.,  {Bamba} A.,  {Hughes} J.~P.,  {Maeda}
  Y.,  {Ishida} M.,   {Fraschetti} F.,  2018, \mn@doi [Astrophys. J.]
  {10.3847/1538-4357/aaa021}, \href
  {https://ui.adsabs.harvard.edu/abs/2018ApJ...853...46S} {853, 46}

\bibitem[\protect\citeauthoryear{{Schlafly} \& {Finkbeiner}}{{Schlafly} \&
  {Finkbeiner}}{2011}]{SFDmap}
{Schlafly} E.~F.,  {Finkbeiner} D.~P.,  2011, \mn@doi [\apj]
  {10.1088/0004-637X/737/2/103}, \href
  {https://ui.adsabs.harvard.edu/abs/2011ApJ...737..103S} {737, 103}

\bibitem[\protect\citeauthoryear{{Schlegel}}{{Schlegel}}{1990}]{1990MNRAS.244..269S}
{Schlegel} E.~M.,  1990, Mon. Not. Roy. Astron. Soc., \href
  {https://ui.adsabs.harvard.edu/abs/1990MNRAS.244..269S} {244, 269}

\bibitem[\protect\citeauthoryear{{Schure}, {Achterberg}, {Keppens}  \&
  {Vink}}{{Schure} et~al.}{2010}]{2010MNRAS.406.2633S}
{Schure} K.~M.,  {Achterberg} A.,  {Keppens} R.,   {Vink} J.,  2010, \mn@doi
  [Mon. Not. Roy. Astron. Soc.] {10.1111/j.1365-2966.2010.16857.x}, \href
  {https://ui.adsabs.harvard.edu/abs/2010MNRAS.406.2633S} {406, 2633}

\bibitem[\protect\citeauthoryear{{Shappee} et~al.,}{{Shappee}
  et~al.}{2014}]{2014ApJ...788...48S}
{Shappee} B.~J.,  et~al., 2014, \mn@doi [\apj] {10.1088/0004-637X/788/1/48},
  \href {https://ui.adsabs.harvard.edu/abs/2014ApJ...788...48S} {788, 48}

\bibitem[\protect\citeauthoryear{{Shvartzvald} et~al.,}{{Shvartzvald}
  et~al.}{2023}]{Shvartzvald:2023ofi}
{Shvartzvald} Y.,  et~al., 2023, \mn@doi [arXiv e-prints]
  {10.48550/arXiv.2304.14482}, \href
  {https://ui.adsabs.harvard.edu/abs/2023arXiv230414482S} {p. arXiv:2304.14482}

\bibitem[\protect\citeauthoryear{{Slane}, {Lee}, {Ellison}, {Patnaude},
  {Hughes}, {Eriksen}, {Castro}  \& {Nagataki}}{{Slane}
  et~al.}{2015}]{2015ApJ...799..238S}
{Slane} P.,  {Lee} S.~H.,  {Ellison} D.~C.,  {Patnaude} D.~J.,  {Hughes} J.~P.,
   {Eriksen} K.~A.,  {Castro} D.,   {Nagataki} S.,  2015, \mn@doi [Astrophys.
  J.] {10.1088/0004-637X/799/2/238}, \href
  {https://ui.adsabs.harvard.edu/abs/2015ApJ...799..238S} {799, 238}

\bibitem[\protect\citeauthoryear{{Smith}}{{Smith}}{2017}]{Smith:2016dnb}
{Smith} N.,  2017, in {Alsabti} A.~W.,  {Murdin} P.,  eds, , Handbook of
  Supernovae.
p.~403, \mn@doi{10.1007/978-3-319-21846-5_38}

\bibitem[\protect\citeauthoryear{Smith, Chornock, Li, Ganeshalingam, Silverman,
  Foley, Filippenko  \& Barth}{Smith et~al.}{2008}]{Smith:2008ez}
Smith N.,  Chornock R.,  Li W.,  Ganeshalingam M.,  Silverman J.~M.,  Foley
  R.~J.,  Filippenko A.~V.,   Barth A.~J.,  2008, \mn@doi [Astrophys. J.]
  {10.1086/591021}, 686, 467

\bibitem[\protect\citeauthoryear{Smith, Li, Filippenko  \& Chornock}{Smith
  et~al.}{2011}]{Smith:2010vz}
Smith N.,  Li W.,  Filippenko A.~V.,   Chornock R.,  2011, \mn@doi [Mon. Not.
  Roy. Astron. Soc.] {10.1111/j.1365-2966.2011.17229.x}, 412, 1522

\bibitem[\protect\citeauthoryear{{Stein} et~al.,}{{Stein}
  et~al.}{2023}]{Stein:2022rvc}
{Stein} R.,  et~al., 2023, \mn@doi [Mon. Not. Roy. Astron. Soc.]
  {10.1093/mnras/stad767}, \href
  {https://ui.adsabs.harvard.edu/abs/2023MNRAS.521.5046S} {521, 5046}

\bibitem[\protect\citeauthoryear{{Sturner}, {Skibo}, {Dermer}  \&
  {Mattox}}{{Sturner} et~al.}{1997}]{1997ApJ...490..619S}
{Sturner} S.~J.,  {Skibo} J.~G.,  {Dermer} C.~D.,   {Mattox} J.~R.,  1997,
  \mn@doi [Astrophys. J.] {10.1086/304894}, \href
  {https://ui.adsabs.harvard.edu/abs/1997ApJ...490..619S} {490, 619}

\bibitem[\protect\citeauthoryear{Suzuki, Moriya  \& Takiwaki}{Suzuki
  et~al.}{2020}]{Suzuki:2020qui}
Suzuki A.,  Moriya T.~J.,   Takiwaki T.,  2020, \mn@doi [Astrophys. J.]
  {10.3847/1538-4357/aba0ba}, 899, 56

\bibitem[\protect\citeauthoryear{Tammi \& Duffy}{Tammi \&
  Duffy}{2009}]{Tammi:2008vg}
Tammi J.,  Duffy P.,  2009, \mn@doi [AIP Conf. Proc.] {10.1063/1.3076711},
  1085, 475

\bibitem[\protect\citeauthoryear{Villar, Berger, Metzger  \& Guillochon}{Villar
  et~al.}{2017}]{Villar:2017oya}
Villar V.~A.,  Berger E.,  Metzger B.~D.,   Guillochon J.,  2017, \mn@doi
  [Astrophys. J.] {10.3847/1538-4357/aa8fcb}, 849, 70

\bibitem[\protect\citeauthoryear{Vitagliano, Tamborra  \& Raffelt}{Vitagliano
  et~al.}{2020}]{Vitagliano:2019yzm}
Vitagliano E.,  Tamborra I.,   Raffelt G.,  2020, \mn@doi [Rev. Mod. Phys.]
  {10.1103/RevModPhys.92.045006}, 92, 45006

\bibitem[\protect\citeauthoryear{{Weaver}}{{Weaver}}{1976}]{1976ApJS...32..233W}
{Weaver} T.~A.,  1976, \mn@doi [Astrophys. J.] {10.1086/190398}, \href
  {https://ui.adsabs.harvard.edu/abs/1976ApJS...32..233W} {32, 233}

\bibitem[\protect\citeauthoryear{Yuksel, Kistler, Beacom  \& Hopkins}{Yuksel
  et~al.}{2008}]{Yuksel:2008cu}
Yuksel H.,  Kistler M.~D.,  Beacom J.~F.,   Hopkins A.~M.,  2008, \mn@doi
  [Astrophys. J. Lett.] {10.1086/591449}, 683, L5

\bibitem[\protect\citeauthoryear{Zirakashvili \& Ptuskin}{Zirakashvili \&
  Ptuskin}{2016}]{Zirakashvili:2015mua}
Zirakashvili V.~N.,  Ptuskin V.~S.,  2016, \mn@doi [Astropart. Phys.]
  {10.1016/j.astropartphys.2016.02.004}, 78, 28

\bibitem[\protect\citeauthoryear{Zyla et~al.}{Zyla et~al.}{2020}]{Zyla:2020zbs}
Zyla P.,  et~al., 2020, \mn@doi [PTEP] {10.1093/ptep/ptaa104}, 2020, 083C01

\makeatother
\end{thebibliography}

\appendix


\section{Dependence of the supernova  lightcurve properties on the model parameters}
\label{App:parameters}
In this appendix, we investigate the dependence of the parameters characteristic of the  lightcurve on the SN model properties.
Figure~\ref{Fig: T_rise,L_peak,Erad wind} displays how the rise time $t_{\rm{rise}}$ (defined in Sec.~\ref{Sec:Scaling_relations}) of the bolometric luminosity depends on  the SN parameters of interest. For any fixed combination of $E_{\rm{k}}$, $M_{\rm{ej}}$ and $R_{\rm{CSM}}$, the rise time increases with $M_{\rm{CSM}}$, since a denser CSM extends the photon diffusion time. In the left panel, we see that the larger the kinetic energy, the shorter $t_{\rm{rise}}$. This is explained by the fact that large shock velocities  cause the breakout to happen later and shorten  the time that  photons take to reach the photosphere. The same trend is expected for decreasing $M_{\rm{ej}}$, as shown  in the middle panel, where a mild trend in this direction is noticeable.  Furthermore in the BW regime (which corresponds to the breaks in the curves) we see that $t_{\rm{rise}}$ is independent of $M_{\rm{ej}}$ (the curves for low $M_{\rm{ej}}$ saturate at the same value), a trend confirmed by the numerical simulations presented in~\citep{Suzuki:2020qui}. In the right panel of Fig.~\ref{Fig: T_rise,L_peak,Erad wind},  one can observe that for large CSM masses, there is a transition region shifting towards larger $R_{\rm{CSM}}$ where the trend of $t_{\rm{rise}}$ is reversed. The reason of this inversion is to be found in the dependence of the photospheric radius on $R_{\rm{CSM}}$ (see Eq.~\ref{Eq: R_PH}), which for  fixed $M_{\rm{CSM}}$ increases and saturates at a certain $R_{\rm{CSM}}$, to turn and decrease for larger CSM radii. 

The middle panels  of Fig.~\ref{Fig: T_rise,L_peak,Erad wind} show that an increase in $M_{\rm{CSM}}$ makes $L_{\rm{peak}}$ larger in all cases, since a larger $M_{\rm{CSM}}$ causes more kinetic energy to be dissipated and radiated. This is true as long as the shock is in the FE regime.  
In the BW regime, $L_{\rm{peak}}$ declines with $M_{\rm{CSM}}$.
The left and middle panels show that  the peak luminosity increases with larger ejecta energy and smaller ejecta masses, since both make the shock velocity larger and thus more energetic. In the BW regime, the peak luminosity becomes independent of the ejecta mass, as confirmed by the saturation to the same branch for low $M_{\rm{ej}}$. The right panel shows that  the  brightest lightcurves are obtained when the CSM is more compact, i.e.~for the smallest $R_{\rm{CSM}}$ (apart from the transition region visible for large $M_{\rm{CSM}}$, due to the transition into the BW regime). 

The bottom panels show the trend of $E_{\rm{diss, thick}}$. The dissipated energy in the optically thick part of the CSM increases with $M_{\rm{CSM}}$, is very large for small $M_{\rm{ej}}$ and $R_{\rm{CSM}}$, since the first allows for high shock velocity, and the second for very compact region, and thus high densities.

\begin{figure*}
	\centering
   \includegraphics[width=0.9\textwidth]{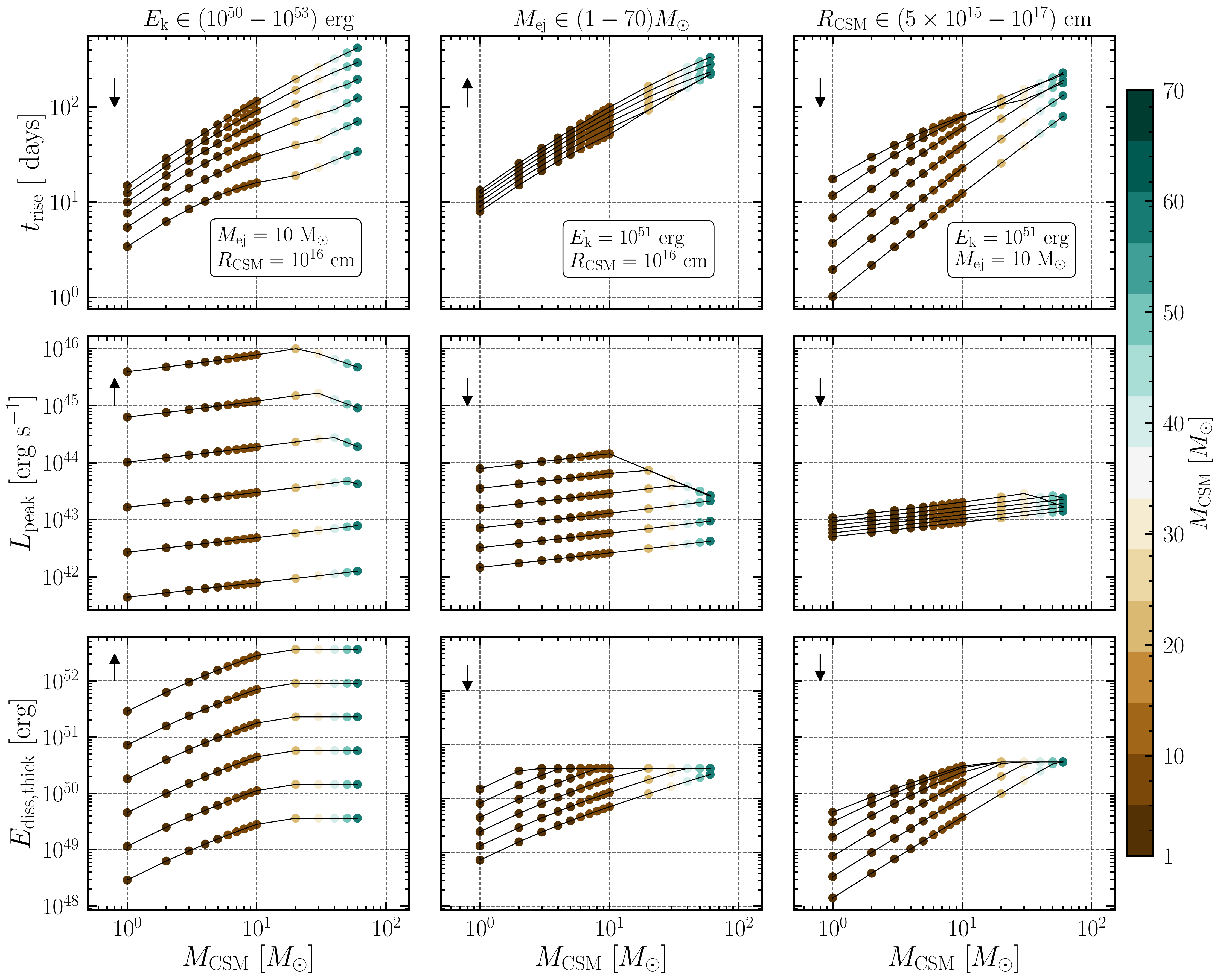}
  
	\caption{Rise time of the bolometric lightcurve (top panels), bolometric peak luminosity (middle panels), dissipated energy in the optically thick part of the CSM envelope (bottom panels)  as  functions of the CSM mass and $E_k$ (left panels, for fixed $M_{\rm{ej}}$ and $R_{\rm{CSM}}$), the ejecta mass (middle panels, for fixed $E_{\rm{k}}$ and $R_{\rm{CSM}}$), and $R_{\rm{CSM}}$ (right panels, for fixed $E_{\rm k}$ and $M_{\rm{ej}}$), respectively. In each panel, the arrow indicates the direction of  increase of the parameter under investigation, e.g.~in the left panel of the first row, $t_{\rm{rise}}$ decreases for increasing $E_{\rm{k}}$, for  fixed $M_{\rm{CSM}}$;
    $t_{\rm{rise}}$ increases with $M_{\rm{CSM}}$ since a denser CSM envelope increases the optical depth and delays the photon escape. From the top left and middle panels, we see that for increasing $E_{\rm{k}}$ or decreasing $M_{\rm{ej}}$, the diffusion time becomes shorter. Indeed both the increase of $E_{\rm{k}}$  and the decrease of $M_{\rm{ej}}$ are responsible for an increase of the  shock velocity, which in turn causes the radius where the photon diffusion velocity exceeds the shock velocity to shift outwards. In the top right panel, we observe that for small $R_{\rm{CSM}}$ an initial increase of $t_{\rm{rise}}$, which declines again for larger CSM radii. This transition region is related to the photosphere dependence on $M_{\rm{CSM}}$ and $R_{\rm{CSM}}$. 
    For what concerns $L_{\rm{peak}}$, in all three middle panels we see that $L_{\rm{peak}}$ initially increases with $M_{\rm{CSM}}$ in the FE regime. When transitioning to the BW regime (indicated by the breaks in the curves), a saturation of the radiated energy occurs and this, together with the increase of $t_{\rm{rise}}$, causes $L_{\rm{peak}}$ to drop as $M_{\rm{CSM}}$ increases. Larger $E_{\rm{k}}$ and smaller	$M_{\rm{ej}}$ are responsible for a larger shock velocity, and thus an increase of $L_{\rm{peak}}$, as it can be seen in the left and middle panels. In the right middle panel, we observe that small $R_{\rm{CSM}}$, for  fixed $M_{\rm{CSM}}$, make the medium denser and therefore it is easier to dissipate the ejecta energy, leading to an increase of $L_{\rm{peak}}$. Similarly, $E_{\rm{diss,thick}}$ increases  with $M_{\rm{CSM}}$, and saturates to a constant fraction of $E_{\rm{k}}$ in the BW regime. The dots are colored according to the $M_{\rm{CSM}}$ value, as shown in the color bar.}
	\label{Fig: T_rise,L_peak,Erad wind}
\end{figure*}

\section{Dependence of the maximum proton energy on the supernova model parameters}
\label{App:maxEp}
In this appendix, we analyze  the dependence  of the maximum $E_{\rm{p, max}}$ on the SN model parameters. To do so, we first highlight  the dependence on the SN parameters of the main timescales entering the problem. From  Eqs.~\ref{Eq: t_cool} and~\ref{Eq: Lambda_cool}, we see that the plasma cooling timescale scales as:
\begin{equation}
t_{\rm{cool}}\propto \frac{1}{n_{\rm{sh}}}\times \begin{cases}
    v_{\rm{sh}}^{16/5}\quad\, \textrm{if}\quad 10^{5} < T \lesssim 4.7\times 10^{7}~\rm{K}\\
    v_{\rm{sh}}\quad \,\, \textrm{if} \quad T> 4.7\times 10^{7}~\rm{K}\ .
    \end{cases}
\end{equation}
For the wind scenario, it becomes
\begin{itemize}
    \item[-] for $R<R_{\rm{dec}}$:
    \begin{equation}
    t_{\rm{cool}}\propto \bigg(\frac{R_{\rm{CSM,w}}}{M_{\rm{CSM,w}}}\bigg)\times \begin{cases}
        R^{54/35}\quad\, \textrm{if}\quad 10^{5} < T \lesssim 4.7\times 10^{7}~\rm{K}\\
        R^{13/7}\quad \quad\quad \textrm{if} \quad T> 4.7\times 10^{7}~\rm{K}\ .
    \end{cases}
    \end{equation}
    \item[-] for $R>R_{\rm{dec}}$:
    \begin{equation}
    t_{\rm{cool}}\propto \bigg(\frac{R_{\rm{CSM,w}}}{M_{\rm{CSM,w}}}\bigg)\times \begin{cases}
        R^{2/5}\quad\, \textrm{if}\quad 10^{5} < T \lesssim 4.7\times 10^{7}~\rm{K}\\
        R^{3/2}\quad \quad\quad \textrm{if} \quad T> 4.7\times 10^{7}~\rm{K}\ .
    \end{cases}
    \end{equation}
\end{itemize}
For the shell scenario, it is
\begin{itemize}
    \item[-] for $R<R_{\rm{dec}}$:
    \begin{equation}
    t_{\rm{cool}}\propto \bigg(\frac{R_{\rm{CSM,s}}^{3}}{M_{\rm{CSM,s}}}\bigg)\times \begin{cases}
        R^{-48/35}\quad\, \textrm{if}\quad 10^{5} < T \lesssim 4.7\times 10^{7}~\rm{K}\\
        R^{-3/7}\quad \quad\quad \textrm{if} \quad T> 4.7\times 10^{7}~\rm{K}\ .
    \end{cases}
    \end{equation}
    \item[-] for $R>R_{\rm{dec}}$:
    \begin{equation}
    t_{\rm{cool}}\propto \bigg(\frac{R_{\rm{CSM,s}}^{3}}{M_{\rm{CSM,s}}}\bigg)\times \begin{cases}
        R^{-24/5}\quad\, \textrm{if}\quad 10^{5} < T \lesssim 4.7\times 10^{7}~\rm{K}\\
        R^{-3/2}\quad \quad\quad \textrm{if} \quad T> 4.7\times 10^{7}~\rm{K}\ .
    \end{cases}
    \end{equation}
\end{itemize}
The acceleration time scales as $t_{\rm{acc}}\propto E_{\rm{p}} v_{\rm{sh}}^{-3}n_{\rm{sh}}^{-1/2}$, given ${B\propto v_{\rm{sh}}n_{\rm{sh}}^{1/2}}$. For the wind scenario it is
\begin{equation}
    t_{\rm{acc}}\propto \bigg(\frac{R_{\rm{CSM,w}}}{M_{\rm{CSM,w}}}\bigg)^{1/2} E_{\rm{p}}\times \begin{cases}
        R^{10/7}\quad\, \textrm{if}\quad R<R_{\rm{dec}}\\
        R^{5/2}\quad \textrm{if} \quad R>R_{\rm{dec}}\ ;
    \end{cases}
\end{equation}
while for the shell scenario, it is
\begin{equation}
    t_{\rm{acc}}\propto \bigg(\frac{R_{\rm{CSM,s}}^{3}}{M_{\rm{CSM,s}}}\bigg)^{1/2} E_{\rm{p}}\times \begin{cases}
        R^{9/7}\quad\, \textrm{if}\quad R<R_{\rm{dec}}\\
        R^{9/2}\quad \textrm{if} \quad R>R_{\rm{dec}}\ .
    \end{cases}
\end{equation}
The proton-proton interaction time $t_{\rm{pp}}=(c n_{\rm{sh}}\sigma_{\rm{pp}})^{-1}$ is
\begin{equation}
    t_{\rm{pp}}\propto \begin{cases}
        \frac{R_{\rm{CSM,w}}}{M_{\rm{CSM,w}}}\times R^{2}\quad \textrm{for the wind}\\
        \frac{R_{\rm{CSM,s}}^{3}}{M_{\rm{CSM,s}}}\quad\quad\quad\ \ \textrm{for the shell}\ .
    \end{cases}
\end{equation}

Using the relations above, we can investigate how $E_{\rm{p,max}}$ depends on the SN model parameters and how it evolves with the shock radius.
If $t_{\rm{cool}}$ is the $\textrm{min}[t_{\rm{cool}},t_{\rm{dyn}},t_{\rm{pp}}]$, the maximum proton energy is determined by $t_{\rm{acc}}=t_{\rm{cool}}$. For the  wind scenario,
\begin{itemize}
    \item[-] for $R<R_{\rm{dec}}$:
    \begin{equation}
    E_{\rm{p,max}}\propto \bigg(\frac{R_{\rm{CSM,w}}}{M_{\rm{CSM,w}}}\bigg)^{1/2}\times \begin{cases}
        R^{4/35}\quad\, \textrm{if}\quad 10^{5} < T \lesssim 4.7\times 10^{7}~\rm{K}\\
        R^{3/7}\quad \,\, \textrm{if} \quad T> 4.7\times 10^{7}~\rm{K}\ ;
    \end{cases}
    \end{equation}
   \item[-] for $R>R_{\rm{dec}}$:
    \begin{equation}
        E_{\rm{p,max}}\propto \bigg(\frac{R_{\rm{CSM,w}}}{M_{\rm{CSM,w}}}\bigg)^{1/2}\times \begin{cases}
        R^{-21/10}\quad\, \textrm{if}\quad 10^{5} < T \lesssim 4.7\times 10^{7}~\rm{K}\\
        R^{-1}\quad \quad\quad \textrm{if} \quad T> 4.7\times 10^{7}~\rm{K}\ .
    \end{cases}
    \end{equation}
\end{itemize}
For the shell scenario, instead, it is
\begin{itemize}
    \item[-] for $R<R_{\rm{dec}}$:
    \begin{equation}
    E_{\rm{p,max}}\propto \bigg(\frac{R_{\rm{CSM,s}}^{3}}{M_{\rm{CSM}}}\bigg)^{1/2}\times \begin{cases}
        R^{-93/35}\,\, \textrm{if}\quad 10^{5} < T \lesssim 4.7\times 10^{7}~\rm{K}\\
        R^{-12/7}\quad \,\, \textrm{if} \quad T> 4.7\times 10^{7}~\rm{K}\ ;
    \end{cases}
    \end{equation}
    \item[-] for $R>R_{\rm{dec}}$:
    \begin{equation}
        E_{\rm{p,max}}\propto \bigg(\frac{R_{\rm{CSM,s}}^{3}}{M_{\rm{CSM}}}\bigg)^{1/2}\times \begin{cases}
        R^{-93/10}\,\, \textrm{if}\quad 10^{5} < T \lesssim 4.7\times 10^{7}~\rm{K}\\
        R^{-6}\quad \quad \textrm{if} \quad T> 4.7\times 10^{7}~\rm{K}\ .
    \end{cases}
    \end{equation}
\end{itemize}

If $t_{\rm{pp}}$ corresponds to the $\textrm{min}[t_{\rm{cool}},t_{\rm{dyn}},t_{\rm{pp}}]$, then the maximum proton energy is determined by $t_{\rm{acc}}=t_{\rm{pp}}$ and can be written for the wind scenario as
\begin{equation}
E_{\rm{p,max}}\propto \bigg(\frac{R_{\rm{CSM,w}}}{M_{\rm{CSM,w}}}\bigg)^{1/2}\times \begin{cases} R^{4/7} \quad\textrm{for}\quad R<R_{\rm{dec}}\\ 
R^{-1/2}\quad\textrm{for}\quad R>R_{\rm{dec}}\ ,
\end{cases}
\end{equation}
and for the shell scenario as
\begin{equation}
E_{\rm{p,max}}\propto \bigg(\frac{R_{\rm{CSM,s}}^{3}}{M_{\rm{CSM,s}}}\bigg)^{1/2}\times \begin{cases} R^{-9/7} \quad\textrm{for}\quad R<R_{\rm{dec}}\\ 
R^{-9/2}\quad\textrm{for}\quad R>R_{\rm{dec}}\ .
\end{cases}
\end{equation}

Finally, if $t_{\rm{dyn}}$ corresponds to $\textrm{min}[t_{\rm{cool}},t_{\rm{dyn}},t_{\rm{pp}}]$, the maximum proton energy is determined by $t_{\rm{acc}}=t_{\rm{dyn}}$ and  for the wind scenario it is
\begin{equation}
    \label{Eq: EP_max_wind_adiabatic}
    E_{\rm{p,max}}\propto \bigg(\frac{M_{\rm{CSM,w}}}{R_{\rm{CSM,w}}}\bigg)^{1/2}\times \begin{cases} R^{-2/7} \quad\textrm{for}\quad R<R_{\rm{dec}}\\ 
R^{-1}\quad\textrm{for}\quad R>R_{\rm{dec}}\ ,
\end{cases}
\end{equation}
while, for the shell scenario, it is
\begin{equation}
        E_{\rm{p,max}}\propto 
        \bigg(\frac{M_{\rm{CSM,s}}}{ R_{\rm{CSM,s}}^{3}}\bigg)^{1/2}\times \begin{cases} R^{1/7} \quad\textrm{for}\quad R<R_{\rm{dec}}\\ 
R^{-2}\quad\textrm{for}\quad R>R_{\rm{dec}}\ .
\end{cases}
\end{equation}
Note that  we assume constant  $\sigma_{\rm{pp}}\sim 3\times 10^{-26}$~cm$^{2}$ for the sake of simplicity in this appendix in order to obtain the above analytical relations.

We immediately see from the relations above that for the wind scenario, independently on the cooling mechanism,  the maximum proton energy has a decreasing trend with $R$ in the deceleration phase ($R>R_{\rm{dec}}$). However, in the ejecta-dominated phase ($R<R_{\rm{dec}}$), the maximum proton energy always increases, except for the case in which the adiabatic cooling is dominant (Eq.~\ref{Eq: EP_max_wind_adiabatic}). Finally,  in the shell scenario,  $E_{\rm{p, max}}$  always decreases, apart from the case where $t_{\rm{cool}}$ and $t_{\rm{pp}}$ are too long compared to the dynamical time, and it  slowly increases in the free-expansion phase.

We define $R_{\rm{cool}}$ as the radius where $t_{\rm{dyn}}=t_{\rm{cool}}$, and $R_{\rm{pp}}$ the radius where $t_{\rm{dyn}}=t_{\rm{pp}}$. The maximum value of $E_{\rm{p,max}}$, denoted as $E^{\rm{\ast}}_{\rm{p,max}}$, can be achieved at any of the following radii: $R_{\rm{bo}}$, $R_{\rm{cool}}$, $R_{\rm{pp}}$, $R_{\rm{dec}}$, or $R_{\rm{CSM}}$. There are various  configurations of such radii. If for example $R_{\rm{bo}}<R_{\rm{cool}}<R_{\rm{pp}}<R_{\rm{CSM}}<R_{\rm{dec}}$, and both $t_{\rm{dyn}}<t_{\rm{cool}}$ for $R>R_{\rm{cool}}$ and $t_{\rm{dyn}}<t_{\rm{pp}}$ for $R>R_{\rm{pp}}$, then the maximum $E_{\rm{p,max}}$ is obtained at $R_{\rm{pp}}$.

\begin{figure*}
	\centering
	\includegraphics[width=1\textwidth]{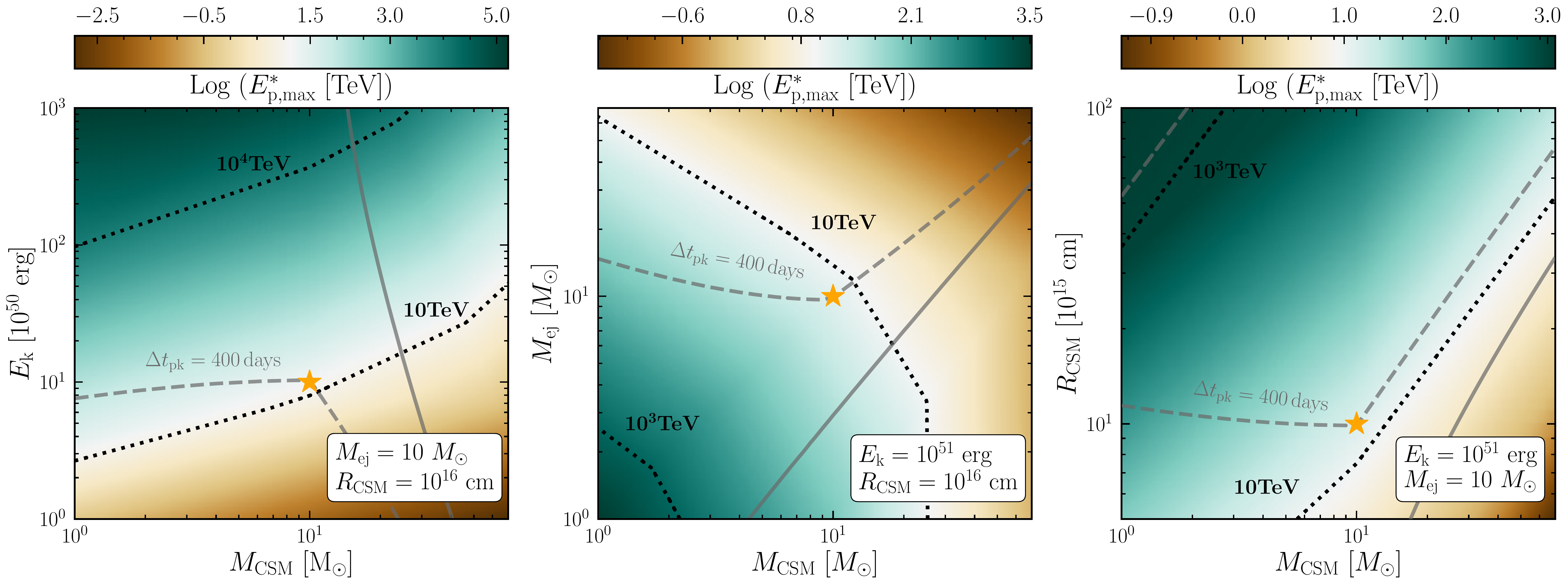}
	\caption{Contour plots of the maximum proton energy $E^{\ast}_{\rm{p,max}}$  reached throughout the  evolution of the shock in the wind scenario, in the plane spanned by $M_{\rm{CSM}}$ and $E_{\rm{k}}$ (left panel), $M_{\rm{ej}}$ (middle panel), and $R_{\rm{CSM}}$ (right panel). The dotted contours mark  isocontours of $E^{\ast}_{\rm{p,max}}$ to guide the eye. The largest proton energies can be achieved with large $E_{\rm{k}}$ and small $M_{\rm{ej}}$, both maximizing $v_{\rm{sh}}$, and thus the acceleration rate; low $M_{\rm{CSM}}$ and/or large $R_{\rm{CSM}}$, both making the CSM less dense, and thus the proton energy losses less severe. For each panel, the gray line represents $\Delta t_{\rm{pk}}=t|_{E^{\ast}_{\rm{p,max}}}-t_{\rm{peak}}$, i.e.~the time at which the maximum proton energy is reached with respect to the bolometric peak of the lightcurve. The solid gray lines correspond to  $\Delta t_{\rm{pk}}=0$. From the dashed  gray line, we can  see that the largest time interval is expected for low $E_{\rm{k}}$, and large $M_{\rm{ej}}$ and $R_{\rm{CSM}}$. The parameter space between the solid and the dashed gray lines leads to $0<\Delta t_{\rm{pk}} < 400$~days, which is the  follow-up time window adopted for SNe. The orange star marks our benchmark scenario (see Table~\ref{Table: Parameters}). 
    }
	\label{Fig:EP_MAX_cool_wind}
\end{figure*}

Note that this procedure serves to inspect the dependence of the maximum proton energy  analytically. However, the total cooling time is the sum of $t_{\rm{dyn}}$ and $t_{\rm{pp}}$ or $t_{\rm{cool}}$ and $t_{\rm{pp}}$; since the energy dependence of  $t_{\rm{pp}}$ increases slightly at higher energies,  the value of $E^{\rm{\ast}}_{\rm{p,max}}$ that we find is underestimated by a few percent in the transition region $t_{\rm{dyn}}\sim t_{\rm{cool}}$ and at very large energies. 
Figure~\ref{Fig:EP_MAX_cool_wind} displays how $E^{\ast}_{\rm{p,max}}$ depends on the SN parameters. The most promising configurations that allow to reach large $E^{\ast}_{\rm{p,max}}$ are  the ones with large $E_{\rm{k}}$ and low $M_{\rm{CSM}}$ (left panel), or low $M_{\rm{ej}}$ and low $M_{\rm{CSM}}$ (middle panel), or high $R_{\rm{CSM}}$ and low $M_{\rm{CSM}}$ (right panel), which maximize the acceleration rate and minimize the energy loss rate.
   
For the fiducial parameters adopted in each panel of Fig.~\ref{Fig:EP_MAX_cool_wind}, total energies $\gtrsim 10^{51}$~erg, relatively low ejecta ($\lesssim 20 M_{\odot}$), CSM masses ($\lesssim 10 M_{\odot}$), and  extended CSM envelopes ($\gtrsim 10^{16}$~cm) are required to obtain protons with $\sim$~PeV energy. Furthermore, as shown through  the gray contour lines, which display $t|_{E^{\ast}_{\rm{p,max}}}-t_{\rm{peak}}$ (where $t|_{E^{\ast}_{\rm{p,max}}}$ is the time at which the maximum proton energy is reached), the maximum $E^{\ast}_{\rm{p,max}}$ is achieved at relatively late times [$\mathcal{O}(100\,\rm{days})$] with respect to the peak time $t_{\rm{peak}}=t_{\rm{bo}}+t_{\rm{rise}}$. Such longer timescales  are expected for  low kinetic energies of the ejecta, and large $M_{\rm{ej}}$ and $R_{\rm{CSM}}$. Only the configurations with large CSM mass, due to the onset of the decelerating phase, are expected to invert the increasing trend of $E^{\ast}_{\rm{p,max}}$ before the lightcurve reaches its peak.

\section{Constant density scenario}
\label{Appendix: Constant density case}

\begin{figure*}
	\centering
 	\includegraphics[width=0.81\textwidth]{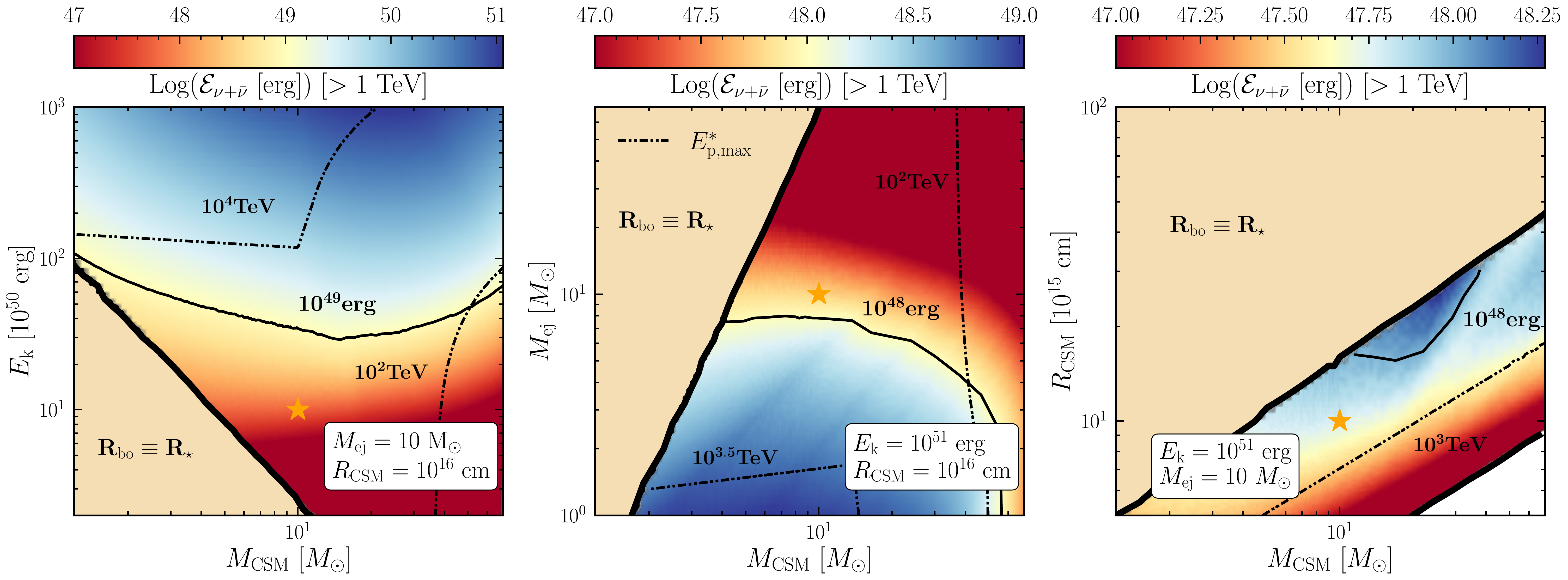}\\
  	\includegraphics[width=0.81\textwidth]{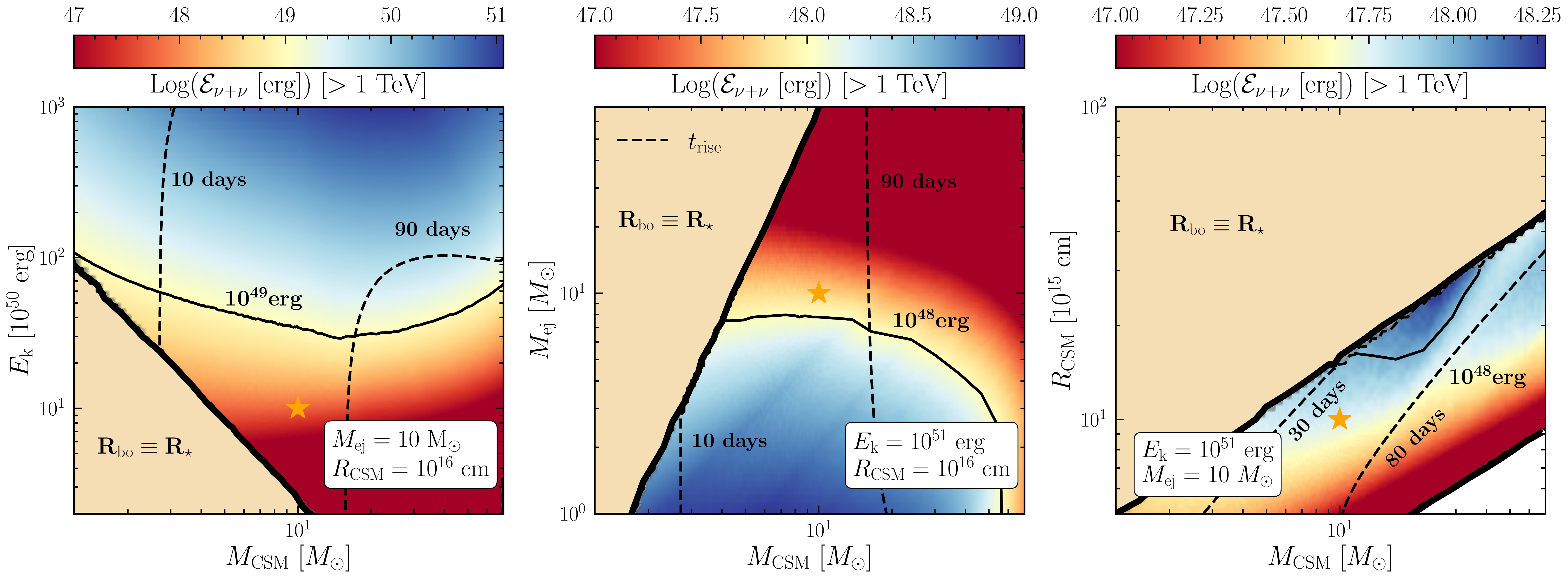}\\
  	\includegraphics[width=0.81\textwidth]{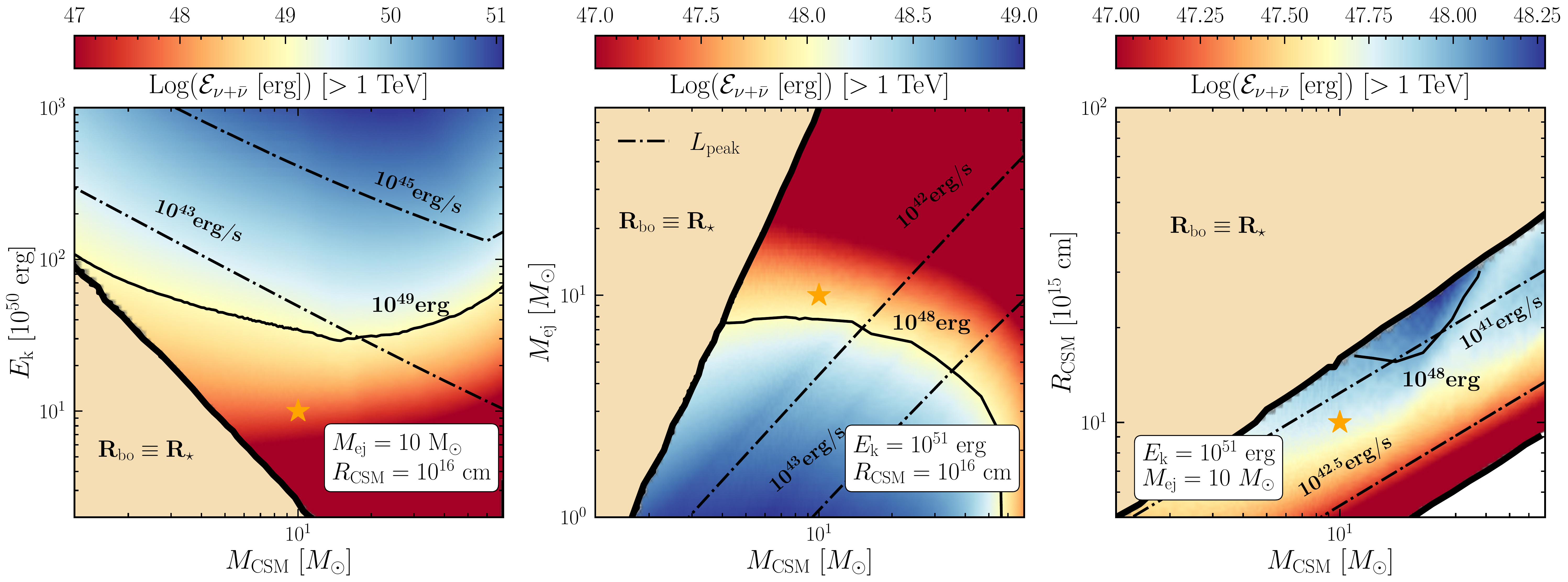}\\

	\caption{The same as in Fig.~\ref{Fig: Etot_wind}, but for the constant density shell scenario. The beige region has been excluded from our investigation since here the breakout of the shock does not occur in the CSM shell, but at the radius of the progenitor star. The white region, visible only in the lower right corner of the third column, has instead been excluded because leading to $E^{\ast}_{\rm{p,max}}<\rm{10 TeV}$.
	The SN configurations leading to the largest outcomes in neutrinos are similar to the ones in the wind case, and are given by large SN kinetic energies ($E_{\rm{k}}\gtrsim 10^{51}$~erg), small ejecta masses ($M_{\rm{ej}}\lesssim 10 \, M_{\odot}$), intermediate CSM masses with respect to $M_{\rm{ej}}$ ($1 M_{\odot}\lesssim M_{\rm{CSM}}\lesssim 30 M_{\odot}$),  and relatively large CSM extent ($R_{\rm{CSM}}\gtrsim 10^{16}$~cm). 
	} 
	\label{Fig: Etot_shell}
\end{figure*}

In this appendix, we explore the dependence of neutrino production  in the scenario of a radially independent CSM mass distribution. We follow a similar approach to the wind-profile case discussed in Sec.~\ref{Sec. results_neutrinos_wind}. Specifically, we investigate the connection between the total energy in neutrinos  ($\mathcal{E}_{\nu+\bar{\nu}}$, see Eq.~\ref{Eq: E_nu_tot}) with  $ E_{\nu}\geq 1\, \mathrm{TeV}$. 
The results are shown in Fig. ~\ref{Fig: Etot_shell}.

We exclude from our investigation the region of the SN parameter space where the maximum achievable proton energy  is $E^\ast_{\rm{p,max}}\leq 10$~TeV. 
Additionally, we disregard parameters that lead to a shock breakout at the surface of the progenitor star ($R_{\rm{bo}}\equiv R_{\star}$), as indicated by the beige region in the contour plots. In this work, our focus is on the parameter space that results in the shock breakout occurring inside the CSM envelope. This is the first difference with the wind case, where the much higher density at smaller radii cause the shock to occur inside the wind for all the considered parameters.
Isocontours of  $E^{\ast}_{\rm{p,max}}$ (first row), the rise time $t_{\rm{rise}}$ (second row), and the bolometric peak $L_{\rm{peak}}$ (third row) are also displayed on top of the $\mathcal{E}_{\nu+\bar\nu}$ colormap in Fig.~\ref{Fig: Etot_shell}.

The dependence of $\mathcal{E}_{\nu+\bar{\nu}}$  on the SN model parameters is analogous to the wind scenario.  Indeed we see that in all panels of Fig.~\ref{Fig: Etot_shell}, $\mathcal{E}_{\nu+\bar\nu}$ increases with $M_{\rm{CSM}}$, namely with larger target proton numbers, and then saturates once the critical $\rho_{\rm{CSM}}$ is reached. Beyond such critical density, $pp$ interactions or the cooling of thermal plasma  becomes too strong, limiting the maximum achievable proton energy, and thus the neutrino outcome.
From the contour lines in each panel, analogously to the wind case, we see that the optimal configuration for what concerns neutrino production, results from large $E_{\rm{k}}$, $M_{\rm{CSM}}\gtrsim M_{\rm{ej}}$, and $R_{\rm{CSM}}$ larger as $M_{\rm{CSM}}$ increases.

We  see from  Fig.~\ref{Fig: Etot_shell} that we do not have the same  regions of the parameter space excluded as in the wind case (see Fig.~\ref{Fig: Etot_wind}) that lead to $E^\ast_{\rm{p,max}}\leq 10$~TeV. Indeed in a constant density shell the proton maximum energy has a rather different dependence especially on the radius as discussed in Appendix~\ref{App:maxEp}. This leads to overall  higher values of $E^\ast_{\rm{p,max}}$ in the parameter space, as well as the times at which they are achieved during the shock evolution. Most of the parameter space in all panels leads to  $\Delta t_{\rm{pk}}=t|_{E^{\ast}_{\rm{p,max}}}-t_{\rm{peak}} < 0$ (see Fig.~\ref{Fig:EP_MAX_cool_wind} for the wind case). This means that in the constant density scenario most of the energetic neutrinos are produced earlier than the bolometric peak. 

With respect to the wind scenario, another difference lies in  the relation between $t_{\rm{rise}}$ and $L_{\rm{peak}}$, as can be seen from the second and third row of Fig.~\ref{Fig: Etot_shell}. In the case of a constant density shell, the CSM density is considerably lower. Consequently, the shock breakout tends to occur  earlier than in the wind scenario, resulting in significantly smaller peak luminosities across a significant portion of the parameter space. 
Nonetheless, the lower CSM density leads to larger deceleration radii compared to the wind case. As a result, a  larger $M_{\rm{CSM}}$ is required to enter the decelerating regime, delaying the transition to the decreasing trend of $L_{\rm{peak}}$ with $M_{\rm{CSM}}$ in the blast-wave regime (as observed in the wind case in Fig.~\ref{Fig: T_rise,L_peak,Erad wind}).
As for $t_{\rm{rise}}$, lower CSM densities result in longer photon mean free paths, enabling faster diffusion through the CSM envelope. Furthermore, as shown in the second row of Fig.~\ref{Fig: Etot_shell}, $t_{\rm{rise}}$ increases with $M_{\rm{CSM}}$, but remains independent on $M_{\rm{ej}}$ and $E_{\rm{k}}$ for most of the parameter space. This is explained because $R_{\rm{bo}}$ is significantly smaller than $R_{\rm{ph}}$, making the diffusion time unaffected by the shock velocity.

In summary, similar to the wind scenario, large $\mathcal{E}_{\nu+\bar\nu}$ is expected for 
large SN kinetic energy ($E_{\rm{k}}\gtrsim 10^{51}$~erg), small ejecta mass ($M_{\rm{ej}}\lesssim 10 \, M_{\odot}$), and large CSM radii, $R_{\rm{CSM}}\gtrsim 10^{16}$~cm. Unlike in the wind case, a  larger range of $M_{\rm{CSM}}$ leads to comparable predictions, even if scenarios with $M_{\rm{CSM}}\gg M_{\rm{ej}}$ would  limit  neutrino production. Such parameters  imply large bolometric luminosity peak  ($ L_{\rm{peak}}\gtrsim 10^{43}$--$10^{44}$~erg) and relatively long rise times ($t_{\rm{rise}}\gtrsim 10$--$90$~days). In the shell case, large $t_{\rm{rise}}$ do not necessarily correspond to  low $\mathcal{E}_{\nu+\bar\nu}$, as it is the case for the wind scenario. Furthermore,  energetic neutrinos are produced at early times. Hence, if neutrinos should be observed from long-rising optical lightcurves relatively soon with respect to the optical peak, this might hint towards a constant density of the CSM envelope.

\bsp	
\label{lastpage}

\end{document}